\documentclass{aa}
\usepackage[varg]{txfonts}
\usepackage{enumitem}
\usepackage{graphicx}
\usepackage{natbib}
\bibpunct{(}{)}{;}{a}{}{,} 
\usepackage{color}
\usepackage{float}
\usepackage{hyperref}
\usepackage{rotating}
\usepackage{pdflscape}
\usepackage{soul}
\usepackage{subcaption}
\usepackage{comment}
\usepackage{multirow}
\usepackage{afterpage}
\usepackage{soul}

\usepackage{amsmath}	
\usepackage{amssymb}	
\usepackage{caption}
\usepackage{subcaption}
\usepackage{comment}
\usepackage{arydshln}
\usepackage{setspace}
\usepackage{tocloft}
\usepackage{float}
\usepackage{lineno}
\usepackage{rotating} 

\usepackage{xcolor}

\newcommand{\Lya}{\mbox{Ly$\alpha$}}

\newcommand{\udft}{\textsf{UDF-10}}
\newcommand{\mosaic}{\textsf{MOSAIC}}
\newcommand{\mxdf}{\textsf{MXDF}}
\newcommand{\origin}{\textsf{ORIGIN}}
\newcommand{\odhin}{\textsf{ODHIN}}
\newcommand{\nbext}{\textsf{NBEXT}}

\newcommand{\photutils}{\texttt{photutils}}
\newcommand{\gold}{\texttt{GOLD}}
\newcommand{\silver}{\texttt{SILVER}}
\newcommand{\bronze}{\texttt{BRONZE}}

\begin{document}

\title{The MUSE Extremely Deep Field:}
\subtitle{Classifying the Spectral Shapes of \Lya\ Emitting Galaxies.}

\author{Elo\"ise Vitte\inst{\ref{inst1},\ref{inst2}\thanks{e-mail: \href{mailto:eloise.vitte@gmail.com}{eloise.vitte@gmail.com}}}
\and Anne Verhamme\inst{\ref{inst1}}
\and Pascale Hibon\inst{\ref{inst2}}
\and Floriane Leclercq\inst{\ref{inst3}}
\and Belén Alcalde Pampliega\inst{\ref{inst2}}
\and Josephine Kerutt\inst{\ref{inst4}}
\and Haruka Kusakabe \inst{\ref{inst1},\ref{inst5}}
\and Jorryt Matthee\inst{\ref{inst6},\ref{inst7}}
\and Yucheng Guo\inst{\ref{inst8}}
\and Roland Bacon\inst{\ref{inst8}}
\and Michael Maseda\inst{\ref{inst9}}
\and Johan Richard\inst{\ref{inst8}}
\and John Pharo\inst{\ref{inst10}}
\and Joop Schaye\inst{\ref{inst11}}
\and Leindert Boogaard\inst{\ref{inst12}}
\and Themiya Nanayakkara\inst{\ref{inst13}}
\and Thierry Contini\inst{\ref{inst14}}
}

\institute{
Observatoire de Genève, Université de Genève, Chemin Pegasi 51, 1290 Versoix, Switzerland \label{inst1}
\and
ESO Vitacura, Alonso de Córdova 3107,Vitacura, Casilla 19001, Santiago de Chile, Chile \label{inst2}
\and 
Department of Astronomy, The University of Texas at Austin, 2515 Speedway, Stop C1400, Austin, TX 78712-1205, USA \label{inst3}
\and
Kapteyn Astronomical Institute, University of Groningen, P.O. Box 800, NL-9700 AV Groningen, the Netherlands \label{inst4}
\and
National Astronomical Observatory of Japan (NAOJ), 2-21-1 Osawa, Mitaka, Tokyo, 181-8588, Japan \label{inst5}
\and
Department of Physics, ETH Zürich, Wolfgang-Pauli-Strasse 27, Z\"urich, 8093, Switzerland \label{inst6}
\and
Institute of Science and Technology Austria (IST Austria), Am Campus 1, Klosterneuburg, Austria \label{inst7}
\and
Univ. Lyon, Univ. Lyon1, ENS de Lyon, CNRS, Centre de Recherche Astrophysique de Lyon UMR5574, 69230 Saint-Genis-Laval, France \label{inst8}
\and
Department of Astronomy, University of Wisconsin-Madison, 475 N. Charter St., Madison, WI53706, USA \label{inst9}
\and
Leibniz-Institut fur Astrophysik Potsdam (AIP), An der Sternwarte 16, 14482 Potsdam, Germany \label{inst10}
\and
Leiden Observatory, Leiden University, P.O. Box 9513, 2300 RA Leiden, The Netherlands \label{inst11}
\and
Max Planck Institute for Astronomy, K\"onigstuhl 17, D-69117 Heidelberg, Germany \label{inst12}
\and
JWST Australian Data Centre, Centre for Astrophysics and Supercomputing, Swinburne University of Technology, Victoria 3122, Australia \label{inst13}
\and
Institut de Recherche en Astrophysique et Planétologie (IRAP), Université de Toulouse, CNRS, UPS, CNES, 31400 Toulouse, France \label{inst14}
}

\date{Accepted day/month/year}


\abstract {
The Hydrogen Lyman-alpha (\Lya) line, the brightest rest-frame UV-line of high-redshift galaxies, shows a large variety of shapes which is caused by factors at different scales, from the interstellar medium to the intergalactic medium (IGM).
} 
{
This work aims to provide a systematic inventory and classification of the spectral shapes of \Lya\ emission lines to better understand the general population of high-redshift \Lya\ emitting galaxies (LAEs).
}
{
Using the unprecedented deep data from the MUSE eXtremely Deep Field (\mxdf, up to 140-hour exposure time), we select 477 galaxies observed in the $\sim$ 2.8 $-$ 6.6 redshift range, fifteen of them having a systemic redshift from nebular lines.
We develop a method to classify \Lya\ emission lines in four spectral and three spatial categories, by combining a pure spectral analysis with a narrow-band image analysis. We measure spectral properties, such as the peak separation and the blue-to-total flux ratio (B/T) for the double-peaked galaxies.
}
{
To ensure a robust sample for statistical analysis, we define two unbiased subsets, inclusive and restrictive, by applying thresholds for signal-to-noise ratio, peak separation, and \Lya\ luminosity, yielding a final unbiased sample of 206 galaxies.
Our analysis reveals that between 32\% and 51\% of the galaxies exhibit double-peaked profiles, with peak separations ranging from 150 km~s$^{-1}$ to nearly 1600 km~s$^{-1}$.
The fraction of double-peaks seems to evolve dependently with the \Lya\ luminosity, while we don't notice a severe decrease of this fraction with redshift, as expected due to the IGM attenuation at high reshift. An artificial increase of the number of double-peaks at the highest redshifts may cause the observation of a plateau instead of a decrease.
A notable amount of these double-peaked profiles shows blue-dominated spectra, suggesting unique gas dynamics and inflow characteristics in some high-redshift galaxies. The consequent fraction of blue-dominated spectra needs to be confirmed by obtaining new systemic redshift measurements.
Among the double-peaked galaxies, 4\% are spurious detections, i.e. the blue and red peaks do not come from the same spatial location.
Around 20\% out of the 477 sources of the parent sample lie in a complex environment, meaning there are other clumps or galaxies at the same redshift within a distance of 30 kpc.
} 
{
Our results suggest that the \Lya\ double-peak fraction may trace the evolution of IGM attenuation, but faintest galaxies are needed to be observed at high redshift.
We also need more data to confirm the trend seen at low redshift.  
In addition, it is crucial to obtain secure systemic redshifts for LAEs to better constrain the nature of the double-peaks. 
Statistical samples of double-peaks and triple-peaks are a promising probe of the evolution of the physical properties of galaxies across cosmic time.
}

\keywords{galaxies: high-redshift -- galaxies: formation -- galaxies: evolution -- cosmology: observations }

\maketitle


\section{Introduction}
\label{sec:1}

The Lyman-alpha (\Lya, $\lambda 1216$ \AA) line of Hydrogen, as the brightest UV-line of star-forming galaxies \citep{PartridgePeebles1967}, is a key spectral feature in the observation of high-redshift galaxies, and often the only detected signal \citep[e.g.,][]{rhoads2004,malhotra2004,Maseda2018}.  
A remarkable characteristic of this line is the wide diversity of spectral shapes that have been reported in the literature, at every redshift \citep[e.g.,][]{Kulas2012,henry2015,Yang2016,leclercq2017,kerutt2022}.
The archetypical \Lya\ shape, making it easily identifiable, is a single red asymmetric line profile \citep[e.g.,][]{Shapley2003, Tapken2007}. But over the last decade, double-peaked \Lya\ lines \citep{henry2015,hu2016, matthee2018, Songaila2018, meyer2021, hayes2021} as well as triple-peaked ones \citep[e.g.][]{vanzella2016,naidu2017,vanzella2018,Izotov2018b,rivera-thorsen2019} have been observed.
The observation of this wide diversity of line shapes has become possible thanks to the emergence of instruments such as the Cosmic Origins Spectrograph onboard the Hubble Space Telescope \citep[HST/COS,][]{Green2012} which observed low-$z$ \Lya\ emitting galaxies, the Multi-Unit Spectroscopic Explorer at the Very Large Telescope \citep[VLT/MUSE,][]{Bacon2010}, which has unveiled a large population of faint star-forming galaxies at $z = 3-6$ \citep{bacon2023}, and the Near Infrared Spectrograph onboard the James Webb Space Telescope (JWST/NIRSpec), which is already pushing the limits of \Lya\ line observation towards higher redshifts \citep{Bunker2023}. The wavelength coverage of these instruments, ranging from UV to IR, enables the scientific community to observe the evolution of galaxies over a time span of approximately 13 Gyr, using the same indicator: the \Lya\ line.

The observed diversity of spectral shapes arises from the resonant nature of the \Lya\ line \citep[see e.g.,][for a review on \Lya\ radiation transfer effects in galaxies]{dijkstra2017}. 
Therefore, depending on the sightline, the shape of the \Lya\ line profile may vary drastically \citep{blaizot2023}. The observed \Lya\ line profiles encode information on the gas velocity and its density distribution as the \Lya\ photons travel through the interstellar medium (ISM).
For double-peaks in particular, \cite{verhamme2015} suggest that the separation between the two peaks of the \Lya\ line correlates with the neutral Hydrogen column density.
Indeed, a relationship between the \Lya\ peak separation and the Lyman-continuum (LyC) escape fraction has been found empirically for low-$z$ LyC leakers \citep{Verhamme2017, Izotov2021, Flury2022}  and has been used at higher redshift on samples of \Lya\ emitting galaxies (LAEs) to select LyC leaking candidates \citep{naidu2022, kramarenko2024}. Nevertheless, \cite{kerutt2024} could not confirm this relationship with their LAE sample at $z = 3-4$.
The value of the blue-to-total flux ratio (B/T), another spectral property containing physical information, characterises the gas exchanges between the galaxy and its surrounding environment.
In simulations, expanding shells produce red-dominated \Lya\ lines \citep[i.e. B/T < 0.5,][]{Verhamme2006}, whereas blue-dominated spectra are seen when gas falls into the galaxy \citep{blaizot2023}. The brightest \Lya\ phases of galaxies seem to be outflow phases \citep{blaizot2023}.
In observations, more red-dominated spectra have been observed than blue-dominated ones \citep{Kulas2012, trainor2015}. A detailed analysis has only been done for a few blue-dominated objects \citep{mukherjee2023, furtak2022, marques-chaves2022}. Finally, the intergalactic Medium (IGM) transmission decreases with increasing redshift, preferentially suppressing flux on the blue part of the \Lya\ line \citep{Laursen2011, Garel2021, hayes2021}.
To quantify the inflow/outflow phases, constrain the duty cycle of galaxies and to study LyC leakers across different redshifts, it is important to measure the peak separation and the B/T flux ratio for a large number of galaxies, over a large redshift range.

Until now, the diversity of the \Lya\ line profile has been quantified by dedicated surveys targeting previously detected  LAEs. \cite{Kulas2012} found 30\% of double-peaks in their UV-selected galaxy sample at $z = 2-3$ and \cite{Yamada2012} found a fraction of double-peaks of 50\%  for $z = 3.1$ equivalent width selected LAEs. \cite{kerutt2022} found 33\% of double-peaks for LAEs with $z < 4$ in the MUSE-Wide blind survey \citep{urrutia2019}.
The way the double-peaks are identified among a population of LAEs introduces biases. Indeed, most of the time the double-peaks are visually identified \citep[e.g.][]{sobral2018} or a combination of a detection algorithm and a visual inspection is used \citep{Kulas2012}.
In this study we use for the first time an automatic algorithm on a blind survey without pre-selection, to quantify the diversity of the \Lya\ lines, both spectrally and spatially.

This paper is structured as follows. In Sect.~\ref{sec:2}, we describe the MUSE-Deep data and the sample selection. Sect.~\ref{sec:3} is dedicated to the description of the method developed for identifying and characterising the spectral profiles of the \Lya\ lines of our sample of galaxies. 
In Sect.~\ref{sec:4}, we present the results of the classification and the universal fraction of double-peaks. The results are discussed in Sect.~\ref{sec:5}. Finally, the summary and conclusions are given in Sect.~\ref{sec:6}.

Throughout this paper, we assume a $\Lambda$CDM cosmology with $\Omega_{m}$=0.3, $\Omega_{\Lambda}$=0.7, and H$_{0}$=67.4 km~s$^{-1}$ Mpc$^{-1}$.


\section{Data and sample definition}
\label{sec:2}

We constructed a sample of $z$-selected galaxies from the Data Release 2 (DR2) catalogue \citep[hereafter B23]{bacon2023} \defcitealias{bacon2023}{B23} of the MUSE eXtremely Deep Field (hereafter \mxdf), to identify and characterise their \Lya\ emission line $\lambda 1216$ \AA\ based on publicly available spectra\footnote{\url{https://amused.univ-lyon1.fr/}}. 

Our data set is described in Sect.~\ref{sec:21}. Sect.~\ref{sec:22} is dedicated to our sample selection and finally, the choice of spectral extraction is discussed in Sect.~\ref{sec:23}. 

\subsection{Data set}
\label{sec:21}

\begin{figure}
\centering
   \resizebox{\hsize}{!}{\includegraphics{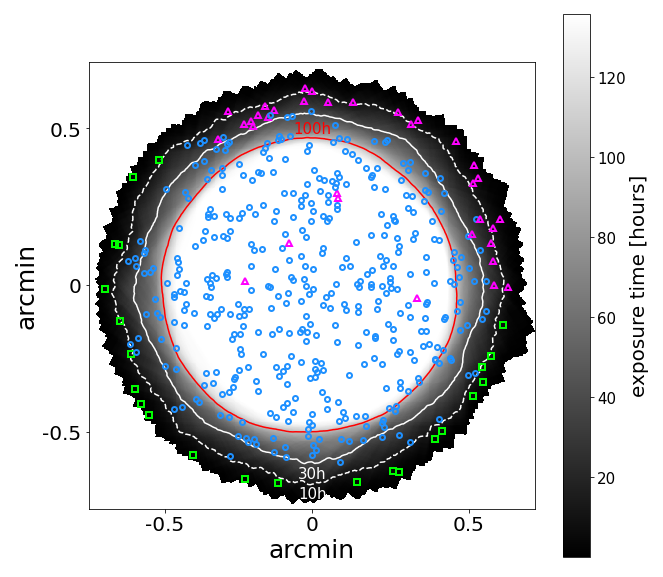}}
    \caption{Exposure time map of the \mxdf. The blue circles are the \mxdf-selected objects, the pink triangles indicate the \udft-selected targets and the green squares the \mosaic-selected objects (see Sect.~\ref{sec:22}). The field is coloured by exposure time in hours, from 0h to 140h. The red contour represents the 100h limit of the field. The solid and dashed white contours show the 30h and 10h exposure time of the \mxdf, respectively. The \udft-selected objects present in the deepest part of the field have their \Lya\ line in the AO gap. 
    }
    \label{fig:expmap}
\end{figure}

The \mxdf\ data were taken as part of the MUSE guaranteed time observations (GTO) program between August 2018 and January 2019. All observations were performed using the MUSE ground-layer adaptive optics (GLAO) mode. Details about the \mxdf\ data processing and the production of the source catalogues can be found in the survey paper \citetalias{bacon2023}.
In summary, a single pointing with the deepest exposure time of 140 hours was achieved. The final \mxdf\ field of view has a circular shape designed to minimise systematics, with the following centre coordinates: 53\textdegree.16467, -27\textdegree.78537 (J2000 FK5). The exposure time of the field exceeds 100 hours in the inner 31\arcsec\ radius, which represents an area of 0.84 arcmin$^{2}$, while it reaches around 10 hours at a radius of 41\arcsec\ (area of 1.47 arcmin$^{2}$, Fig.~\ref{fig:expmap}). The 50\% \Lya\ detection completeness, in the deepest 140-hour area, is achieved for an AB magnitude of 28.7 (F775W) at $z = 3.2 - 4.5$.
The field overlaps partially with the other MUSE Hubble Ultra Deep Fields \citep[HUDF,][]{Bacon2017}  of 31-hour (1 $\times$ 1 arcmin$^{2}$) and 10-hour (3 $\times$ 3 arcmin$^{2}$) total exposure times, called \udft\ and \mosaic, respectively. The 50\% completeness is reached at 26.5 mag and 25.5 mag ($z \sim 4$) in HST F775W, in \udft\ and \mosaic, respectively. The \mxdf\ is two orders of magnitude deeper than \udft.

The \mxdf\ data cover a wavelength range from 4700 \AA\ to 9350 \AA\, excluding an adaptive optics (AO) gap from 5800 \AA\ to 5966.25 \AA\ due to the notch filter that blocks the bright light from the four sodium laser guide stars. Both the \udft\ and \mosaic\ fields were observed without AO and therefore do not have any AO gap in wavelength. In the \mxdf, the full-width at half maximum (FWHM) of the Moffat point spread function (Moffat PSF, \citeauthor{Moffat1969} \citeyear{Moffat1969}) is 0.6\arcsec\ at 4700 \AA\ and 0.4\arcsec\ at 9350 \AA. The line spread function (LSF) of the \mxdf\ is constant in the field and larger in the outer parts. The mean \mxdf\ LSF over the wavelength range is 2.6 \AA\ which corresponds to a LSF of $\approx$ 150 km~s$^{-1}$ at $z = 3$ and $\approx$ 90 km~s$^{-1}$ at $z = 6$ (see details in Sect.~4.2.2 of \citetalias{bacon2023}).

The data reduction process performed on the DR2 survey is similar to the one for the Data Release 1 survey and is described in \cite{Bacon2017}. However, some improvements have been made, especially in the sky-subtraction process (Appendix~B of \citetalias{bacon2023}).
The output data of the \mxdf\ is a 3D datacube with dimensions of 3721 $\times$ 470 $\times$ 470 pixels, meaning that for each spatial pixel of $0\farcs2 \, \times \, 0\farcs2$, a spectrum divided into 3721 pixels of 1.25~\AA\ width is available.
On top of the signal datacube, a variance datacube is provided. 

\subsection{Catalogue and data set selection}
\label{sec:22}

The DR2 catalogue compiles all sources observed in the three MUSE Ultra Deep Fields: \mosaic\ (9 arcmin$^{2}$), \udft\ (1 arcmin$^{2}$) and \mxdf\ (1.47 arcmin$^{2}$) (see Fig.~2 of \citetalias{bacon2023}). This catalogue contains 2221 sources, from nearby galaxies ($z < 0.25$) to high-redshift galaxies (up to $z$ $\approx$ $6.64$).
The \Lya\ line is observable by MUSE in the redshift range $z = 2.87 - 6.64$.  While a total of 1308 DR2 sources have their redshift in this range, not all have a strong \Lya\ line. Indeed, the \Lya\ emission can be faint, not detected, or in absorption.

For this study, we decided to restrict our analysis to the deepest data available, i.e. the \mxdf\ field, to develop our method. The analysis of \mosaic\ and \udft\ will be the subject of future work.
The first step in building our parent sample consists of selecting the objects detected in the 1.47 arcmin$^{2}$ \mxdf\ field of view, with a spectroscopic redshift above 2.87. This reduces the number of galaxies from 1308 to 504.
We did use the \mosaic\ or \udft\ data sets of a given source if (i) the data depth is deeper than in the \mxdf\ data set at the location of the source (i.e., higher signal-to-noise ratio, hereafter S/N) which corresponds to 53 sources, or (ii) the \Lya\ line falls in the \mxdf\ AO gap (five sources, see Fig.~\ref{fig:expmap}).  
After a cautious examination, 26 objects were removed from the parent sample because of misclassification in the DR2 catalogue. In addition, a last source has also been removed because the segmentation map used to extract the spectrum is on another galaxy. 

The distribution of our 477 galaxies is as follows (see Table~\ref{table:data set_zconf}):

\begin{itemize}[noitemsep,topsep=0pt]
    \item 419 galaxies have their deepest observations in the \mxdf\ data set (blue circles in Fig.~\ref{fig:expmap}).
    \item 34 sources have their deepest observations in the \udft\ data set (with a maximum of 30 hours of observations, pink triangles in Fig.~\ref{fig:expmap}).
    \item  24 have their deepest observations in the \mosaic\ field (green squares in Fig.~\ref{fig:expmap}), accumulating a total of 10 hours of integration.
\end{itemize}

\noindent In the DR2 catalogue, the redshift confidence parameter, ZCONF, indicates the reliability of the redshift solution (see \citetalias{bacon2023}). This parameter can range from 0 to 3, 0 being the least confident redshift solution, and 3 the most secure one (see Sect.~5.3.7 of \citetalias{bacon2023}).
Generally, when a source with a \Lya\ line is assigned a ZCONF = 2 or 3, it means that the S/N of the \Lya\ line is above 5 or 7, respectively. If a source can be matched to an HST counterpart, it adds confidence to the detection. In addition, if the photometric redshift of \cite{rafelski2015} is reliable and matches with the MUSE redshift, this also increases the confidence of the redshift measurement. The presence of other nebular lines (such as \ion{C}{IV}, \ion{He}{II}, [\ion{O}{II}] and [\ion{O}{III}]) with reliable S/N in the spectrum also increases the confidence level of the redshift.
If a source is assigned a redshift confidence level of 1, there can be various reasons. It could be either due to a low S/N of the lines, noisy observations, other potential redshift solutions, or the presence of additional lines with good enough S/N in the spectrum but unexplained by the proposed redshift solution.
In the specific case of low S/N spectra with a single emission line, to avoid misclassifications, \citetalias{bacon2023} estimated the expected fraction of \Lya\ and [\ion{O}{II}] emitters as other emission lines are much less likely because of the small accessible volume. 
The likelihood of misclassified lines is less than 10\%, even for the faintest galaxies (i.e. fainter than F775W = 28.5, \citetalias{bacon2023}).
Out of the 477 sources:
\begin{itemize}[noitemsep,topsep=0pt]
    \item 170 have a secure redshift, ZCONF = 3
    \item 161 sources have a confident redshift, ZCONF = 2
    \item 146 sources do not have a reliable spectroscopic redshift solution, ZCONF = 1
\end{itemize}

\noindent To be inclusive, we did not make any selection based on the redshift confidence level. Only three of the sources with ZCONF = 1 passed the selection to be part of the unbiased sample (Sect.~\ref{sec:413}). Two of them are single-peak and one is a double-peak. Hence their inclusion in the unbiased parent sample does not impact our findings on the universal fraction of double-peaked LAEs, as detailed in Sect.~\ref{sec:431}. 
We still verified the impact of the ZCONF=1 objects on the different distributions shown in Fig.~\ref{fig:Histos_all} and confirmed that their inclusion or rejection does not modify these distributions. 

\begin{table*}
\caption{Data sets and redshift confidence (ZCONF) levels of our parent sample (see Sect.~\ref{sec:22})}              
\label{table:data set_zconf}      
\centering                                      
\begin{tabular}{l r r r r}          
\hline\hline                        
data set (\# of sources) & ZCONF 1 & ZCONF 2 & ZCONF 3  \\    
\hline                                    
    \mosaic\ (24) & 9 & 9 & 6 \\ 
    \udft\ (34) & 8 & 15 & 11 \\ 
    \mxdf\ (419) & 129 & 137 & 153 \\ 
    ALL (477) & 146 & 161 & 170 \\ 
\hline                                             
\end{tabular}
\end{table*}

\subsection{Spectral extraction}
\label{sec:23}

Since MUSE is an integral-field unit spectrograph, the produced data format is a 3D datacube, from which spectra can be extracted in different ways. \cite{bacon2023} used three different methods to extract the spectra of their MUSE sources (see their Sect.~5.8.1) that we briefly describe below:
\begin{itemize}[noitemsep,topsep=0pt]
    \item \odhin\ (Optimal Deblending of Hyperspectral ImagiNg, \cite{Bacher2017}, see also Appendix~C of \citetalias{bacon2023}): HST-prior spectral extraction. This is a source de-blending method using HST broadband images, three different HST catalogues, and the MUSE datacube. However, this method misses flux if the \Lya\ emission extends far beyond the detection in the HST broadband \citep[e.g,][]{leclercq2017}. \odhin\ is not optimised for LAE detection as it is blind to any source undetected by HST.
    \item \origin\ \citep[detectiOn and extRactIon of Galaxy emIssion liNes,][]{Mary2020}: blind source detection software performing optimal spectral extraction. This method automatically detects spatial-spectral emission signatures and is particularly efficient at detecting faint \Lya\ emitters in the MUSE datacube. The produced spectra are optimized in S/N. It has been proven to be the method with the most secure identification of sources and the most reliable estimate of purity.  
    \item \nbext\ (Narrow-Band EXTraction method, \citetalias{bacon2023}): an alternative to the \origin\ extraction method. It is used for a few objects in particular cases, especially when \origin\ is not able to distinguish between two sources. This method of extraction usually provides lower S/N. We refer the reader to \citetalias{bacon2023} for more detailed explanations.
\end{itemize}

\noindent Each source has been assigned a reference extraction in \citetalias{bacon2023}, preferentially \origin\ for high-$z$ galaxies because of the higher S/N of the spectra but, in case of strong contamination, \odhin\ was preferred, even if it does not capture all the \Lya\ emission.
For this study focusing on the classification of the \Lya\ line profile of high-$z$ galaxies, we need spectra with the best S/N so we use the extracted spectra selected in \citetalias{bacon2023} as the reference ones (hereafter, REF spectra). 

We will now investigate and describe the diversity of the \Lya\ profiles among our parent sample of 477 objects.


\section{Identification and Classification Method}
\label{sec:3}

\begin{figure}
\centering
    \resizebox{\hsize}{!}{\includegraphics{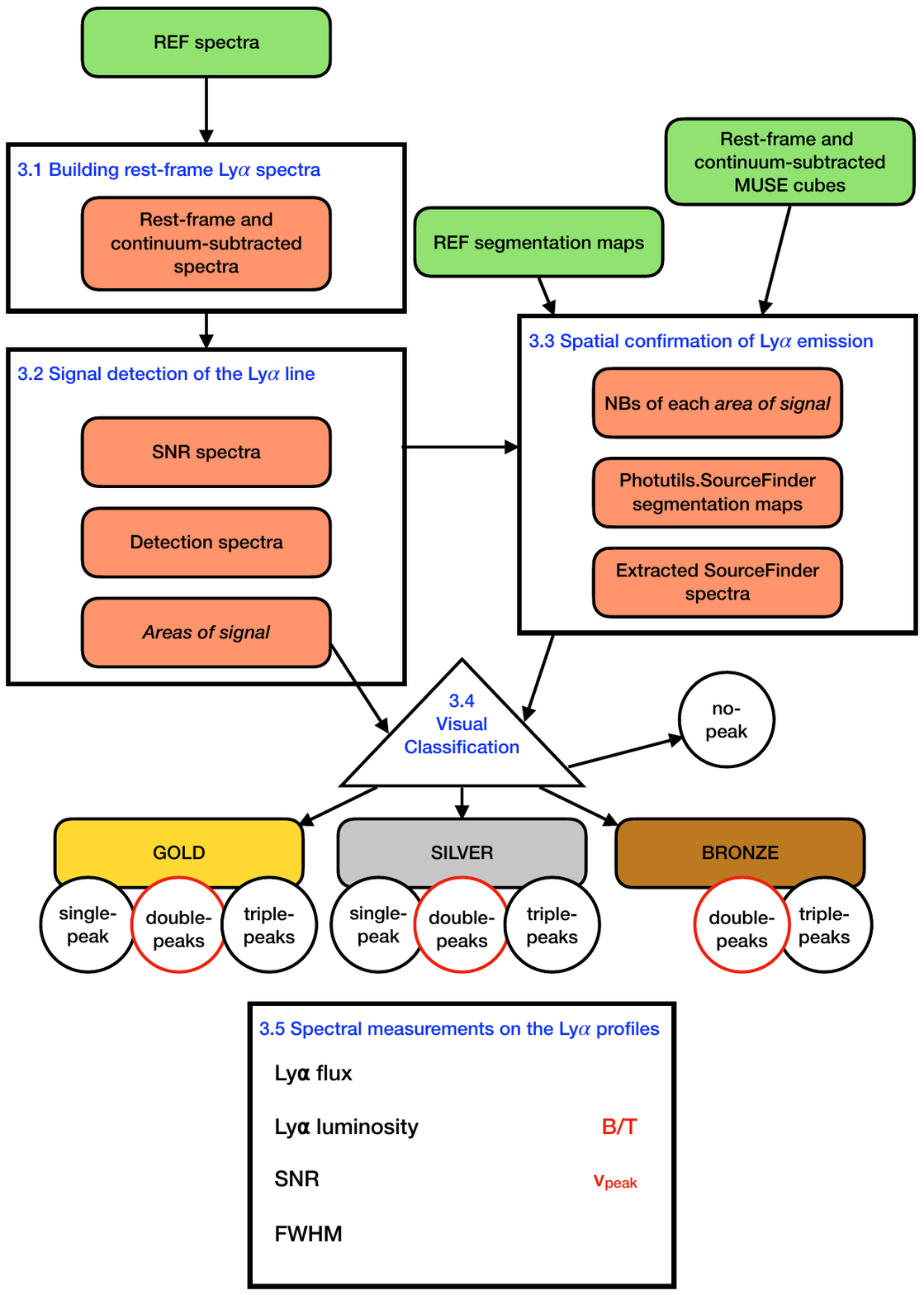}}
    \caption{Flowchart of the method. The input data are indicated by green boxes at the top of the figure. Black boxes refer to the different steps of the method. The corresponding sections of the paper are shown in blue. The orange boxes inside the black ones indicate the output data.
    The visual classification is shown by a triangle (see Sect.~\ref{sec:34}). The \gold, \silver\ and \bronze\ categories are illustrated by a gold, a grey and a brown rectangle, respectively. The spectral shapes of the \Lya\ emission are indicated by circles, black for single- and triple-peaks and red for double-peaks. The box at the bottom presents the \Lya\ spectral parameters (black) and the spectral parameters measured for double-peak objects (red), detailed in Sect.~\ref{sec:35}.}
    \label{fig:flowchart}
\end{figure}

When exploring the spectral properties of the \Lya\ emitters observed with MUSE, we noticed the wide diversity of profiles in the \Lya\ line. We thus developed a method aiming at characterising their spectral profiles. This section is dedicated to its step-by-step description. An overview of the method can be seen in Fig.~\ref{fig:flowchart}.

\subsection{\texorpdfstring{Building rest-frame \Lya\ spectra}{Building rest-frame Lya spectra}}
\label{sec:31}

To characterise the whole \Lya\ emission of each source in our parent sample, we consider a $\pm$ 2000 km~s$^{-1}$ window around the \Lya\ line peak wavelength. Indeed, previous studies have shown multi-peaked \Lya\ lines with peak separations larger than 1000 km~s$^{-1}$ \citep{kerutt2022}, even reaching values around 1800 km~s$^{-1}$ \citep{Kulas2012}. 
We use the following formula to convert wavelengths into velocities (in km~s$^{-1}$): 

\begin{equation}
\rm V = \frac{\rm \lambda}{(1+z)\times1215.67 \, \text{\AA}} \times \frac{c}{1000},
\end{equation}

\noindent where $\lambda$ is the wavelength in vacuum in \AA, $z$ is the redshift given by the \Lya\ wavelength in the DR2 catalogue and $c$ is the speed of light in vacuum in m~s$^{-1}$ given by astropy.constants\footnote{\url{https://docs.astropy.org/en/stable/constants/index.html}}. Before using this equation, our data are corrected using the air-to-vacuum correction function\footnote{\url{https://mpdaf.readthedocs.io/en/latest/api/mpdaf.obj.vactoair.html\#mpdaf.obj.vactoair}} of the Python package \texttt{MPDAF} \citep[MUSE Python Data Analysis Framework,][]{piqueras2019}.
 
Among the 477 sources of the parent sample, fifteen have their \Lya\ line located near a MUSE wavelength edge, which limits the information we have to correctly process them:
\begin{itemize}[noitemsep,topsep=0pt]
    \item 5 objects have their \Lya\ line at the blue edge of the instrumental spectral range ($\lambda \approx $ 4700 \AA, corresponding to $z \approx 2.8$)
    \item 10 of them have their \Lya\ line close to the AO gap (5800 \AA\ -- 5966.25 \AA, corresponding to redshifts between 3.8 to 3.9)
\end{itemize}
We still run the classification on these 15 objects, but we flag them as potentially missing information for reliable classification (Sect.~\ref{sec:413}).

The extracted spectra provided by the \origin, \odhin\ or \nbext\ methods are not continuum subtracted and some sources show a stellar continuum. To accurately analyse the \Lya\ emission line using our 
classification method, we subtract the continuum following the protocol used in \cite{kusakabe2022} adapted to the DR2 data (top left black box in Fig.~\ref{fig:flowchart}). In brief, the continuum spectra are estimated from spectral median filtering of the original spectra in a 100-pixel spectral window.

At the end of this step, we have built a continuum-subtracted rest-frame spectrum, centred on the DR2 spectroscopic redshift, over the same rest-frame wavelength window, for all 477 objects of the parent sample. 

\subsection{\texorpdfstring{Signal detection of the \Lya\ line}{Signal detection of the Lya line}}
\label{sec:32}

The goals of this step are, for each source, to (i) generate bootstrapped spectra to obtain their S/N spectra, (ii) run the classification method on the S/N spectra to obtain a detection spectrum, and (iii) determine the \emph{areas of signal} by applying a threshold of N = 40 on the detection spectrum, as summarised in Fig.~\ref{fig:flowchart}. 

\begin{figure*}[!h]
\begin{minipage}{0.7\textwidth}
\centering
    \resizebox{\hsize}{!}{\includegraphics{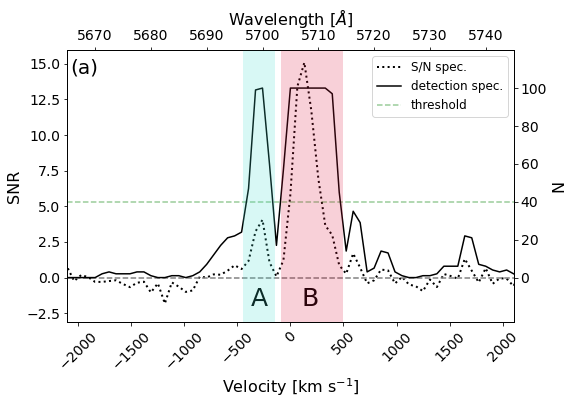}}
\end{minipage}
\begin{minipage}{0.3\textwidth}    
     \centering
         \centering
             \includegraphics[width=\textwidth]{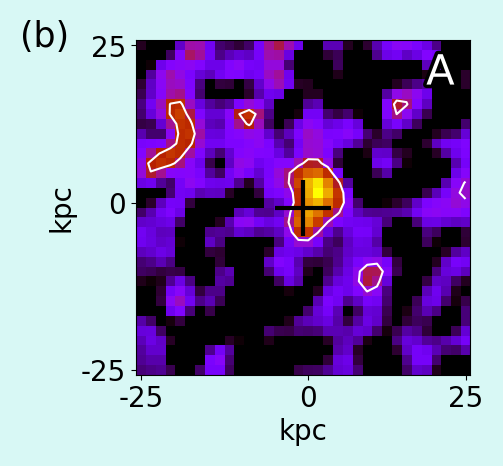}
     \hfill
         \centering
         \includegraphics[width=\textwidth]{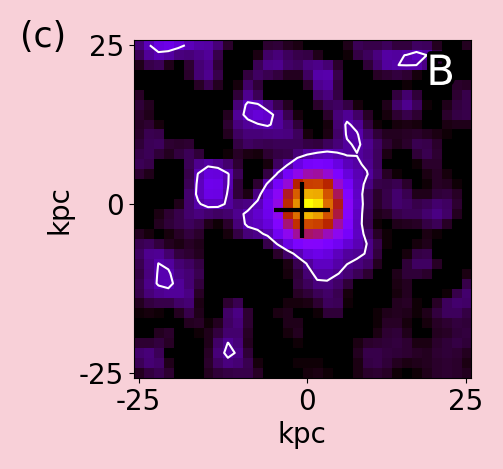}
\end{minipage}
\begin{minipage}{0.5\textwidth}    
     \centering
         \centering
             \includegraphics[width=\textwidth]{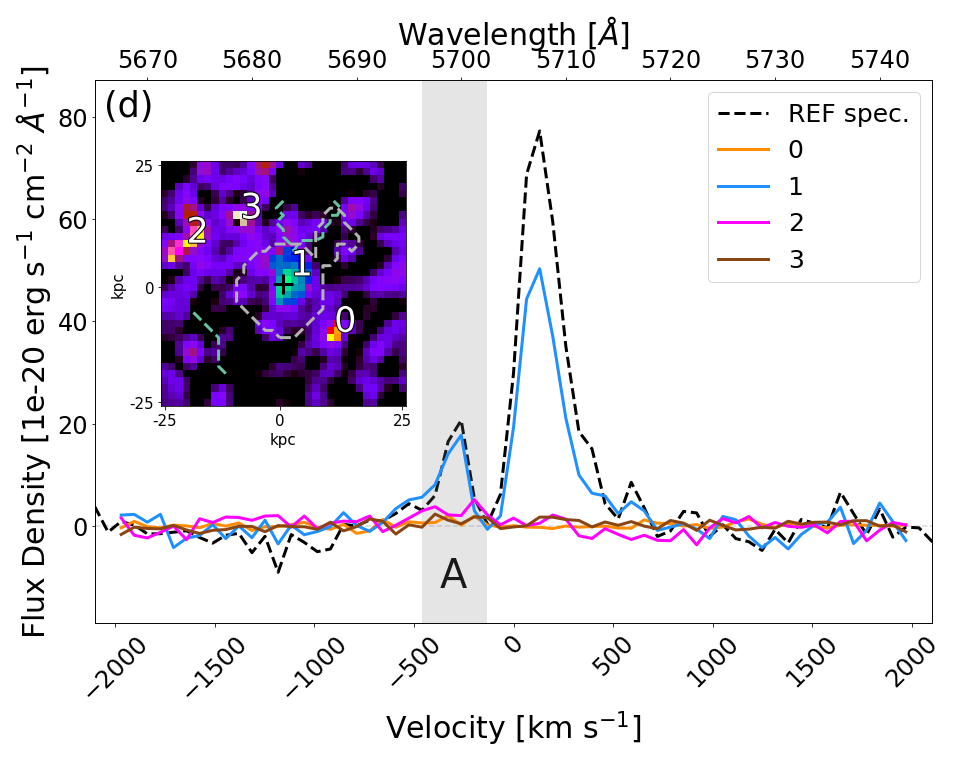}
\end{minipage}
\begin{minipage}{0.5\textwidth}    
     \centering
         \centering
             \includegraphics[width=\textwidth]{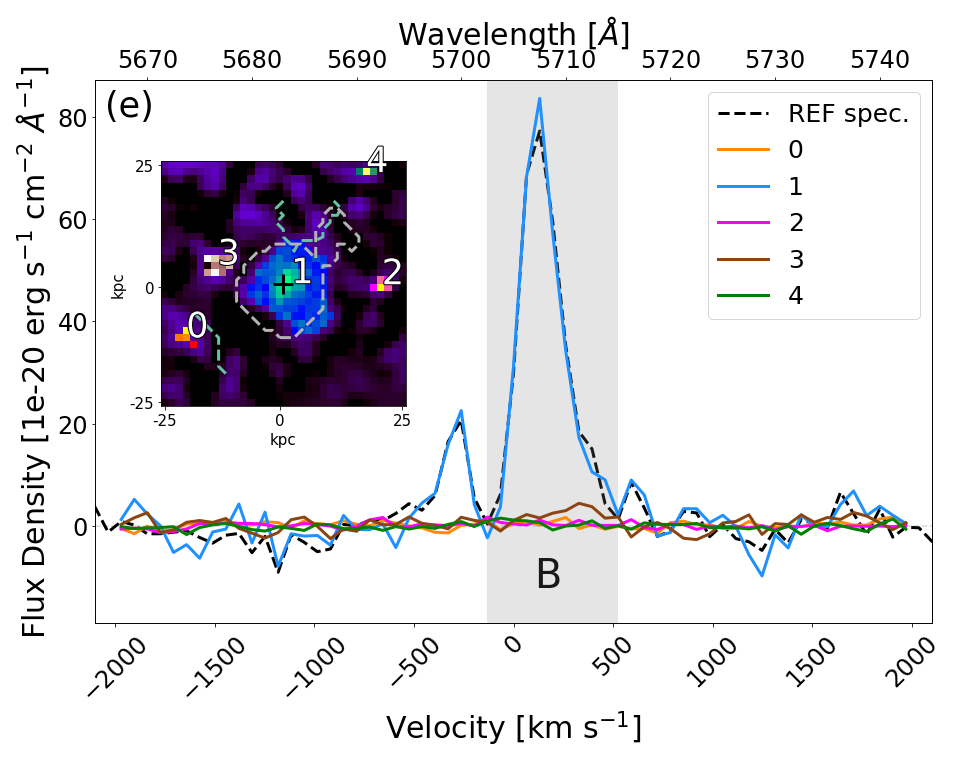}
\end{minipage}
    \caption{ID 3240, double-peak, \gold\ category. \textit{(a)}: Example of a detection spectrum in black obtained from the 100 realizations of the original spectrum (Sect.~\ref{sec:32}). The S/N spectrum of the original spectrum is plotted in dotted black. The horizontal dashed green line shows our detection threshold of N = 40 defining \emph{areas of signal}. The blue A and red B shaded areas correspond to the \emph{areas of signal} obtained by the crossing of the detection spectrum with the threshold line of N = 40. 
    \textit{(b)} and \textit{(c)}: 50 $\times$ 50 kpc$^2$ NB images of the \emph{area of signal} A and the \emph{area of signal} B, respectively. The black cross represents the centre coordinates of the source. The white contours correspond to a S/N level of 2. 
    \textit{(d)} and \textit{(e)}: Spectra extracted from the \photutils.\texttt{SourceFinder} segmentation maps (Sect.~\ref{sec:33}). The colour of each spectrum matches the NB image segmentation map colour inserted in the plot. Only the blue spectrum extracted from the blue \texttt{SourceFinder} segmentation map number 1 contributes to the \Lya\ line. The reference spectrum (Sect.~\ref{sec:23}) in black dashed line is displayed as a reference. The dashed contours on the NB images represent the reference segmentation maps used in \citetalias{bacon2023}. The grey dashed contour is the segmentation map of the targeted source, the green ones correspond to other objects.
    }
    \label{fig:ID3240_GOLD}          
\end{figure*}

For each source, we generate 100 realizations of the 1D-extracted spectrum where the flux value of each pixel is randomly drawn from a normal distribution centred on its original flux value with a standard deviation given by its variance.
Then, we determine their S/N spectra by dividing the flux by the square root of the variance provided by the variance datacube. For each of the one hundred S/N spectra, a peak is detected when the S/N value per pixel is above 1 for at least 2 adjacent pixels.
Assuming that the noise of adjacent pixels is not correlated, the false-positive detection rate for an S/N = 2 pixel or for 2 adjacent pixels with an SN = 1 is similar ($\sim$ 2.3\% and $\sim$ 2.5\%).
We choose to use a criterion of SN $\geq$ 1 for at least 2 adjacent pixels to avoid noise spikes of 1-pixel width and to allow the detection of faint (S/N $\approx$ 2) \Lya\ emission.
For each pixel of the 100 generated spectra, a value of 1 is assigned if this pixel belongs to a
peak, otherwise, a value of 0 is set. For every generated spectrum, we thus have a list of 0 and 1.
We then sum these lists for the 100 spectra and we obtain a single list with values between 0 and 100, resulting in the detection spectrum (see solid line in panel \textit{(a)} of Fig.~\ref{fig:ID3240_GOLD}). The detection spectrum reaches 100 when a pixel belongs to a peak in 100\% of the generated spectra and 0 when a pixel never belongs to a peak.

Finally, to select the final \emph{areas of signal}, we apply a detection threshold of N = 40 on those detection spectra (solid spectrum above the horizontal dashed green line in panel \textit{(a)} of Fig.~\ref{fig:ID3240_GOLD}) and each pixel above this threshold is considered as real signal (shaded areas in panel \textit{(a)} of Fig.~\ref{fig:ID3240_GOLD}). The value of N = 40 has been chosen empirically and is discussed in App.~\ref{ap:A}.\\
The number of peaks of each \Lya\ line corresponds to the number of \emph{areas of signal}. 

In certain instances, an \emph{area of signal} might encompass a double-peak with some flux present in the trough between the peaks. Consequently, the detection spectrum in the \emph{signal area} does not fall below N = 40 (for example see Fig.~\ref{fig:ID399_satellite} in Sect.~\ref{sec:51}). To detect such double-peaks, we conduct a flux variation analysis.
This analysis consists in comparing the flux value pixel per pixel, starting from the highest value of the \emph{area of signal}, i.e. the peak of the line. The method analyses the flux variation on both sides of the peak until reaching the edges of the \emph{area of signal}. If the flux value of the pixel $n+1$ is higher than the value of the pixel $n$, then a secondary peak is detected by the method, as illustrated in panel \textit{(a)} of Fig.~\ref{fig:ID399_satellite}. This flux variation analysis is performed when there is only one \emph{area of signal} detected or when, after the spatial confirmation of \Lya\ emission, only one \emph{area of signal} remained. In the latter case, the flux variation analysis is performed and if a secondary peak is detected, this object goes through the spatial confirmation of \Lya\ emission phase to confirm this secondary peak.

\subsection{\texorpdfstring{Spatial confirmation of \Lya\ emission}{Spatial confirmation of Lya emission}}
\label{sec:33}

The nature of the MUSE data allows us to investigate the spatial distribution of the \Lya\ peaks detected as described in the previous section (Sect.~\ref{sec:32}). To confirm that the emission is coming from the targeted source and to discard "fake" multi-peaks caused by neighbours, we proceed to a narrow-band image inspection (see right side of Fig.~\ref{fig:flowchart}).

For each \emph{area of signal}, we extract from the MUSE cubes a  $50 \times 50$ kpc$^2$ NB image with the same width as the  \emph{area of signal}, so 2 pixels minimum, as shown in panels \textit{(b)} and \textit{(c)} of Fig.~\ref{fig:ID3240_GOLD}. 
We then use the \photutils.\texttt{SourceFinder} class to detect and deblend sources in our images, using a threshold of 2-$\sigma$ and a minimum number of connected pixels of 3 for the signal search. Panels \textit{(d)} and \textit{(e)} of Fig.~\ref{fig:ID3240_GOLD} show the NB images extracted from the two \emph{areas of signal} A and B. We also notice the detections of \texttt{SourceFinder} through the segmentation maps in coloured pixels, numbered from 0.
As explained in Sect.~\ref{sec:23}, we use reference spectra from \citetalias{bacon2023} in this study. Those spectra are extracted from the MUSE datacubes using specific segmentation maps for each type of extraction (see Fig.~18 of \citetalias{bacon2023}). 
Thus, the location of the peak of the \texttt{SourceFinder} segmentation map (i.e. the brightest coloured pixels of \texttt{SourceFinder} segmentation map "1" in Fig.~\ref{fig:ID3240_GOLD}) inside the reference segmentation map of \citetalias{bacon2023} confirms the detection of the peak from the \emph{area of signal} and has been taken into consideration during the spectral extraction. 
When the \texttt{SourceFinder} segmentation maps are located outside the reference segmentation map, they are not considered because they do not contribute to the MUSE reference spectrum. 

This NB image verification is a crucial step to discard false peak detections, i.e., peaks without coherent spatial counterparts.
We visually inspect the NB images and the extracted spectra to determine whether the \emph{area of signal} is emitting inside the segmentation map used to extract the reference spectrum in \citetalias{bacon2023} or if it is noise or simply if the emission is too faint to be detected by \texttt{SourceFinder} with our criteria.
In the example given in Fig.~\ref{fig:ID3240_GOLD}, panels \textit{(d)} and \textit{(e)}, we see that only one \texttt{SourceFinder} segmentation map is located inside the reference segmentation map. These blue segmentation maps labelled with number 1 give the blue spectra on each panel and have the same shape as the reference spectrum in black. We also notice that the other \texttt{SourceFinder} segmentation maps (labelled 0, 2 and 3) are located outside the reference segmentation map and their spectra are noise. The choice of using a threshold of 2-$\sigma$ is discussed in App.~\ref{ap:A}.

Out of 708 \emph{areas of signal}, 89 have been discarded (i.e., 12.6\%) because their \texttt{SourceFinder} segmentation map was outside the reference segmentation map or because nothing was detected on the NB image.
This false-positive detection rate value is higher than the 2.5\% false-positive detection rate expected in an ideal case, i.e. when the source is isolated, without any contamination, and when the pixels are not correlated (see Sect.~\ref{sec:32}).
This higher false-positive detection rate is thus expected given that the noise in MUSE data is correlated \citep{weilbacher2020, bacon2023} and galaxies are rarely isolated in \mxdf.

This step enables us to classify the galaxies in three different categories following the spatial distribution of their \Lya\ emission, as explained in the following Sect.~\ref{sec:34}.

\subsection{Final classification}
\label{sec:34}

This final step aims at providing a trustworthy classification of each \Lya\ line profile, taking into account the spatial distribution.
 
\begin{figure}
\centering
   \resizebox{\hsize}{!}{\includegraphics{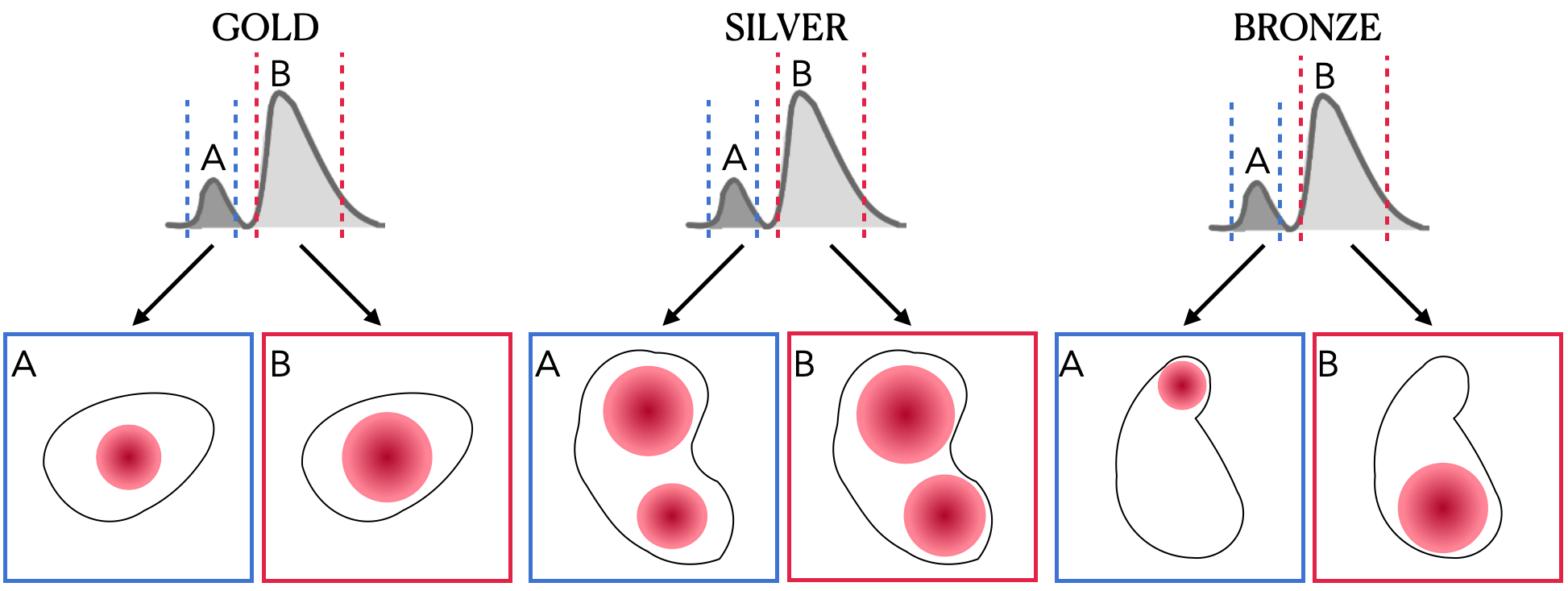}}
    \caption{\textit{Left panel:} Illustration of a \gold\  double-peaked galaxy. The emission of each peak (red circle) is located inside the segmentation map (black contour) at the same spatial location. \textit{Middle panel:} \silver\ category. Each peak of the double-peaked \Lya\ line is coming from two different locations inside the segmentation map. \textit{Right panel:} Case of a \bronze\ galaxy. The blue peak (A) is emitted in a certain region and the red peak (B) is emitted in a different region of the segmentation map.}
    \label{fig:categories_G_S_B}
\end{figure}

We divide the parent sample into three qualitative spatial categories:
\begin{itemize}
    \item \gold: when all \emph{areas of signal} have only one emission located inside the reference segmentation map, and at the same spatial location. 
    Illustration of a \gold\ galaxy can be seen in panels \textit{(d)} and \textit{(e)} of Fig.~\ref{fig:ID3240_GOLD} and in left panel of Fig.~\ref{fig:categories_G_S_B}.
    \item \silver: when the peak emission arises from several distinct regions contained within the segmentation map, as illustrated in the middle panel of Fig.~\ref{fig:categories_G_S_B} and shown in App.~\ref{ap:B}.
    \item \bronze:  for double- and triple-peaked galaxies when peaks are emitted distinctly from different visual locations of the reference segmentation map. An example of this category is shown in Fig.~\ref{fig:categories_G_S_B}, right panel.
\end{itemize}

\noindent Discarding \emph{areas of signal} after spatial confirmation (Sect.~\ref{sec:33}) results in the declassification of double-peaks and triple-peaks into single-peaks or double-peaks. That way, the spectral classification obtained in Sect.~\ref{sec:32} is different from this classification taking into account NB images. The fact that the classification changes highlights how not doing the NB verification can lead to misclassifying the line profiles. 
This also highlights the importance of a careful analysis of the spatial distribution of the \Lya\ emission to unveil the complexity of its spatial emission on top of the spectral one (see Sect.~\ref{sec:51} and Sect.~\ref{sec:52}).

\subsection{\texorpdfstring{Spectral measurements on the \Lya\ profiles}{Spectral measurements on the Lya profiles}}
\label{sec:35}

For all categories, the following parameters are measured on the reference spectra:

\begin{itemize}[noitemsep,topsep=0pt]
    \item \Lya\ flux (F$_{\rm Ly\alpha}$): obtained by summing the pixel flux values inside each \emph{area of signal}. If the \Lya\ line is composed of several \emph{areas of signal}, the flux of each \emph{area of signal} is summed to obtain the total \Lya\ flux of the line. The error on the flux is determined using the variance of the spectra (i.e. the square root of the sum of the variance).
    \item \Lya\ luminosity (L$_{\rm Ly\alpha}$): derived from the total flux of the \Lya\ line (F$_{\rm Ly\alpha, \rm tot}$) and the spectroscopic redshift. The error on the \Lya\ luminosity is derived from the \Lya\ flux errors.
    \item Integrated S/N: \Lya\ flux divided by the square root of the total variance over all the \emph{areas of signal} and in between.
    \item FWHM: we locate the maximum of the \emph{area of signal}, we take half of this maximum value and find the two points on the spectrum corresponding to it, rounded to the nearest pixel. The error estimations of the FWHM values are determined by generating one hundred noise spectra (as explained in Sect.~\ref{sec:32}) for each source. Then, the standard deviation of the one hundred FWHM measurements is used to estimate the error on the FWHM. We note that the FWHM of each \Lya\ peak is measured.
\end{itemize}
For the double-peak objects, specific spectral measurements are determined. 
We call the peak on the left, i.e. at the bluer wavelength, the blue peak and the one on the right, i.e. at redder wavelengths, the red peak (but see Sect.~\ref{sec:53} for discussion):

\begin{itemize}[noitemsep,topsep=0pt]
    \item Blue-to-total flux ratio (B/T): the integrated flux of the blue peak is divided by the total integrated flux of the \Lya\ line. The error on the flux ratio is derived using the \Lya\ flux errors.
    \item Peak separation ($\rm v_{\rm sep}$): the velocity distance between the maxima of the blue and red peaks. The unit is km~s$^{-1}$. The error on the peak separation is determined by generating one hundred noise spectra for each double-peak source. Then, the standard deviation of the one hundred peak separations is taken to be the error on the peak separation.
\end{itemize}
Tables containing the measurements described above can be found in App.~\ref{ap:C}.

\subsection{Test of the method on background spectra}
\label{sec:36}

In order to test the reliability of the classification method on the detection of noise spikes as signal, we applied the method on 100 spectra at $z = 3$ and 100 at $z = 6$. 
We selected 100 random places in the continuum-subtracted \mxdf\ cube and extracted the spectra in a 2 arcseconds diameter circular aperture. 
We selected two zones on each spectrum, one zone of $\pm$ 2000 km~s$^{-1}$ window around the \Lya\ line peak wavelength as if it were emitted at $z = 3$, and another one as if the \Lya\ line peaked at $z = 6$.
We then reproduced the exact same procedure described in Sect.~\ref{sec:32} and Sect.~\ref{sec:33}. A summary of the results is given in Table~\ref{table:bkg_spectra}.

At $z = 3$, over the whole selection of background spectra, 91 of them do not have any peak detected by the method. 
First, 52 do not pass the first part of the method (Sect.~\ref{sec:32}). No peaks are detected on the detection spectra.
For the remaining 48 background spectra, 12 of them are discarded throughout the NB image verification (Sect.~\ref{sec:33}). 
Finally, the last step of extracting the spectra of the \texttt{SourceFinder} segmentation maps enables to eliminate 27 background spectra. 
Only 6 background spectra remain with a noise peak detected, 1 with two noise peaks. One background spectrum is clearly contaminated by a neighbouring galaxy and one last spectrum presents a clear emission line (the SNR of the peak of the line peaks above 5).
In total, the method detects noise peaks in 7\% of the background spectra at $z = 3$.

At $z = 6$, over the whole selection of background spectra, 87 of them do not have any peak detected by the method. 
During the first step (Sect.~\ref{sec:32}), 24 spectra do not pass. 
Then, 16 more background spectra are discarded throughout the NB image verification. 
Finally, the last step of extracting the spectra of the \texttt{SourceFinder} segmentation maps enables to eliminate 47 background spectra.
Overall, the spectra are much noisier than at $z = 3$.
A total of 11 background spectra remain with a noise peak detected, 1 with two noise peaks. One background spectrum is clearly contaminated by a neighbouring galaxy.
In total, the method detects noise peaks in 12\% of the background spectra at $z = 6$.

As a result of this exercise, we estimate that around 10\% of the peaks detected are spurious peaks but we decide to not make the method more selective: 10\% of spurious peaks ($\sim$ 76 over 760 detected peaks) is the price to pay for an inclusive method. We will keep this in mind when discussing trends in Sect.~\ref{sec:4} and Sect~\ref{sec:5}. 

\begin{table*}
    \caption{Results obtained by the method applied on 100 background spectra (see Sect.~\ref{sec:36})}              
    \label{table:bkg_spectra}      
    \centering                                      
    \begin{tabular}{l |c|c|c|c|c|c}          
    \hline\hline                        
    \multirow{2}{*}{Redshift} &
      \multicolumn{3}{c}{No peak detected} &
      \multicolumn{3}{|c}{Peak(s) detected} \\
    & detection spectrum (\ref{sec:32}) & NB image verification (\ref{sec:33}) & extracted spectra (\ref{sec:33}) & 1 noise peak & 2 noise peaks & others \\
    \hline                                   
    $z = 3$ & 52 & 12 & 27 & 6 & 1 & 2 \\    
    $z = 6$ & 24 & 16 & 47 & 11 & 1 & 1 \\
    \hline                                            
    \end{tabular}
\end{table*}

\subsection{Method limitations}
\label{sec:37}

The method described has been designed to analyse the MUSE data (spectrum $+$ image) and the thresholds used have been fixed following the characteristics of the \mxdf.
We discuss our threshold choices and characterise their impact on our results in App.~\ref{ap:A}.

In special cases when one of the two peaks is located on the tail of the main \Lya\ peak, such as illustrated in Fig.~\ref{fig:ID399_satellite} below, the spectral measurements done on the faintest peak contain part of the flux of the main \Lya\ peak. Deciphering the flux of each peak is a difficult task as we do not know the proportion of flux that is belonging to a peak or another in each pixel. In this paper, we do not attempt to deblend the peaks and we assume that each \emph{area of signal} contains only one peak. 
Modeling such configurations could help estimate the flux ration belonging to each peak but is out of the scope of this paper.


\section{Results}
\label{sec:4}

The first part of this section is devoted to the definition of an unbiased sample, cleaned from observational limitations.
Then, we describe the physical parameter distributions of the unbiased sample regarding their \Lya\ line shapes. 
Finally, we determine the universal fraction of double-peaks and consider its evolution with \Lya\ luminosity and redshift.

\subsection{Unbiased samples definitions}
\label{sec:41}

\subsubsection{Classification of the parent sample}
\label{sec:411}

As explained in the previous section, we classified our galaxies in two steps. 
We first assign a spectral classification to ease the comparison with previous classifications that were done on spectroscopic data only \citep{Yamada2012, Kulas2012, hashimoto2015}. We then present our final classification refined by inspecting the NB images.

\noindent \textbf{Spectral classification} \\
The distribution of the different \Lya\ line shapes emergent from the signal detection of the \Lya\ line (Sect.~\ref{sec:32}), over the full parent sample, is:
\begin{itemize}[noitemsep,topsep=0pt]
    \item No-peak: \Lya\ line is not detected, i.e., not distinguishable from the noise. Seven objects fall in this category, representing $\approx 1.5\%$ of the parent sample. Three of them are \Lya\ absorbers \citep[one of them, MID 103, is presented in][]{kusakabe2022}. Six have a low confidence (ZCONF=1) redshift \citepalias{bacon2023}.
    \item Single-peak: 155 objects, 32.5\% of the parent sample is composed of single-peaked galaxies.
    \item Double-peaks: 271 have a double \Lya\ line. The proportion of double-peaks among the parent sample is 57\%. This is the most common category.
    \item Triple-peaks: 44 objects have three peaks, representing  $\approx$ 9\% of the parent sample.
\end{itemize}

\begin{figure*}
\centering
   \resizebox{\hsize}{!}{\includegraphics{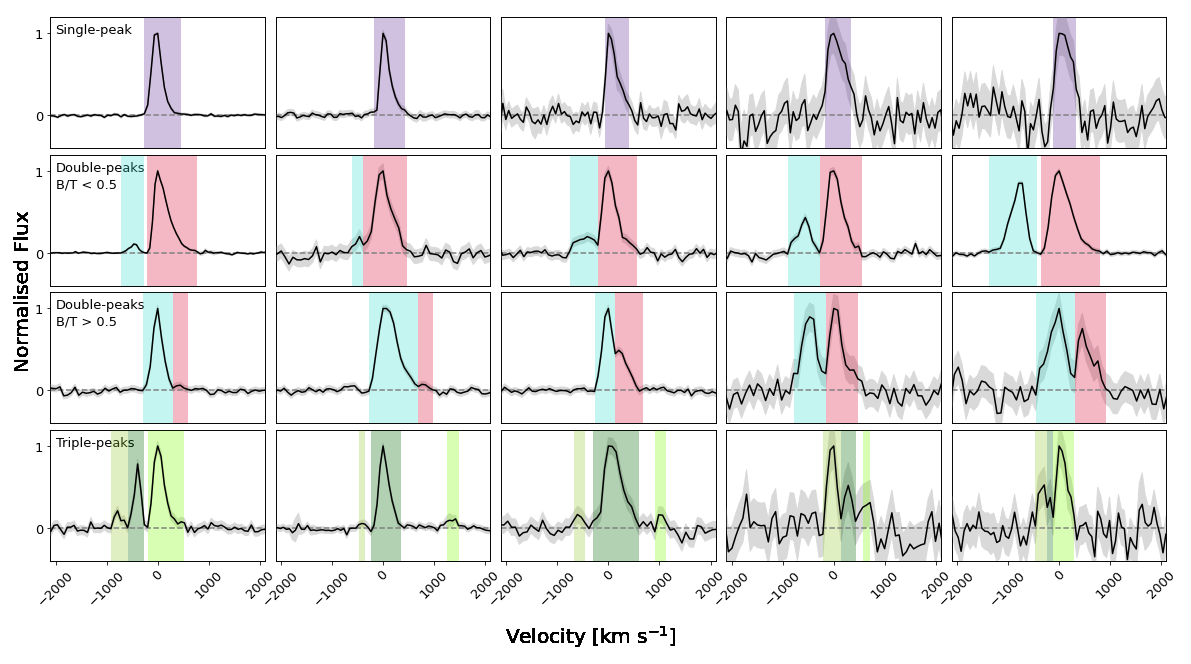}}
    \caption{Examples of the different spectral shape categories (Sect.~\ref{sec:41}), except the "No-peak" category (see App.~\ref{ap:D}). \textit{First row}: spectra of single-peak galaxies. \textit{Second and third rows}: spectra from sources belonging to the double-peak category, red and blue dominated spectra, respectively. \textit{Last row}: spectra showing triple-peak \Lya\ lines.
    We note that the \textit{y}-axis shows the flux normalised to the maximum of each line.}
    \label{fig:ex_lya_line}
\end{figure*}

\noindent \textbf{Final classification} \\
The final step of the classification is the visual classification (Sect.~\ref{sec:34}) performed on each object of the parent sample taking into consideration the spatial distribution of the \Lya\ emission peaks (Sect.~\ref{sec:33}). 
The final classification is shown in Table~\ref{table:final_classification}. \\

\begin{table}[h!]
\caption{Overview of the final classification}              
\label{table:final_classification}     
\centering                                     
\begin{tabular}{c c c c | c}         
\hline
\hline
 & \gold\  & \silver\ & \bronze\ & Total \\
\hline                       
Single-peak &  190 & 8 & -- & 198 (41\%) \\
Double-peaks &  210 & 28 & 10 & 248 (52\%) \\
Triple-peaks &  19 & 2 & 1 & 22 (5\%) \\
No-peak &  -- & -- & -- & 9 (2\%)\\
\hline     
Total & 419 & 38 & 11 & 477 \\
 & (88\%) & (8\%) & (2\%) &  \\
\hline
\end{tabular}
\end{table}

\noindent Two objects have been added to the No-peak category, both with a ZCONF of 1 (i.e. redshift measurement not reliable). The spectral classification being inclusive, hence not fully optimised for individual cases, their NB image shows noise spikes inside the reference segmentation map resulting in extracted spectra not showing any significant peak. 
Spectral examples of each category are shown in Fig.\ref{fig:ex_lya_line}, except for the nine "No-peak" objects, shown in Fig.~\ref{fig:parent_sample_0peak}. All the single-peaks, double-peaks, and triple-peaks are shown in App.~\ref{ap:D}. \\

\subsubsection{Observational limitations}
\label{sec:412}

Our study aims at quantifying the diversity of LAE spectral shapes, and in particular, to give a universal fraction of double-peaked spectra among a population of LAEs. However, our parent sample suffers from several observational limitations, preventing us from detecting double-peaks accurately depending on their spectral characteristics (e.g. close peak separations or extreme B/T values, are harder to detect). In Sect.~\ref{sec:431}, we attempt to determine the universal fraction of double-peaked LAEs based on a sample cleaned from the observational biases presented below. We also do not take into consideration double-peaked \bronze\ objects since the two peaks are coming from two different spatial locations, i.e. the double-peak \Lya\ line is not produced by radiation transfer processes (see Sect.~\ref{sec:34}). We will thus only consider the \gold\ and \silver\ double-peaks to determine the universal double-peak fraction.

\begin{figure}[!h]
\centering
   \resizebox{\hsize}{!}{\includegraphics{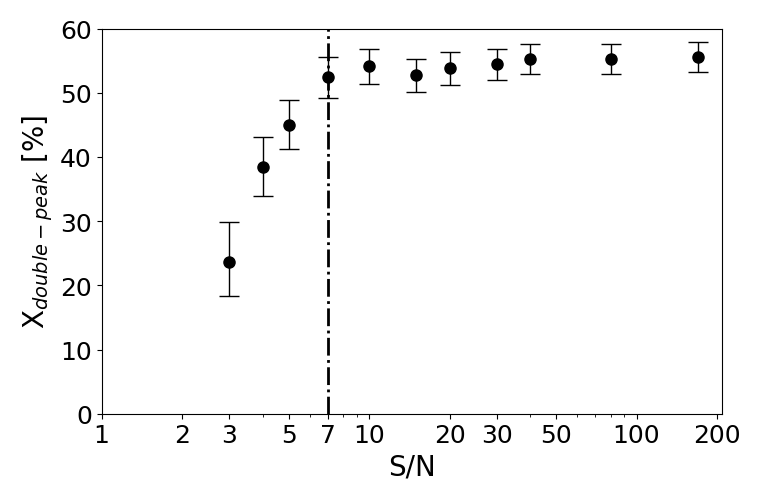}}
    \caption{Cumulative fraction of double-peaks with increasing S/N. The vertical dash-dotted line represents the S/N cut used to prevent missing the detection of double-peaks at lower S/N as the fraction of double-peaks drops below this value.}
    \label{fig:Xdp_vs_SN}
\end{figure}

\noindent\textbf{Minimum S/N of the line}\\
Our ability to detect double peaks is expected to strongly depend on the S/N. We show on Fig~\ref{fig:Xdp_vs_SN} the cumulative fraction of double-peaked objects with increasing S/N. Above SNR = 7, the fraction of detected double-peaks reaches a plateau of 53\%, but for \Lya\ lines with SNR < 7, the fraction of detected double-peaks is lower (at SNR = 5, the fraction is 42\%). A minimum S/N of 7 is therefore required to be able to detect most of the double-peaks.
We discuss below in this section the consequences of this cut on measurable B/T flux ratios. 

\begin{figure*}[!h]
\centering
    \resizebox{\hsize}{!}{\includegraphics{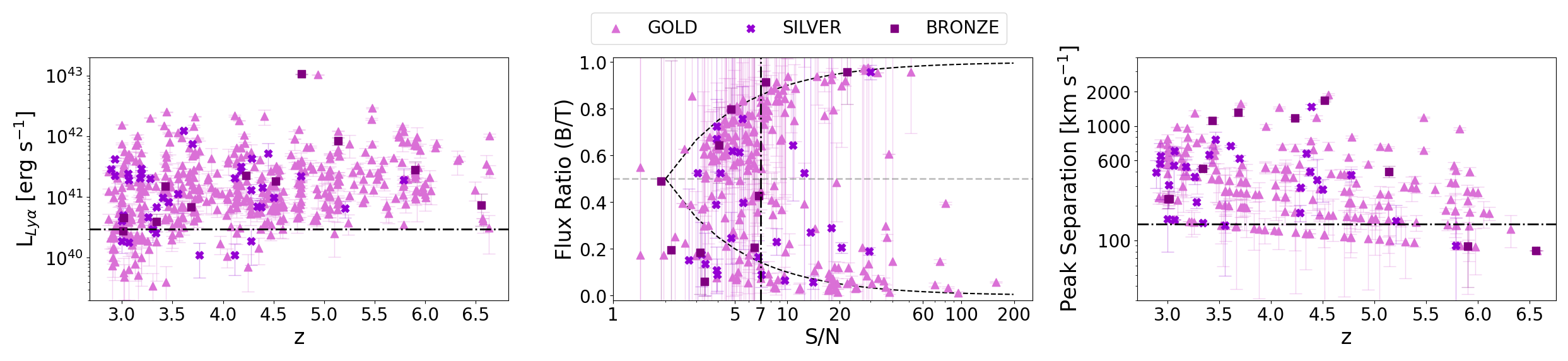}}
    \caption{\gold, \silver\ and \bronze\ objects are represented by triangles, crosses, and squares symbols, respectively. 
    \textit{Left}: \Lya\ luminosity as a function of redshift for the parent sample. The horizontal black dash-dotted line represents the luminosity cut ($L_{Ly\alpha} = 3 \times 10^{40} $ [erg s$^{-1}$]) used to mitigate the redshift dependence. 
    \textit{Middle}: B/T flux ratio as a function of S/N (calculated as described in Sect.~\ref{sec:35}) for the whole double-peak sample. B/T $\approx$ 0 objects can be seen at the beginning of Fig.~\ref{fig:dp_sample_1}. They have a very small blue peak. The horizontal grey dashed line at B/T = 0.5 marks the blue-dominated versus red-dominated dividing line. The vertical black dash-dotted line represents the S/N cut (S/N = 7) made to balance the demands of being able to detect extreme B/T and keeping a representative sub-sample of double-peaks. The dashed lines mark boundaries of 1/(S/N) < B/T < 1 - 1/(S/N).
    Three objects have a B/T > 1 because the total flux is smaller than the flux in the blue peak because the continuum is negative between the peaks (see the last three spectra in Fig.~\ref{fig:dp_sample_3}).
    \textit{Right}: Peak separation as a function of redshift for the whole double-peak sample. The horizontal black dash-dotted line represents the v$_{\rm sep}$ cut (v$_{\rm sep}$ = 150 km~s$^{-1}$) used to account for the spectral resolution of MUSE.
    For both the \Lya\ luminosity and the v$_{\rm sep}$, the observational sensitivity and spectral resolution curves are not straight lines. The ones shown in the plots are our selection limits.}
    \label{fig:All_Obs_Lim}
\end{figure*}

\noindent\textbf{Minimum \Lya\ luminosity}\\
Because of redshift dimming, only the brightest galaxies can be observed among the population of most distant galaxies, but thanks to the depth of the \mxdf, we detect much fainter LAEs than in previous GTO surveys (\citetalias{bacon2023}) which allows us to investigate the evolution of \Lya\ properties with redshift with better statistics. We show in the left panel of Fig.~\ref{fig:All_Obs_Lim} the \Lya\ luminosity distribution of our parent sample as a function of redshift. The lowest luminosity reached at $z > 6.5$ is $L_{Ly\alpha} = 3 \times 10^{40} $ erg s$^{-1}$. Applying a \Lya\ luminosity cut at $L_{Ly\alpha} = 3 \times 10^{40} $ erg s$^{-1}$ enables us to minimise the redshift dependence on the luminosity. 
A total of 414 LAEs match this \Lya\ luminosity constraint, corresponding to about 86\% of the parent sample.\\
\noindent We should also keep in mind that the MUSE + VLT total efficiency rapidly declines after $7500-8000 \: \text{\AA}$ which makes it more challenging to detect the highest redshift objects\footnote{\url{https://www.eso.org/sci/facilities/paranal/instruments/muse/inst.html}}. 

\noindent\textbf{Detectable range of B/T flux ratio}\\
Our ability to detect B/T ranges is expected to strongly depend on the S/N (calculated as described in Sect.~\ref{sec:35}) of the \Lya\ spectra, as illustrated on the middle panel of Fig.~\ref{fig:All_Obs_Lim}. In principle, we can detect extreme B/T values only for spectra with high S/N, or, for a given S/N, we can measure B/T values in the range: 1/(S/N) < B/T < 1 - 1/(S/N). 
We choose empirically a S/N of 7, allowing the detection of double-peaks in the range 0.15 < B/T < 0.85, illustrated by a black dash-dotted vertical line in the middle panel of Fig.~\ref{fig:All_Obs_Lim}, as a compromise between being able to detect extreme B/T and keeping a statistically significant sub-sample. The number of galaxies having a total S/N above 7 is 200, i.e. 46\% of the parent sample.

\noindent\textbf{Minimum peak separation}\\
The MUSE spectral resolution varies with wavelength, ranging from $\approx$ 150 km~s$^{-1}$ to $\approx$ 90 km~s$^{-1}$, at the blue edge (4700 \AA) and the red end (9350 \AA), respectively. Hence, a narrow blue peak will be less contrasted and possibly not detected at low-$z$ while the same peak could be identified at higher redshift, where the resolution is at its maximum\footnote{\url{https://www.eso.org/sci/facilities/paranal/instruments/muse/inst.html}}.  
We see the smallest peak separation values evolving with redshift: when the redshift increases, the minimum peak separation decreases, from $\approx$ 150 km~s$^{-1}$ at $z \approx 3$ to $\approx 90$ km~s$^{-1}$ at $z > 6$. We clearly see the instrumental and redshift effects on the minimum measurable peak separation. 
The minimum peak separation measurable at all redshifts is v$_{\rm sep}$ = 150 km~s$^{-1}$. The fraction of \gold\  and \silver\ double-peak LAEs having a \Lya\ peak separation above this value represents $\sim$ 76\% of the double-peak sample (186/248).

\subsubsection{Unbiased samples}
\label{sec:413}

If we take into account the observational limitations presented in the previous section (see Sect.~\ref{sec:412}), the parent sample is reduced to 214 objects. These galaxies have a \Lya\ luminosity above L$_{Ly\alpha}$ = 3 $\times$ 10$^{40}$ erg s$^{-1}$ and a S/N above 7. Among these 214 sources, 108 have a double-peaked \Lya\ line with a peak separation above 150 km~s$^{-1}$. These double-peaked LAEs are either \gold\ or \silver\ as we discarded the \bronze\ ones due to the nature of their \Lya\ emission (Sect.~\ref{sec:34}). 
We also removed eight sources that have their \Lya\ line close to a spectral edge (either the AO gap or the blue edge of the MUSE spectral range, see details in Sect.~\ref{sec:31}), including two double-peaked objects, due to the uncertainties on their line shapes.
From 214, the sample is reduced to 206 objects, 105 of them being double-peaked.
The 206 remaining galaxies form the unbiased sample ($U$) and the 105 double-peak objects are called the inclusive unbiased double-peak sample ($UDP_{I}$), as summarised in Table~\ref{table:normal_vs_unbiased_samples}.

We also define a restrictive unbiased double-peak sample $UDP_{R}$ in which only double-peaks with a significant secondary peak are kept as double-peaks, the ones discarded being considered as single-peaks (only in this section and they are considered as double-peaks in the rest of the paper).
As described at the end of Sect.~\ref{sec:32}, some double-peaks are contained in only one \emph{area of signal} and flux variation analysis is applied on it to detect such double-peaks. This method being very basic and including no conditions on the width or the strength of the secondary peak in order to be inclusive, 145 objects in the double-peak sample (i.e., over 248, see Table~\ref{table:normal_vs_unbiased_samples} for reference) have been identified with this flux variation analysis. In the $UDP_{I}$, this number rises to 61. \\
In order to test the significance of the secondary peaks detected with the flux variation analysis, we implemented the following condition:

\begin{equation}
    \Delta F = F_{peak} - F_{trough} > \sigma_{\Delta F} = \sqrt{\sigma_{F_{peak}}^2 + \sigma_{F_{trough}}^2},
\end{equation}

 $F_{peak}$ is the flux value at the maximum of the secondary peak, $F_{trough}$ is the flux value of the trough between the two peaks and $\sigma_{\Delta F}$ represents the uncertainty on $\Delta F$. Fluxes are in 1e-20 erg s$^{-1}$ cm$^{-2}$ $\AA^{-1}$.
We applied this condition on the $UDP_{I}$ and obtain a reduced number of double-peaks of 66, i.e. $UDP_{R}$. With this condition, only the obvious double-peaks remain. The discarded double-peaks are thus considered as single-peaks and stay in the unbiased sample $U$. An overview of the samples described above can be seen in Table~\ref{table:normal_vs_unbiased_samples}.

The inclusive and restrictive samples will be used to get an upper and a lower limit of the fraction of double-peaks, respectively (see Sect.~\ref{sec:43}).

\begin{table}
\caption{Overview of the samples used in this paper and the number of sources contained in each of them. See Sect.~\ref{sec:413} for details about the unbiased sample $U$, the inclusive unbiased double-peak sample $UDP_{I}$ and the restrictive unbiased double-peak sample $UDP_{R}$.}          
\label{table:normal_vs_unbiased_samples}      
\centering                    
\begin{tabular}{c c}          
\hline\hline
Name of the Sample & \# of sources \\
\hline                       
parent sample & 477 \\
double-peak sample & 248 \\
\hline
$U$ & 206 \\
$UDP_{I}$ & 105  \\
$UDP_{R}$ & 66  \\  
\hline                                             
\end{tabular}
\end{table}

\subsection{Physical parameter distributions}
\label{sec:42}

In this section, we describe the distributions of the different physical properties measured on the \Lya\ lines of the unbiased sample $U$, presented in Fig.~\ref{fig:Histos_all}. For a better comparison, the distributions show the unbiased double-peak samples $UDP_{I}$ and $UDP_{R}$ as defined in Sect.~\ref{sec:413} and the unbiased sample without the unbiased double-peak samples, which will be called $U\smallsetminus\left\{ UDP_{I} \right\}$ and $U\smallsetminus\left\{ UDP_{R} \right\}$ hereafter.
    
\subsubsection{\texorpdfstring{\Lya\ luminosity distribution}{Lya luminosity distribution}}
\label{sec:421}
    
Panel \textit{(a)} of Fig.~\ref{fig:Histos_all} shows the \Lya\ luminosity distribution for the $U\smallsetminus\left\{ UDP \right\}$ and $UDP$ for both the restrictive (in orange) and inclusive (in black) samples. The \Lya\ luminosity of the samples ranges from log(L$_{Ly\alpha}$ [erg s$^{-1}$]) = 40.57 to  43.03.
Thanks to the unprecedented depth of the \mxdf\ data, our sample reaches one order of magnitude lower in terms of \Lya\ luminosity compared to the LAE sample of \cite{kerutt2022} containing MUSE-Wide and MUSE-Deep (\mosaic\ and \udft) LAEs.
The two samples $U\smallsetminus\left\{ UDP_{I} \right\}$ (dotted black histogram) and $U\smallsetminus\left\{ UDP_{R} \right\}$ (dashed orange histogram) show a similar distribution, with more objects in the $U\smallsetminus\left\{ UDP_{R} \right\}$ as the number of double-peaks in $UDP_{R}$ is smaller than in $UDP_{I}$.
For the unbiased double-peak samples, $UDP_{R}$ in solid orange and $UDP_{I}$ in solid black, the distributions also show similarities. They span over the same luminosity range. The mean value of $UDP_{R}$ is log(L$_{Ly\alpha}$ [erg s$^{-1}$]) = 41.83 while the mean value of $UDP_{I}$ is log(L$_{Ly\alpha}$ [erg s$^{-1}$]) = 41.78. Despite the condition applied to get the restrictive unbiased double-peak sample $UDP_{R}$, the \Lya\ luminosity distribution of $UDP_{R}$ and $UDP_{I}$ show similar trends.
Due to the strong constraint on \Lya\ luminosity applied on the parent sample to obtain the unbiased samples, the distributions of the $U\smallsetminus\left\{ UDP \right\}$ and the $UDP$, for both the inclusive and restrictive samples, are very similar, especially their mean \Lya\ luminosity.
Nevertheless, the luminosity distributions of the $U\smallsetminus\left\{ UDP_{I} \right\}$ and the $U\smallsetminus\left\{ UDP_{R} \right\}$ samples tend to slightly peak at a fainter luminosity than their respective unbiased double-peak samples $UDP_{I}$ and $UDP_{R}$.

\subsubsection{Full width at half maximum distribution}
\label{sec:422}
    
The FWHM is broadened by radiation transfer effects and has been proposed as a proxy for the peak shift of the \Lya\ line that can be used to recover systemic redshift when only \Lya\ is detected \citep{verhamme2018}. 
Panel \textit{(b)} of Fig.~\ref{fig:Histos_all} displays the FWHM distributions of the peak of the \Lya\ line with the strongest flux for the different unbiased samples.
The two distributions $U\smallsetminus\left\{ UDP_{I} \right\}$ and $U\smallsetminus\left\{ UDP_{R} \right\}$ peak between 200 and 300 km~s$^{-1}$ while the two double-peak samples, $UDP_{I}$ and $UDP_{R}$, peak between 300 and 400 km~s$^{-1}$. Double-peaked \Lya\ lines tend to have wider profiles than the other \Lya\ lines of the unbiased samples.
The mean FWHM values of $UDP_{I}$ (332 $\pm$ 9 km~s$^{-1}$) and $UDP_{R}$ (347 $\pm$ 11 km~s$^{-1}$) are similar, taking into account the errors. These measurements are consistent with other results obtained for LAEs observed with MUSE \citep{kerutt2022,leclercq2017}.

\subsubsection{Blue-to-total flux ratio distribution}
\label{sec:423}

Concerning the unbiased double-peak samples only, we measure the B/T flux ratio (see Sect.~\ref{sec:35}) where the distributions of $UDP_{I}$ and $UDP_{R}$ are shown in panel \textit{(c)} of Fig.~\ref{fig:Histos_all}. 
For the inclusive unbiased double-peak sample $UDP_{I}$, a bit more than half of the sample (54\%, 57/105) is red peak dominated, i.e., B/T < 0.5, while the other half is blue peak dominated.
We observe 82 objects with extreme B/T values as defined in \citet{blaizot2023}, i.e., between 0 < B/T $\leq$ 0.1 and 0.6 $\leq$ B/T < 1, and 23 objects between B/T = [0.1 $-$ 0.6]. We use the term "extreme" for B/T > 0.6 because of the rareness of such objects in the literature.
We show in the middle panel of Fig.~\ref{fig:All_Obs_Lim} the B/T flux ratio versus the S/N. The extreme B/T values have high S/N, which seems to indicate that the shape of this distribution is not due to noise.
This distribution is surprising since so far, very few blue-peak-dominated galaxies have been reported in the literature, from LAEs \citep{Kulas2012,wofford2013,erb2014,trainor2015,izotov2020,kerutt2022, furtak2022,marques-chaves2022,mukherjee2023} and for other types of sources such as AGNs or extended \Lya\ nebulae \citep{martin2015,vanzella2017,ao2020,daddi2021,li2022}. 
Blue-dominated spectra are also rare according to simulations, as described in \cite{blaizot2023}. Indeed, they found that less than 20\% of the \Lya\ lines are blue-dominated in their work. We discuss our high B/T values in Sect.~\ref{sec:53}.
For the restrictive unbiased double-peak sample $UDP_{R}$, we observe a drastic decrease of the number of blue peak dominated: from 48 for $UDP_{I}$ to 15 for $UDP_{R}$. The fraction of the sample which is red peak dominated is 23\%, which is close to the fraction measured in \cite{blaizot2023}.
For B/T flux ratios between 0.1 and 0.6, the distributions of $UDP_{R}$ and $UDP_{I}$ remain the same.
For extreme B/T values, the distributions are different, $UDP_{R}$ having much less of this kind of values. This difference is explained by the condition applied to get $UDP_{R}$, as explained in Sect.~\ref{sec:413}. We discarded double-peaks with a non significant secondary peak compared to the trough between the two peaks. By definition, double-peaked \Lya\ lines with extreme B/T values tend to have more non significant secondary peaks compared to the double-peaks with intermediate B/T values ([0.1 $-$ 0.6]).
    
\subsubsection{Peak separation distribution}
\label{sec:424}
    
\begin{figure*}[h]
\centering
   \resizebox{\hsize}{!}{\includegraphics{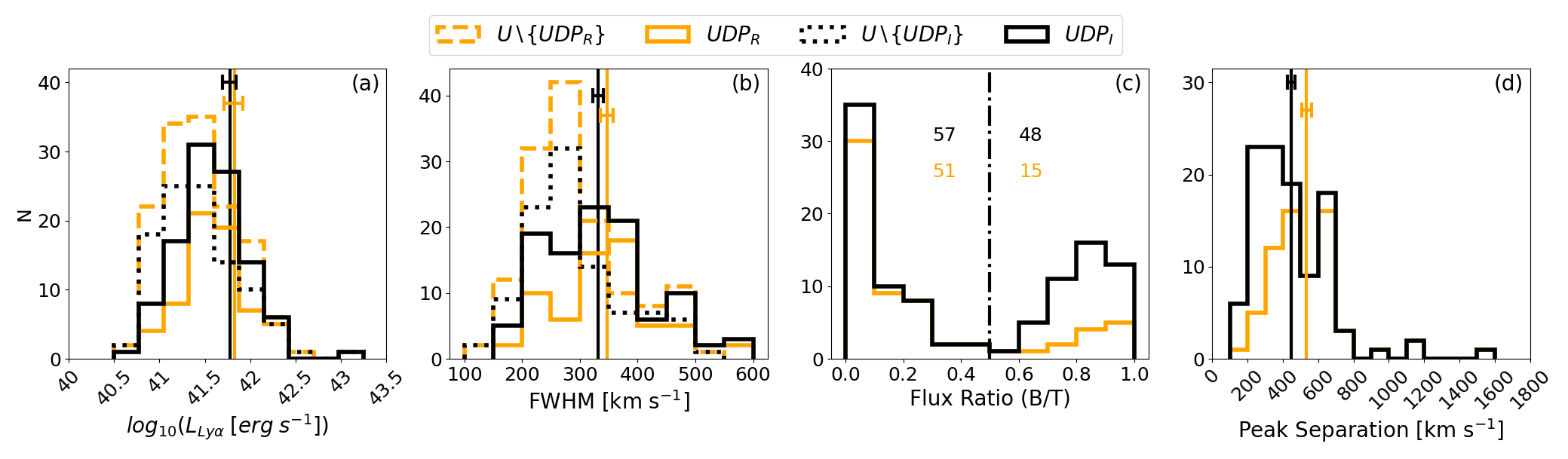}}
    \caption{\textit{(a)}: Logarithmic \Lya\ luminosity distribution. 
    \textit{(b)}: FWHM distribution of the peak of the \Lya\ line with the strongest flux. 
    \textit{(c)}: B/T distribution measured on the \Lya\ emission lines of the double-peaked objects. B/T = 0.5 is represented by a black dash-dotted line. The numbers written in black and orange correspond to the number of galaxies having a B/T value below 0.5 (N $=$ 57 for $UDP_{I}$ and N $=$ 51 for $UDP_{R}$) and above 0.5 (N $=$ 48 for $UDP_{I}$ and N $=$ 15 for $UDP_{R}$). 
    \textit{(d)}: Peak separation distribution of the double-peak samples.
    In orange are represented the restrictive samples: $U\smallsetminus\left\{ UDP_{R} \right\}$ in dashed lines and $UDP_{R}$ in solid line. In black are represented the inclusive samples: $U\smallsetminus\left\{ UDP_{I} \right\}$ in dotted lines and $UDP_{I}$ in solid line.}
    \label{fig:Histos_all}
\end{figure*}

\begin{figure*}[h]
\centering
   \resizebox{0.9\hsize}{!}{\includegraphics{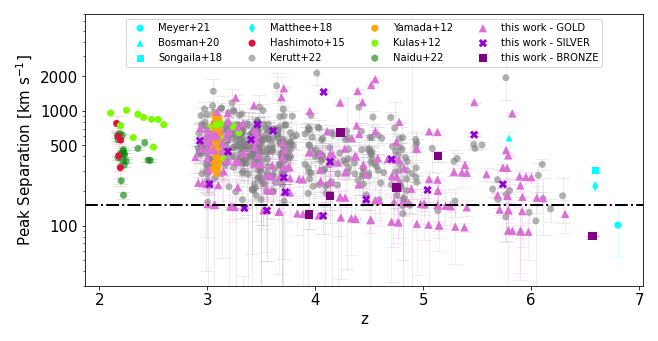}}
    \caption{Peak separation plotted against the redshift for our double-peak sample as well as the literature \citep{Kulas2012, Yamada2012, hashimoto2015, matthee2018, Songaila2018, bosman2020, meyer2021, naidu2022}. Our double-peak sample represented in purple shows three different symbols for the three categories \gold\  (triangle), \silver\ (cross), and \bronze\ (square). A horizontal black dash-dotted line at 150 km~s$^{-1}$ represents the peak separation threshold made in Sect.~\ref{sec:412} to obtain the universal fraction of double-peaks. The peak separations are in discrete lines due to the fact that the minimal peak separation measured is 2 spectral bins and depends on redshift.
    }   \label{fig:peaksep_vs_z_lit}
\end{figure*}

The peak separation distribution of the unbiased double-peak samples (Fig.~\ref{fig:Histos_all}, panel \textit{(d)}) ranges from 150 km~s$^{-1}$ to almost 1600 km~s$^{-1}$.
The mean value of $UDP_{I}$ (in black) is 447 $\pm$ 22 km~s$^{-1}$ and the mean value of $UDP_{R}$ (in orange) is 534 $\pm$ 28 km~s$^{-1}$, which is coherent with the mean peak separation measured by \cite{kerutt2022} for MUSE-Wide and MUSE-Deep LAEs (481 $\pm$ 244 km~s$^{-1}$), $z \approx 2.2$ LAEs from \cite{hashimoto2015} (500 $\pm$ 56 km~s$^{-1}$) or even $z \approx 2$ galaxies from \cite{matthee2021} (500 km~s$^{-1}$). 
We notice that the restrictive sample ($UDP_{R}$) has \Lya\ double-peaks with wider peak separations than the inclusive sample. 
A drop of the number of peak separations below 400 km~s$^{-1}$ is observed for the $UDP_{R}$ sample compared to the $UDP_{I}$ one. This is explained by the condition (described in Sect.~\ref{sec:413}) applied on the double-peaks detected with the flux variation analysis. This flux variation analysis is applied inside one \emph{area of signal} (see Sect.~\ref{sec:32}), which results most of the time on the detection of close by peaks. Thus, \Lya\ double-peaks with small peak separations ($<$ 400 km~s$^{-1}$) have more non significant secondary peaks than \Lya\ double-peaks with bigger peak separations.
The distributions of both $UDP_{R}$ and $UDP_{I}$ are similar above 700 km~s$^{-1}$ with a severe drop above 800 km~s$^{-1}$, similarly to \cite{Kulas2012}, \cite{trainor2015} and \cite{kerutt2022}.
The results are consistent with the overall literature, as we can see in Fig.~\ref{fig:peaksep_vs_z_lit} that compiles the peak separation measurements and redshifts of double-peaked \Lya\ lines reported in the literature.

\subsection{Towards a determination of a double-peak fraction}
\label{sec:43}

\subsubsection{Fractions of double-peak}
\label{sec:431}

The universal double-peak fraction ($X_{\rm DP}$), for the inclusive and restrictive samples are:
\begin{equation}
X_{\rm DP \, I} = N^{UDP_I} / N^{U} = 51\pm 4\%
\end{equation}
and
\begin{equation}
X_{\rm DP \, R} = N^{UDP_R} / N^{U} = 32\pm 3\%,
\end{equation}
respectively.

Our fraction of double-peaked galaxies $X_{\rm DP \, I}$ is an upper limit given the probability of spurious detections of 10\% (see Sect.~\ref{sec:36}), while $X_{\rm DP \, R}$ gives a lower limit due to the restrictive nature of the sample $UDP_{R}$. 
This range of the fraction of double-peaks (32\% $\leq X_{\rm DP} \leq$ 51\%) is consistent with most of the fractions reported in the literature.
Indeed, \cite{Kulas2012} and \cite{sobral2018} find fractions of 30\% and 25\%, for $z = 2 - 3$ LAEs, \cite{cao2020} find an average fraction of 20\% for lensed galaxies and \cite{kerutt2022} find 33\% for the MUSE-Wide and MUSE-Deep LAEs. Moreover, \cite{trainor2015} with a fraction of 40\% at redshift $\sim$ 2.7 or \cite{Yamada2012} finding 50\% of double-peaks for their LAEs at $z = 3.1$. 

It is important to note that the fractions given in the literature are not corrected for observational biases which could lead to an underestimation of the double-peak fraction. 
Additionally, double-peaks in the literature are detected either visually \citep{Yamada2012, kerutt2022} or with an algorithm \citep{trainor2015} completed by a visual inspection only on the spectra \citep{Kulas2012}, the spatial data being unavailable for most studies.

\subsubsection{\texorpdfstring{Fraction of double-peak evolution with \Lya\ luminosity}{Fraction of double-peak evolution with Lya luminosity}}
\label{sec:432}

\begin{figure}[!h]
\centering
    \resizebox{\hsize}{!}{\includegraphics{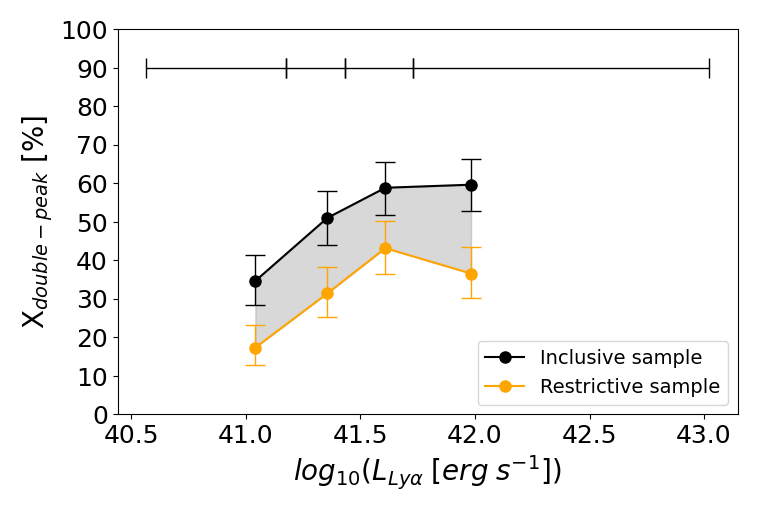}}
    \caption{Fraction of double-peaked LAEs plotted against the logarithmic \Lya\ luminosity. The unbiased sample $U$ has been divided into 4 luminosity bins with the same number of objects (51 or 52). The fraction of double-peaks has been derived in each bin for both $UDP_{I}$ and $UDP_{R}$. The results are positioned at the median \Lya\ luminosity of each bin. The fractions of inclusive double-peaks are represented in black colour. The $X_{\rm DP \, R}$ of the restrictive sample are in orange. The horizontal black line at the top of the figure shows the size of each \Lya\ luminosity bin.}
    \label{fig:Xdp_per_LumBin_Inclusive_Restrictive}
\end{figure}

To investigate if the fraction of double-peaks varies with luminosity, we divide our unbiased samples into four \Lya\ luminosity bins with the same number of objects. Figure~\ref{fig:Xdp_per_LumBin_Inclusive_Restrictive} shows $X_{\rm DP \, I}$ and $X_{\rm DP \, R}$ for each of the four luminosity bins.
The fraction of double-peaks from the inclusive sample evolves from around 34\% for the faintest luminosities to nearly 60\% for the brightest bin (41.7 < log(L$_{Ly\alpha}$ [erg s$^{-1}$]) < 43).
Concerning the evolution of the fraction of double-peaks for the restrictive sample, we observe a similar trend as for the inclusive sample, meaning an increase towards the brighter luminosities, except for the brightest bin in which $X_{\rm DP \, R}$ is smaller than in the previous bin. The fraction evolves from 17\% to 37\% with a peak at 43\% for the third bin(41.4 < log(L$_{Ly\alpha}$ [erg s$^{-1}$]) < 41.7).
As both samples are the lower and upper limits of our study, we might consider the real fraction of double-peaks per bin of \Lya\ luminosity between the two trends shown in Fig.~\ref{fig:Xdp_per_LumBin_Inclusive_Restrictive} delimited by the grey area.
Moreover, as the trends remain the same whatever the restrictions applied on the double-peak sample, they are robusts.
Brighter galaxies seem to have more double-peaked \Lya\ lines, as seen in the Lensed Lyman-Alpha MUSE Arcs Sample of Claeyssens et al. (in prep.). However, since it is easier to detect double-peaks for bright galaxies, this trend may still be due to observational biases. We will explore the bright end of the MUSE GTO samples in forthcoming work by applying our classification method to wider surveys.
The evolution of the fraction of double-peaks per spatial category (\gold\ and \silver), in the unbiased sample, is described in App.~\ref{ap:E}.

\begin{figure}[!h]
\centering
    \centering
    \resizebox{\hsize}{!}{\includegraphics{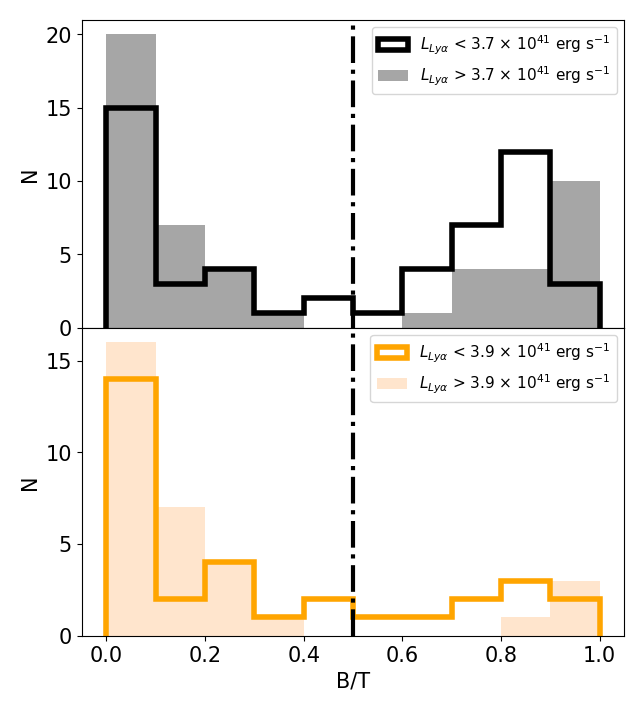}}
    \caption{\textit{Top panel:} B/T flux ratio distributions of the inclusive unbiased double-peak sample ($UDP_{I}$) divided into 2 \Lya\ luminosity bins with the same number of objects (53 and 52). The faint sub-sample (L$_{Ly\alpha}$ < 3.7 $\times$ 10$^{41}$ erg s$^{-1}$) is in black and the bright one (L$_{Ly\alpha}$ > 3.7 $\times$ 10$^{41}$ erg s$^{-1}$) is in grey.
    \textit{Bottom panel:} B/T flux ratio distributions of the restrictive unbiased double-peak sample ($UDP_{R}$) divided into 2 \Lya\ luminosity bins with the same number of objects (33 and 33). The faint sub-sample (L$_{Ly\alpha}$ < 3.9 $\times$ 10$^{41}$ erg s$^{-1}$) is in stepped orange and the bright one (L$_{Ly\alpha}$ > 3.9 $\times$ 10$^{41}$ erg s$^{-1}$) is in orange.
     B/T = 0.5 is represented by a black dash-dotted line in both panels. }
    \label{fig:BT_Distri_per_LumBin}
\end{figure}

\cite{blaizot2023} find an anti correlation between the \Lya\ luminosity and the B/T flux ratio of the \Lya\ line in their simulations. The higher the luminosity (i.e. face-on galaxy), the lower the B/T is. 
In the top panel of Fig.~\ref{fig:BT_Distri_per_LumBin}, we plot the B/T distribution for the inclusive unbiased double-peak sample ($UDP_{I}$) split into two luminosity bins with the same number of objects. The sample is cut at L$_{Ly\alpha}$ = 3.7 $\times$ 10$^{41}$ erg s$^{-1}$. We notice that the bright sub-sample (grey histogram in top panel of Fig.~\ref{fig:BT_Distri_per_LumBin}) populates more the red peak dominated regime (i.e. B/T < 0.5) than the blue peak dominated one (i.e. B/T > 0.5). On the contrary, the faint sub-sample (black histogram) is more dominant in the intermediate B/T range and high B/T range. 
Our results are in line with what \cite{blaizot2023} find in their simulations.
The lower panel of Fig.~\ref{fig:BT_Distri_per_LumBin} shows the same plot but for the restrictive unbiased double-peak sample ($UDP_{R}$) and a \Lya\ luminosity cut at cut at L$_{Ly\alpha}$ = 3.9 $\times$ 10$^{41}$ erg s$^{-1}$. The bright sub-sample (orange histogram) shows more clearly a strong presence at the lower B/T values and a small number of objects at B/T $>$ 0.5 compared to $UDP_{I}$. The presence of the faint sub-sample at high B/T values is comparable with the simulations in \cite{blaizot2023}. Nevertheless, an important number of faint galaxies also have smaller values of B/T.

\subsubsection{Fraction of double-peak evolution with redshift}
\label{sec:433}

According to theoretical predictions, the IGM attenuation increases with redshift, absorbing the blue part of the  \Lya\ emission \citep{Laursen2011, Garel2021}.
\cite{hayes2021} indeed find for stacked spectra that the fraction of flux bluewards of the main (red) peak decreases with redshift. This trend has also been reproduced by \cite{kramarenko2024} by stacking MUSE-Wide LAE spectra. 
We therefore expect the double-peak fraction to decrease with redshift.

\begin{figure}[!h]
\centering
   \resizebox{\hsize}{!}{\includegraphics{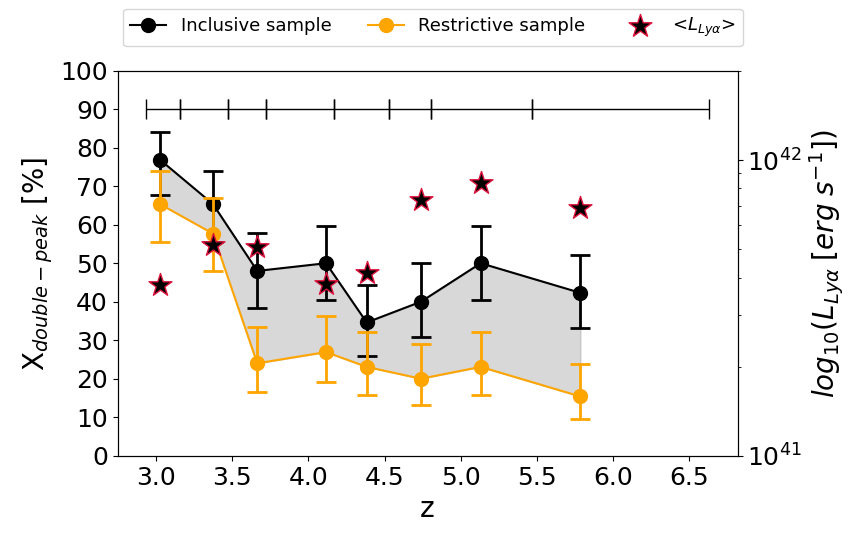}}
    \caption{Fractions of double-peaked LAEs plotted against the redshift. The unbiased sample has been divided into 8 redshift bins with the same number of objects (25 or 26). The fraction of double-peaks has been derived in each bin. The results are positioned at the median redshift of each bin. The fractions of inclusive double-peaks are represented in black colour. The $X_{\rm DP}$ of restrictive objects are in orange. The horizontal black line at the top of the figure shows the size of each redshift bin. The black stars surrounded in red represent the mean \Lya\ luminosity of each bin.}
    \label{fig:Xdp_vs_z_Inclusive_Restrictive}
\end{figure}

Figure~\ref{fig:Xdp_vs_z_Inclusive_Restrictive} shows the evolution of the fraction of double-peaks with redshift for the inclusive (restrictive) sample in black (orange): we report a global decrease from $77^{+7}_{-9}\%$ ($65^{+9}_{-10}\%$) at $z\sim3$ down to $42^{+10}_{-9}\%$ ($15^{+8}_{-6}\%$) at $z > 5.5$, but above $z=4$, the data are compatible with a plateau around 40\% (20\%).
Since we report on Fig.~\ref{fig:Xdp_per_LumBin_Inclusive_Restrictive} a strong evolution of the fraction of double-peaks with luminosity, these plateaux might be caused by the high mean luminosity (represented by stars on Fig.~\ref{fig:Xdp_vs_z_Inclusive_Restrictive}) in the last three bins, artificially raising the fraction of double-peaks. 
At low redshift ($z < 3.5$), we see an interesting increase in the double-peak fraction. If this trend is confirmed with more data, it could indicate an intrinsic evolution of the LAE population towards cosmic noon. If such a possible evolution is confirmed, it could make less pertinent the use of the double-peak fraction to probe the opacity of the IGM at this redshift range. \\

The two evolutions of the fraction of double-peaks, one inclusive and the other one restrictive, give us the lower and upper limits of the $X_{\rm DP}$ range (shaded area on Fig.~\ref{fig:Xdp_vs_z_Inclusive_Restrictive}) our sample truly have.
Additionally, the trend of the evolution of the fraction of double-peaks is conserved regardless of the way the double-peaks are spectrally detected. Our classification is robust.
The evolution of the fraction of double-peaks with redshift per spatial category (\gold\ and \silver), in the unbiased sample, is discussed in App.~\ref{ap:E}.

\section{Discussion}
\label{sec:5}

\subsection{Contaminants to the multi-peak sample}
\label{sec:51}

In this subsection, we will describe the physical interpretation of the \bronze\ category defined in Sect.~\ref{sec:34}.
As a recap, eleven objects belong to the \bronze\ category: ten double-peaks and one triple-peak (Sect.~\ref{sec:411}).

\begin{figure*}
\begin{minipage}{0.65\textwidth}
    \centering
        \includegraphics[width=\textwidth]{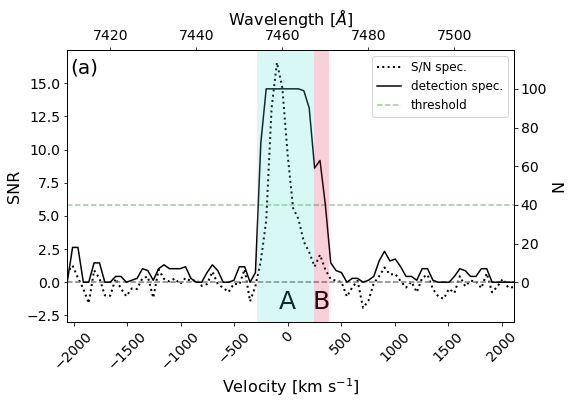} \\
    \centering
        \includegraphics[width=0.9\textwidth]{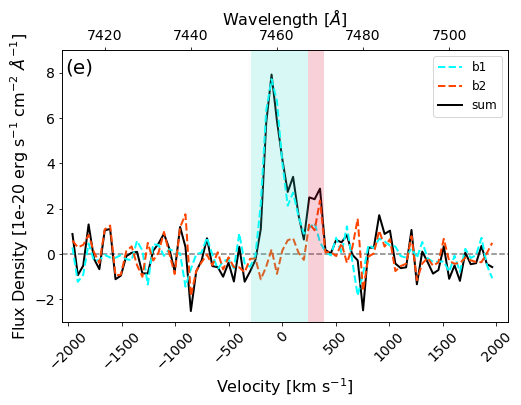}
\end{minipage}
\begin{minipage}{0.3\textwidth}    
     \centering
         \centering
         \includegraphics[width=\textwidth]{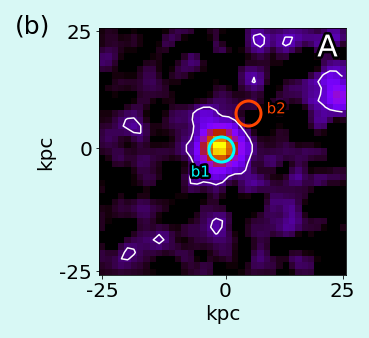}
     \hfill
         \centering
         \includegraphics[width=\textwidth]{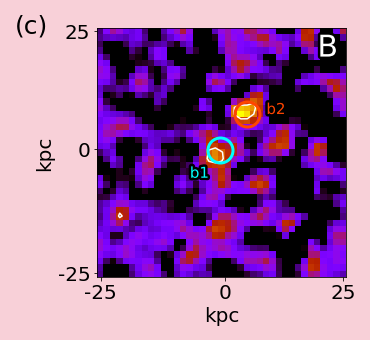}
    \hfill
         \centering
         \includegraphics[width=\textwidth]{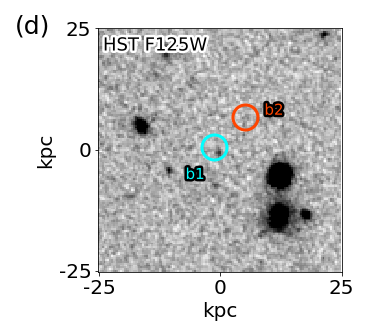}
\end{minipage}
    \caption{ID 399, double-peak, \bronze\ category. Example of a LAE surrounded by a satellite discovered thanks to our method. \textit{(a)}, \textit{(b)} and \textit{(c)}: Same as Fig.~\ref{fig:ID3240_GOLD}. The cyan circle b1 and the orange circle b2 represent the apertures used to extract two spectra at the indicated locations. The size of the circles represent a 0.5\arcsec\ diameter aperture. \textit{(d)}: 50 $\times$ 50 kpc$^2$ HST \textit{F125W} image with the two locations of the extraction (cyan and orange circles). \textit{(e)}: Spectra extracted from the two circles. The orange dash-dotted line corresponds to the spectrum extracted at the position of b2. The spectrum extracted from b1 is shown as a cyan dashed line. The black line is the summed spectrum of the blue and orange spectra. The spectrum extracted at the position b2, i.e. the position of the satellite, contributes mainly to the peak B.}
    \label{fig:ID399_satellite}          
\end{figure*}

Figure~\ref{fig:ID399_satellite} illustrates an interesting case where the NB images show a well centred \Lya\ emission (b1) plus another small compact emission offset from the centre of the central source. This compact emission, named b2 in panels \textit{(b)}, \textit{(c)} and \textit{(d)} is significantly detected on the NB image B and not detected on the NB image A. Interestingly, there is no HST counterpart (panel \textit{(d)} of Fig.~\ref{fig:ID399_satellite}) and no detection in the MUSE DR2 catalogue. The spectra extracted from b1 and b2 apertures show that b2 contributes to most of the red peak of the \Lya\ line. We interpret this compact emission b2 as a satellite.
Among these ten double-peaks, nine of them present a similar configuration like in Fig.~\ref{fig:ID399_satellite}.

\begin{figure*}[h!]
\begin{minipage}{0.7\textwidth}
\centering
    \resizebox{\hsize}{!}{\includegraphics{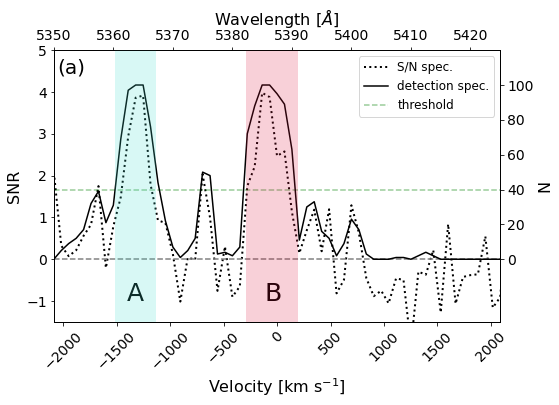}}
\end{minipage}
\begin{minipage}{0.3\textwidth}    
     \centering
         \centering
             \includegraphics[width=\textwidth]{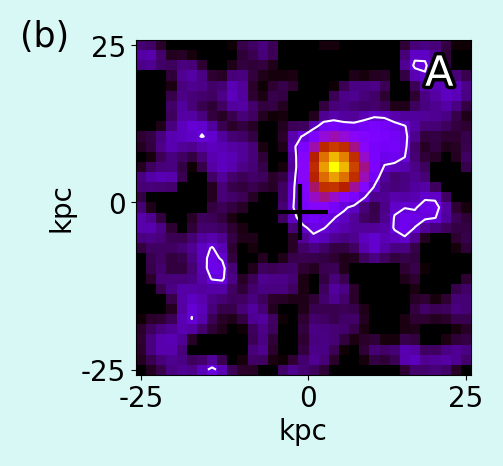}
     \hfill
         \centering
         \includegraphics[width=\textwidth]{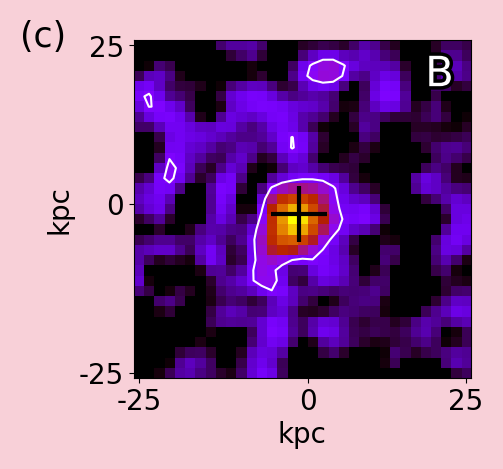}
\end{minipage}
\begin{minipage}{0.5\textwidth}    
     \centering
         \centering
             \includegraphics[width=\textwidth]{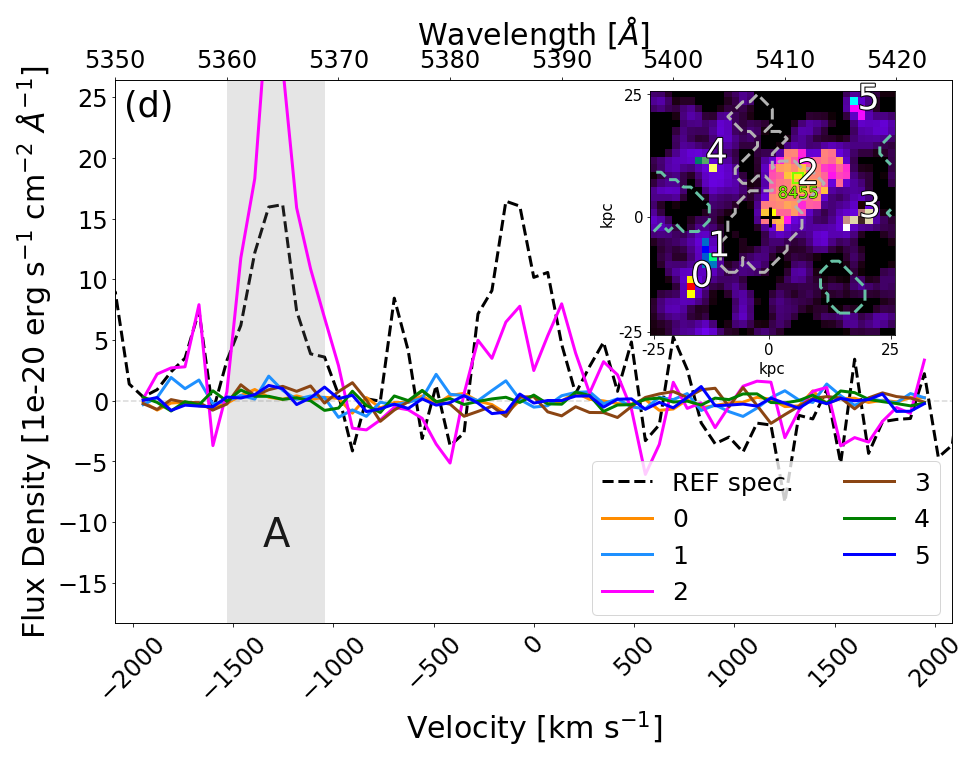}
\end{minipage}
\begin{minipage}{0.5\textwidth}    
     \centering
         \centering
             \includegraphics[width=\textwidth]{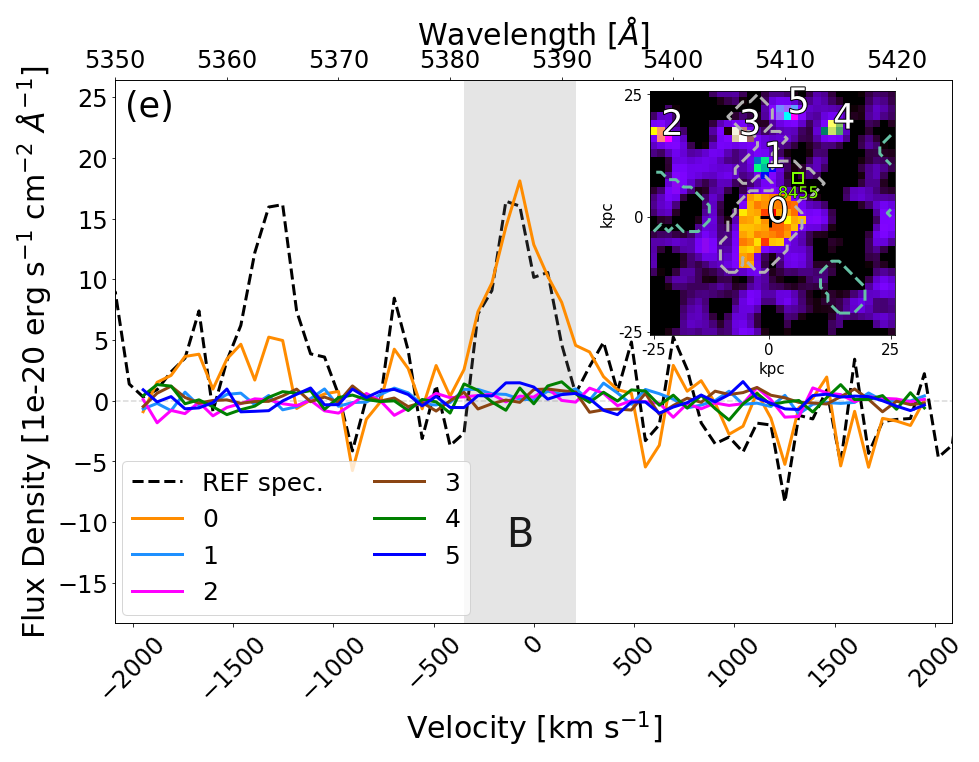}
\end{minipage}
    \caption{ID 8439, double-peak, \bronze\ category. Same as Fig.~\ref{fig:ID3240_GOLD}. Peaks A and B come from two different spatial locations. Peak A is the \Lya\ emission of the halo of ID 8455 (contained in the reference segmentation map of ID 8439). Peak B is the \Lya\ emission coming from the targeted galaxy.}
   \label{fig:ID8439_spectrum_neighbor}          
\end{figure*}

In the remaining \bronze\ double-peaked LAE, the blue peak is artificially created by the \Lya\ halo of a neighboring galaxy, as illustrated in Fig.~\ref{fig:ID8439_spectrum_neighbor}. Panels \textit{(b)} and \textit{(d)} show that peak A is emitted by a neighboring galaxy (ID 8455) that is included in the reference segmentation map of our targeted galaxy. 

Follow-up observations in the infrared are necessary to confirm the hypothesis of the satellites but it may be challenging, given the fact that our targets are very faint.

The \bronze\ category contains galaxies showing a double-peak \Lya\ profile on the spectra but not on the NB images of each peak. Indeed, the NB image of only one of the peaks contains a small compact emission offset from the centre of the galaxy (as shown in panels \textit{(b)} and \textit{(c)} of Fig.~\ref{fig:ID399_satellite}), discarding the fact that these \Lya\ lines originate from radiative transfer processes. 
We call the ten double-peaked \bronze\ sources fake double-peaks. The fraction of fake double-peaks among the total number of double-peaked galaxies (248, Sect.~\ref{sec:411}), X$_{\rm FDP}$, is:
\begin{equation}
X_{\rm FDP} = 10/248 = 4\%
\end{equation}
\noindent Spectral only classification would consider these objects as double-peaked LAEs whereas they are not. 
As a note, even galaxies that are spatially unresolved in MUSE may have multiple unresolved spatial components associated with different spectral components. As such, the distinction between "fake" and "real" multi-peak LAEs may be blurry. Thus, X$_{\rm FDP}$ represents a lower limit.

\subsection{Galaxy pairs identification and prospects}
\label{sec:52}

Interacting galaxies are not limited to the \bronze\ category, as we explain below.
This \bronze\ category contains only galaxies that are in interaction with other objects in a very small area, i.e. inside the reference segmentation map used in DR2 (see Sect.~\ref{sec:33}). These objects do not show a particular trend in their B/T values.

\noindent The \silver\ category (see Sect.~\ref{sec:34} for details) contains objects for which more than one \texttt{SourceFinder} detection is located inside the reference segmentation map. These detections correspond sometimes to well-identified objects in the MUSE DR2 catalogues (10/38), but sometimes not (28/38) as illustrated in Fig.~\ref{fig:ID163_SILVER}. Nevertheless, the \silver\ galaxies are interpreted as being in interaction since very close by clumps are visible. With the data we have in hand, we are not yet able to identify if the \silver\ objects are coming from two clumps of the same source, or from two different objects. 
Additional data detecting other lines could help in this differentiation.

\begin{figure*}
\begin{minipage}{0.7\textwidth}
\centering
    \resizebox{\hsize}{!}{\includegraphics{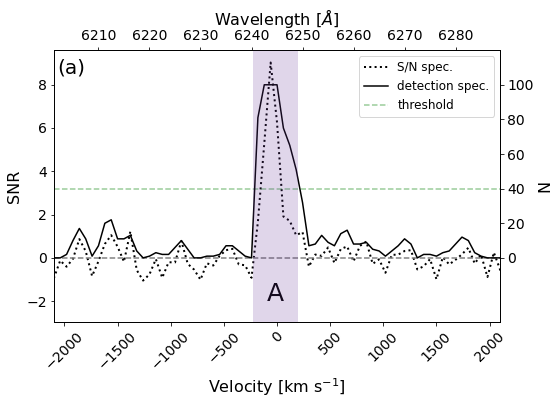}}
\end{minipage}
\begin{minipage}{0.3\textwidth}    
     \centering
         \centering
             \includegraphics[width=\textwidth]{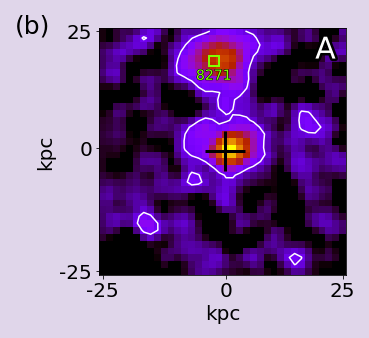}
\end{minipage}
    \caption{ID 7817, single-peak, \gold\ category. Example of a pair of galaxies. The galaxy ID 8271 has the same redshift as ID 7817 ($z \approx 4.14$).}
    \label{fig:ID7817_pair}          
\end{figure*}

\begin{figure*}[!h]
\begin{minipage}{0.7\textwidth}
\centering
    \resizebox{\hsize}{!}{\includegraphics{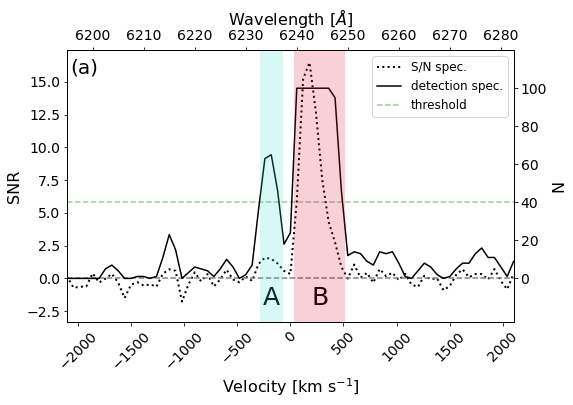}}
\end{minipage}
\begin{minipage}{0.3\textwidth}    
     \centering
         \centering
             \includegraphics[width=\textwidth]{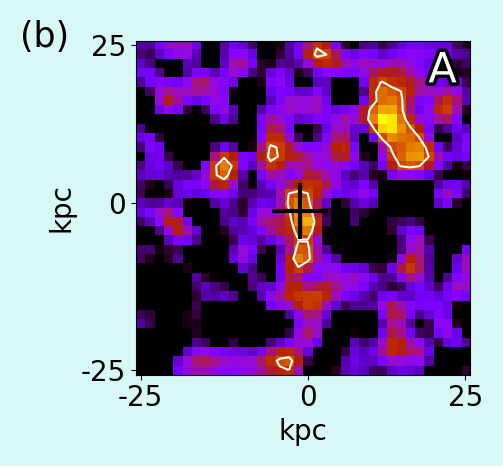}
     \hfill
         \centering
         \includegraphics[width=\textwidth]{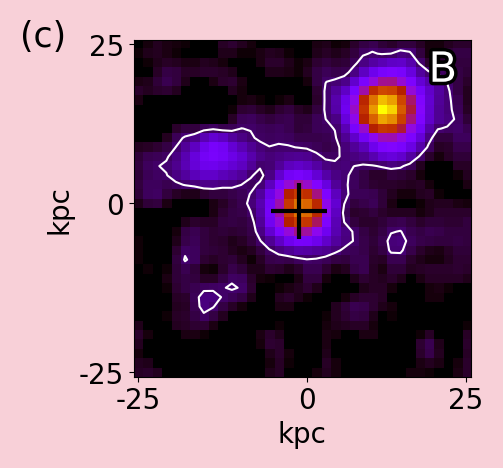}
\end{minipage}
\begin{minipage}{0.5\textwidth}    
     \centering
         \centering
             \includegraphics[width=\textwidth]{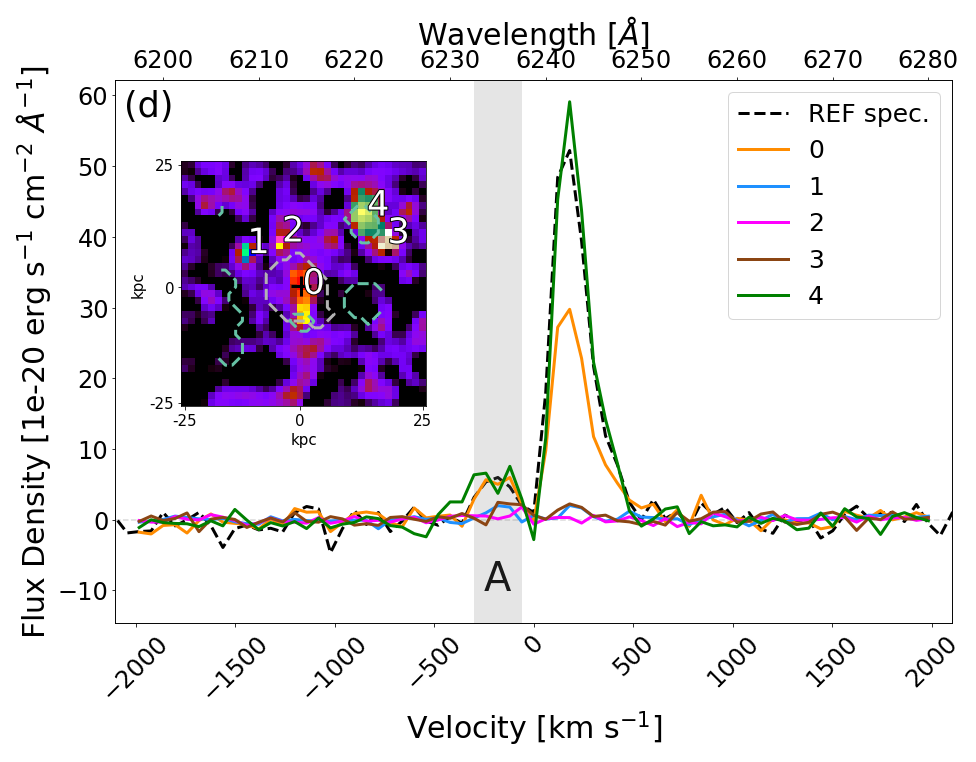}
\end{minipage}
\begin{minipage}{0.5\textwidth}    
     \centering
         \centering
             \includegraphics[width=\textwidth]{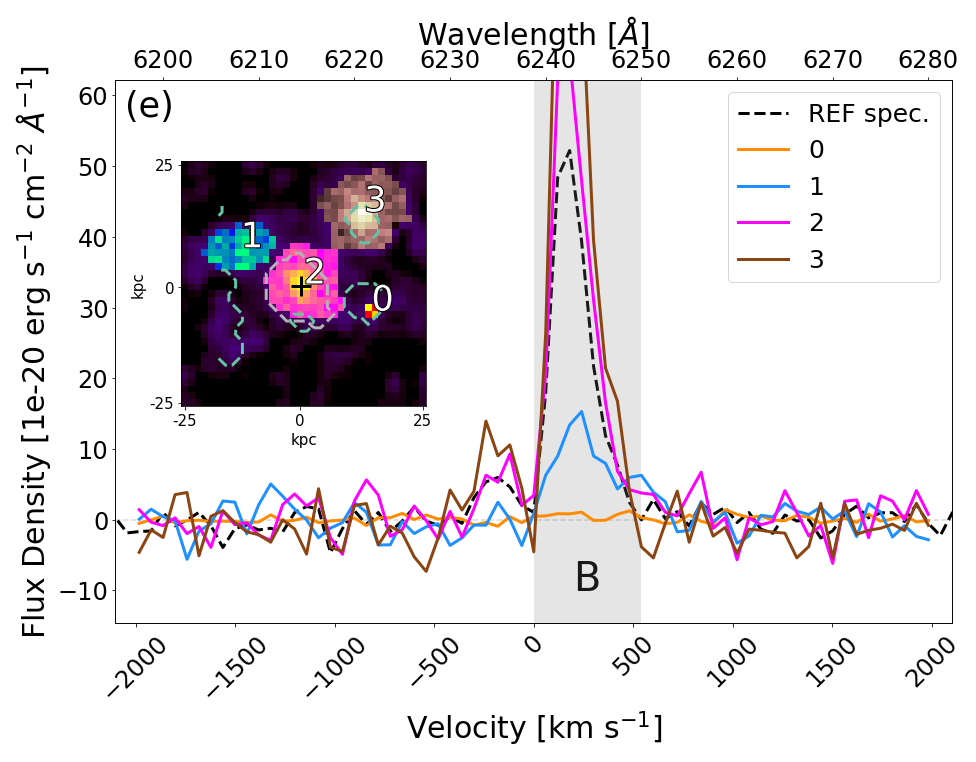}
\end{minipage}
    \caption{ID 412, double-peak, \gold\ category. Same as Fig.~\ref{fig:ID3240_GOLD}.  Example of a triplet system of galaxies (ID 412, 6698 and 8355) at $z \approx 4.13$. Interestingly, in panel \textit{(e)}, two galaxies have a similar red-dominated double-peaked \Lya\ line profile while the third one, labeled 1, shows a blue-dominated double-peak line.}
   \label{fig:ID412_triplet}          
\end{figure*}

Finally, visually inspecting the NB images enabled us to discover complex systems. We discovered 40 galaxies lying in pairs (20 pairs), i.e. two galaxies at a similar redshift and very close (less than 25 kpc, see Fig.~\ref{fig:ID7817_pair} for example). Among those pairs, 10 galaxies are \silver.
Moreover, three other systems have been identified. In all cases, three galaxies of similar redshifts are located within a distance of 25 kpc. We call them triplet systems. One if the triplet system is shown in Fig.~\ref{fig:ID412_triplet}.
The pairs and triplet systems are flagged in App.~\ref{ap:C}.
Given their proximity, all these systems are consistent with interacting systems. 

As a summary, in our parent sample, the \bronze\ objects, the \silver\ ones, the 40 galaxies in pairs as well as the 9 galaxies in triplet systems are interpreted as being in interaction. The fraction of such systems, called X$_{\rm interaction}$, is:

\begin{equation}
X_{\rm interaction} = 88/477 \sim 19\%
\end{equation}

\noindent Interacting galaxies are usually discarded from \Lya\ studies although they represent a significant fraction of our sample.
A more detailed analysis of the effect of the environment on the spectral diversity of LAEs is beyond the scope of this paper. The pairs and the three triplet systems will be the subject of another study (Vitte et al. in prep.).

\subsection{Lack of systemic redshift}
\label{sec:53}

The \Lya\ line is resonant, and radiative transfer effects can result in a double-peaked profile, with peaks on each side of the resonance frequency \citep{neufeld1990, dijkstra2006, dijkstra2017}.  
Based on this assumption, methods have been developed to retrieve the systemic redshift of a galaxy using the \Lya\ line shape \citep{verhamme2018}, or the escape fraction of the ionising radiation using the \Lya\ peak separation \citep{verhamme2015}. Moreover, the B/T measurements can provide information about the gas kinematic configuration (inflows/outflows), as demonstrated in \cite{blaizot2023}.

Throughout this study, we call "blue peak" the peak with the shortest wavelength and "red peak" the one at the longer wavelength, although we do not know the systemic redshift of most of our galaxies. 
Our population of double-peaked LAEs may differ from what is usually called a double-peak in the sense that the blue (red) peak might not be bluer (redder) than the resonant frequency of the central object.
In the same way, single-peaked lines are by default presumed being on the red side of the resonant frequency, since we do not know their systemics redshift. But this assumption for single-peaked lines does not impact the results presented in this paper.

\begin{figure}[!h]
\centering
    \resizebox{\hsize}{!}{\includegraphics{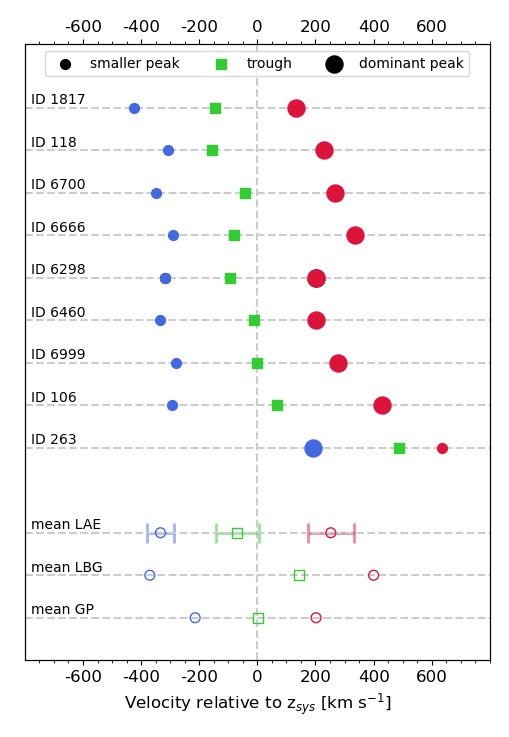}}
    \caption{Velocities of every peak and trough for the double-peaked galaxies having a systemic redshift (Sect.~\ref{sec:53}). Each line corresponds to a source, where the blue (red) circle shows the bluer (redder) peak and the green square indicates the position of the trough. The larger circle represents the peak with the highest flux. The first line of the bottom part of the figure shows the mean positions of the trough and the blue and red peaks of the objects shown above (without taking into account ID 263). The mean LBG and mean Green Pea (GP) are drawn from the literature, \cite{Kulas2012} and \cite{orlitova2018}, respectively.}
    \label{fig:lack_zsys_Kulas12}
\end{figure}

In this section, we investigate the spectral parameters of a subsample of galaxies with known systemic redshift whose spectra are shown in App.~\ref{ap:F}. Fifteen objects have a secure systemic redshift measured from nebular lines detected in their MUSE spectra. One of those (ID 53), was studied in detail in \cite{matthee2022} and its systemic redshift is well constrained. Five others were studied in \cite{kusakabe2022} (MID 106, MID 118, MID 149, MID 6700 and MID 7089). One last (ID 6666) has its redshift measured by VLT/KMOS in \cite{boogaard2021}. Eight sources of this subsample with systemic redshift are double-peaked, three are single-peaked and the last four are triple-peaked. They all belong in the \gold\ category. 
The luminosities of those galaxies do not go lower than log(L$_{Ly\alpha}$ [erg s$^{-1}$]) = 40 but they populate the bright end of the distribution with a mean logarithmic \Lya\ luminosity of 42.25 (panel \textit{a}). This is expected as fainter galaxies usually only have their \Lya\ line detected.
Most of the objects have a FWHM of the peak of the \Lya\ line with the strongest flux within the range 250 $-$ 550 km~s$^{-1}$.
Only one galaxy has a small FWHM of 129 km~s$^{-1}$.
Only one galaxy has a B/T above 0.5 and it is as well the only one having both the blue and red peaks on the red side of the systemic (see ID 263 in App.~\ref{ap:F} and Fig.~\ref{fig:lack_zsys_Kulas12}). Four of the double-peaks have a small B/T (less than 0.1). The three last galaxies of this subsample have a B/T that does not exceed 0.4. 
Concerning the peak separation distribution of the double-peaked LAEs having a systemic redshift, the distribution is concentrated in a small range, between 400 and 800 km~s$^{-1}$. The mean peak separation is 578 km~s$^{-1}$. This is 1.3 times higher than the mean value of the unbiased double-peak sample (447 $\pm$ 29 km~s$^{-1}$).
Figure~\ref{fig:lack_zsys_Kulas12} shows a graphical summary of peak velocities for the eight double-peaked galaxies having a systemic redshift. The mean values of our sample are:
\begin{itemize}[noitemsep,topsep=0pt]
    \item Blue peak: - 333 $\pm$ 47 km~s$^{-1}$
    \item Trough: - 70 $\pm$ 74 km~s$^{-1}$
    \item Red peak: 253 $\pm$ 79 km~s$^{-1}$
\end{itemize}
Our sample has a similar mean red peak shift than the LAE sample of \cite{trainor2015} (200 km~s$^{-1}$) although their sample is NB-selected whereas our survey is blind.
It has on average broader peak separations than the Green Peas studied in \cite{orlitova2018} which have a mean peak separation of $\pm$ 400 km~s$^{-1}$ and some are known to leak LyC.
However, compared to the Lyman-break galaxies (LBGs) of \cite{Kulas2012}, our sample has narrower peak separations. The mean peak separation of the LBGs is 770 km~s$^{-1}$. LBGs are more massive galaxies \citep[$10^{10-11} M_{\odot}$,][]{carilli2008} than LAEs \citep[$10^{8-9} M_{\odot}$,][]{ouchi2020}, thus it can explain the large peak separations measured by \cite{Kulas2012} as, on average, the \Lya\ photons have to scatter more through the LBGs to escape.
Looking at the position of the trough, we notice it is mostly located bluewards with respect to the systemic velocity. Together with the red dominated characteristic, this is expected in the framework of radiation transfer through an outflowing medium \cite[Fig.~14 in][]{Verhamme2006}. \cite{orlitova2018} have a mean position of the trough centred on the systemic velocity while \cite{Kulas2012} have a mean trough at 143 km~s$^{-1}$. 

Only one galaxy has the bluer peak dominant (ID 263, B/T = 0.98, Fig.~\ref{fig:lack_zsys_Kulas12}). This is also the only galaxy with the two peaks located on the red side of the systemic redshift.
Such a configuration has already been observed in \cite{Kulas2012} and \cite{trainor2015}. In this case, the double-peak profile does not emerge from radiative transfer processes, thus it is not possible to retrieve the escape fraction of ionising radiation from the peak separation, nor to interpret the bluer peak dominance as inflowing gas \citep{zheng2002,dijkstra2006,Verhamme2006, blaizot2023}. 
To be able to retrieve these parameters, we need to know the systemic redshift as highlighted in \cite{blaizot2023}.
A dedicated KMOS proposal (113.2682.001(A), PI: Vitte) has been approved and will observe during Period 113 a total of 21 galaxies from MUSE-Deep \citep{bacon2023, inami2017} and MUSE-Wide \citep{urrutia2019} fields to precisely measure their systemic redshift. 

\section{Summary and conclusion}
\label{sec:6}

In this paper, we presented the diversity of the \Lya\ line shapes among a population of MUSE LAEs.
We constructed a parent sample of 477 redshift-selected ($z = 2.87 - 6.64$) sources from a blind, extremely deep, MUSE GTO survey, the \mxdf\ \citep[up to 140 hours of integration time,][]{bacon2023}. 
The main results of our work are summarised as follows:
\begin{itemize}[noitemsep,topsep=0pt]
    \item We developed a method to detect and characterise the \Lya\ emission lines in both spectral and spatial dimensions (Sect.~\ref{sec:3}). 
    \item The \Lya\ lines from the sources of the parent sample are quantitatively classified into four different spectral categories: no-peak (9), single-peak (198), double-peak (248) and triple-peak (22), see Sect.~\ref{sec:411} and Fig.~\ref{fig:ex_lya_line}.
    \item  The analysis of the spatial distribution of the \Lya\ peaks lead to the division of the parent sample into three different categories: \gold, \silver\ and \bronze\ (Sect.~\ref{sec:34} and Fig~\ref{fig:categories_G_S_B}). An object is classified as \gold\ if all the peaks originate from the same spatial location. When the peak emission arises from several distinct regions contained within the reference segmentation map, the object is thus classified as a \silver\ object. Finally, for double- and triple-peaks, an object can be classified \bronze. In that case, each peak has an emission coming from a distinct spatial location, which means the peaks are coming from different sources. Among the 248 double-peaks, 238 are \gold\ or \silver\ objects (i.e. considered as double-peaks coming from radiative transfer processes). The last ten cannot be considered as such given their spatial distribution (i.e. \bronze\ objects).
    \item We identified several observational limitations: the fainter the source, the harder it is to detect multi-peaks; the minimum peak separation detectable varies with redshift; the detectable range of B/T ratios depends on the S/N (see Sect.~\ref{sec:412}). Taking into account these limitations, we built an unbiased sample of 206 galaxies, with \Lya\ luminosities above 3 $\times$ 10$^{40}$ erg~s$^{-1}$, peak separations larger than 150 km~s$^{-1}$ and S/N above 7.
    \item For each object, we measured spectral properties such as the \Lya\ flux, the \Lya\ luminosity, the S/N, the FWHM, and the peak separation and blue-to-total flux ratio (B/T) for double-peaks. Our study reaches fainter luminosities (log(L$_{Ly\alpha}$ [erg s$^{-1}$]) = 39.5) thanks to the depth of our data.
    Our derived FWHM and the peak separation values from the unbiased sample are comparable to other studies in the literature, whereas the B/T distribution shows an important fraction of of blue-dominated spectra (Sect.~\ref{sec:423}).
    \item We found a fraction of double-peaks X$_{\rm DP}$ oscillating between 32\% and 51\%, obtained by two double-peak unbiased samples, one restrictive giving the lower limit and one inclusive giving the upper limit (Sect.~\ref{sec:431}). This fraction seems to evolve dependently with the \Lya\ luminosity (Sect.~\ref{sec:432}).
    \item This fraction of double-peaks shows a global decrease with redshift, stagnating at high redshift (Sect.~\ref{sec:433}). This plateau might be due to the high mean \Lya\ luminosity of each redshift bin, knowing that brighter galaxies seem to have more double-peaked \Lya\ lines.
    Nevertheless, if the increase of the X$_{\rm DP}$ at lower redshift is confirmed with more data, the use of the double-peak fraction to probe the opacity of the IGM becomes less pertinent. 
    \item The double-peaks identified by the method are not necessarily double-peaks originating from radiative transfer (\bronze\ category, see Sect.~\ref{sec:51}), but could come from satellites. In fact, the NB images of the \Lya\ peaks detected by the method show a wide variety of environments for galaxies. This method unveiled the detection of satellites (companions) of main targeted galaxies and also interactions between galaxies (pairs, triplet systems), independently of the spectral category (Sect.~\ref{sec:52}). A detailed characterisation of the environment of our pairs and triplet galaxies will be the subject of an upcoming study (Vitte et al. in prep.). 
    \item Peak separation and B/T flux ratio have a physical meaning (i.e. they are tracers of the escape of ionising photons for peak separation < 400 km~s$^{-1}$ and infalling gas for B/T > 0.5) only if the systemic velocity is located between the blue and the red peaks. We were able to measure a systemic redshift for a subsample of 15 galaxies from the MUSE data (Sect.~\ref{sec:53}). The only blue dominant peak having a systemic redshift has its two peaks located on the red side of the systemic redshift. This spectrum thus can not be considered as tracing infalling gas. More systemic redshift measurements are needed in order to interpret our distributions further. A recently approved KMOS proposal (113.2682, PI: Vitte, 60hrs) will allow us to increase this subsample of systemic redshifts.
\end{itemize}
This paper aims at describing the method we developed to classify \Lya\ emission lines from high-redshift galaxies and its use on a blind sample from MUSE GTO surveys. A more detailed study of the completeness will be the scope of a future work.
We plan to apply the method developed and used for this work to other data sets which will give us a more general view on the fraction of double-peaks among LAEs and how it evolves with redshift. We will also use existing data \citep[e.g. FRESCO, JADES,][respectively]{oesch2023, Eisenstein2023} and obtain new data (KMOS proposal, 113.2682, PI: Vitte, 60hrs) in order to get the systemic redshift of more galaxies of the \mxdf\ to constrain the distributions we found in this paper.

The infrared instruments such as VLT/KMOS and JWST/NIRSpec allow us to perform a multi-wavelength study to explore the possible physical peculiarities of the different categories of LAEs.
In the longer term, this work can be pursued at lower redshift, e.g. down to $z \approx 2$ thanks to the upcoming BlueMUSE \citep{Richard2019}, allowing us to unveil the evolution of the diversity of the LAE population with time, over 3 more Gyr of galaxy evolution.

\bibliographystyle{aa} 
\bibliography{biblio}


\begin{acknowledgements}
AV acknowledges the support from the SNF grants PP00P2 176808 and 211023. HK acknowledges support from Japan Society for the Promotion of Science (JSPS) Overseas Research Fellowship as well as JSPS Research Fellowships for Young Scientists. JP acknowledges funding by the Deutsche Forschungsgemeinschaft, Grant Wi 1369/31-1. This work is based on observations taken by VLT, which is operated by European Southern Observatory. This research made use of \textsc{Astropy}, which is a community-developed core Python package for Astronomy \citep{astropy:2013, astropy:2018, astropy:2022}, and other software and packages: \textsc{MPDAF} \citep{piqueras2019}, \textsc{PHOTUTILS} \citep{bradley2023}, \textsc{Numpy} \citep{vanderwalt2011}, \textsc{Scipy} \citep{virtanen2020}.
The plots in this paper were created using \textsc{Matplotlib} \citep{hunter2007}.
\end{acknowledgements}


\begin{appendix}

\onecolumn

\section{Method limitations}
\label{ap:A}

\textbf{Threshold determination for detection spectra}

To disentangle the \Lya\ emission peaks from the noise spikes, a detection threshold had to be chosen to apply to detection spectra (Fig.~\ref{fig:ID3240_GOLD}). We chose a threshold of N = 40, meaning that a pixel is assigned as containing \Lya\ emission if it has been at least detected in 40 out of the 100 generated spectra (explained in detail in Sect.~\ref{sec:32}). It is an empirical value based on the data used in this work.

\begin{figure*}[h!]
\begin{minipage}{0.7\textwidth}
\centering
    \resizebox{\hsize}{!}{\includegraphics{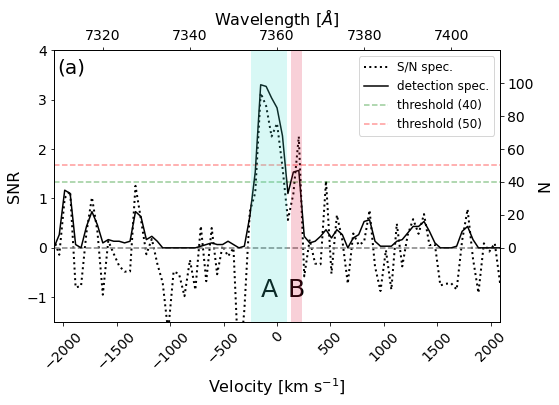}}
\end{minipage}
\begin{minipage}{0.3\textwidth}    
     \centering
         \centering
             \includegraphics[width=\textwidth]{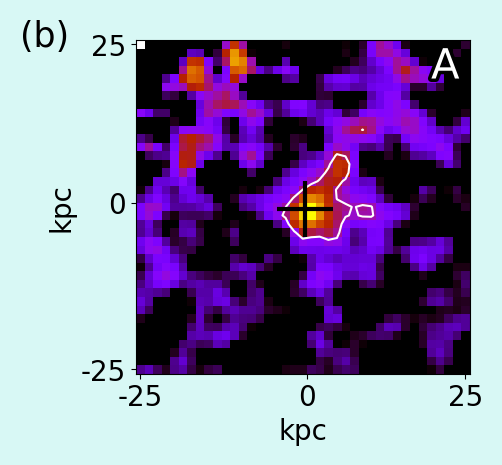}
     \hfill
         \centering
         \includegraphics[width=\textwidth]{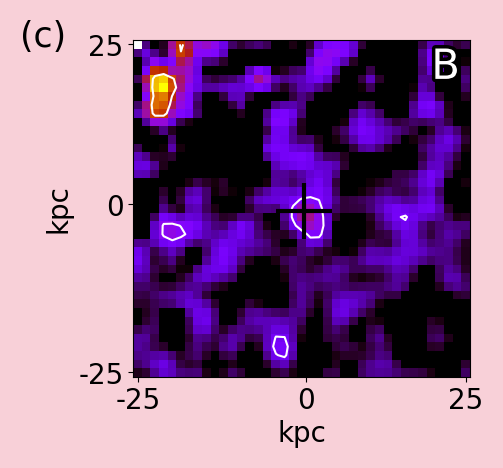}
\end{minipage}
\begin{minipage}{0.5\textwidth}    
     \centering
         \centering
             \includegraphics[width=\textwidth]{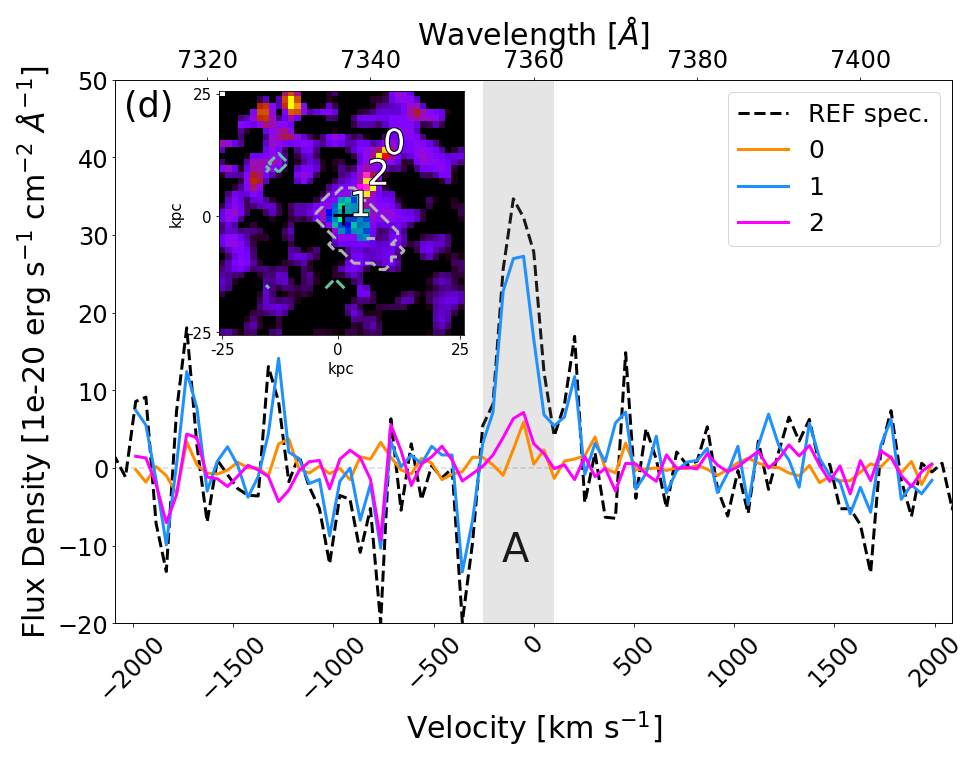}
\end{minipage}
\begin{minipage}{0.5\textwidth}    
     \centering
         \centering
             \includegraphics[width=\textwidth]{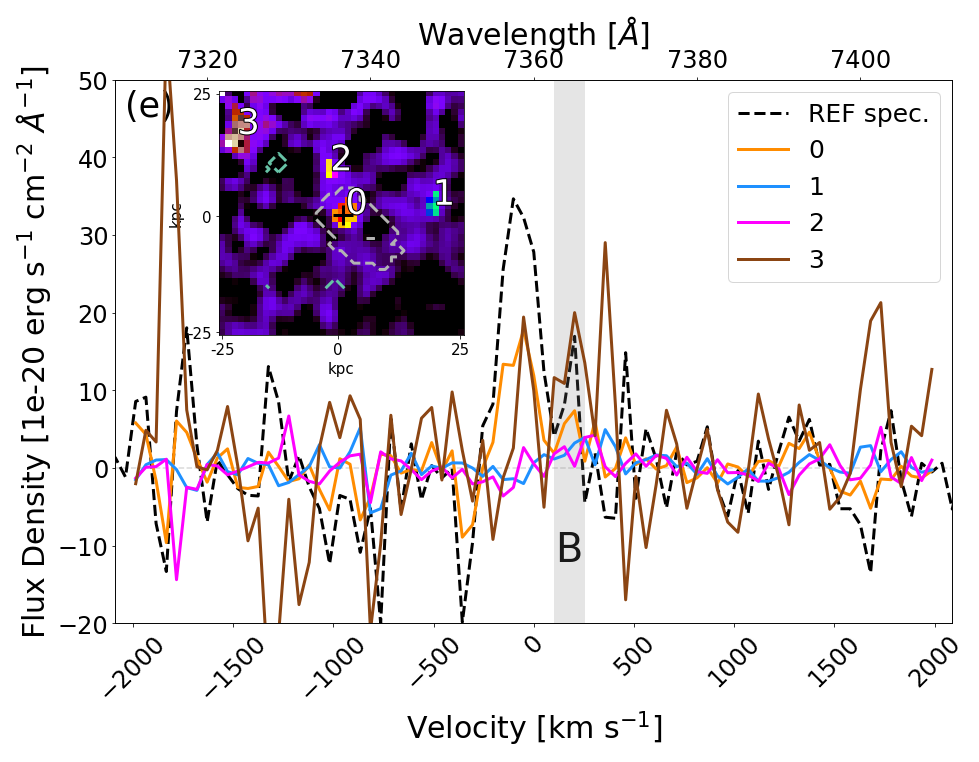}
\end{minipage}
    \caption{ID 7299, double-peak, \gold\ category. Same as Fig.~\ref{fig:ID3240_GOLD}. A threshold of 50 (red dashed line in panel \textit{(a)}) discards peak B while panel \textit{(e)} shows a significant detection in the NB image (labeled 0) as well as an extracted spectrum in orange having the same shape as the reference spectrum.}
   \label{fig:ID7299_threshold50}          
\end{figure*}

To test this threshold, we also performed the classification method with a threshold of N = 30 and N = 50. The corresponding classification results, before and after the NB image verification, are given in Table~\ref{table:thresholds_spectra_before_photutils} and Table~\ref{table:thresholds_spectra_after_photutils}, respectively. 
In Table~\ref{table:thresholds_spectra_before_photutils}, we notice strong differences between the fraction of the different line shapes. The number of single-peaks more than doubles between both extreme threshold values while the fraction of triple-peaks increases by five between a threshold of N = 50 and N = 30.
The NB image verification done in Sect.~\ref{sec:33} tends to homogenise the fractions between the thresholds, especially between N = 30 and N = 40, as seen in Table~\ref{table:thresholds_spectra_after_photutils}. 
It appears that a threshold of N = 30 selects noise peaks on the spectra, which is well seen in the fraction of triple-peaks and double-peaks of Table~\ref{table:thresholds_spectra_before_photutils} and the NB image verification performed cleans these noise peaks to reach fraction values similar to the ones of the N = 40 threshold.
On the contrary, a threshold of N = 50 is too restrictive and tends to discard small bumps, as shown in Fig.~\ref{fig:ID7299_threshold50}. Indeed, the \textit{area of signal} B would be discarded with a threshold of N = 50 (red dashed line in panel \textit{(a)} of Fig.~\ref{fig:ID7299_threshold50}). Specifically, the NB image corresponding to the \textit{area of signal} B displayed panel \textit{(e)} of Fig.~\ref{fig:ID7299_threshold50} shows \texttt{SourceFinder} detection at the centre with a significant orange extracted spectrum (see Sect.~\ref{sec:33}).

Finally, we measured spectral parameters on the \Lya\ lines with the different thresholds. With a threshold of N = 30, we tend to measure more flux but with lower S/N, which is expected as we increase the width of the \emph{areas of signal}. With a threshold of N = 50, the spectral measurements are very similar to the ones at N = 40. The values measured at N = 30 and N = 50 are inside the errorbars of the measurements done with the N = 40 threshold.
As a result of these tests, a detection threshold of N = 40 has been chosen.

\begin{table}[!h]
\caption{Comparison of the classification results with different threshold values (N = 30, 40 and 50) before performing the NB image verification described in Sect.~\ref{sec:33}.}         
\label{table:thresholds_spectra_before_photutils}     
\centering                                     
\begin{tabular}{l c c c c}         
\hline\hline                        
 & N = 30  & N = 40  & N = 50    \\    
\hline                                   
    No-peak & 4 (< 1\%) & 7 (1\%) & 19 (4\%) \\   
    Single-peak & 71 (14\%) & 155 (35\%) & 204 (43\%) \\ 
    Double-peaks & 298 (62\%) & 271 (55\%) & 235 (49\%) \\ 
    Triple-peaks & 104 (22\%) & 44 (9\%) & 19 (4\%) \\ 
\hline                                             
\end{tabular}
\end{table}

\begin{table}[!h]
\caption{Comparison of the classification results with different threshold values (N = 30, 40 and 50) after performing the NB image verification described in Sect.~\ref{sec:33}.}              
\label{table:thresholds_spectra_after_photutils}    
\centering                                      
\begin{tabular}{l c c c c}          
\hline\hline                        
 & N = 30  & N = 40  & N = 50    \\    
\hline                                   
    No-peak & 9 (2\%) & 9 (2\%) & 20 (4\%) \\ 
    Single-peak & 181 (38\%) & 198 (42\%) & 243 (51\%) \\ 
    Double-peaks & 250 (52\%) & 248 (51\%) & 204 (43\%) \\ 
    Triple-peaks & 37 (8\%) & 22 (5\%) & 10 (2\%) \\ 
\hline                                             
\end{tabular}
\end{table}

\textbf{Threshold determination on NB images}

In order to detect emission on NBs, a threshold of 2$\sigma$ on three connected pixels has been chosen (Sect.~\ref{sec:33}). This choice has been made for two reasons. 
The first reason is to be coherent with the criterion used to detect areas of signal on the spectrum (Sect.~\ref{sec:32}): we used a minimum S/N of 1 per pixel for at least two adjacent pixels, which is similar to only one pixel with a S/N of 2.  
The second reason why we did not choose more strict parameters to avoid false detections, like the ones in Fig.~\ref{fig:ID3240_GOLD} (detections labeled 0, 2 and 3), is to be able to detect compact and faint objects such as in the \bronze\ category. 
Using a threshold of 2$\sigma$ on the NB is thus a valid choice as we want to be consistent and inclusive.

We tested two different configurations on the parent sample: without the deblending function and with the default parameters of the deblending function\footnote{\url{https://photutils.readthedocs.io/en/stable/api/photutils.segmentation.SourceFinder.html\#photutils.segmentation.SourceFinder}} of \texttt{SourceFinder}.
While not using the deblending function, emission from other galaxies present on the NB image and close enough to our source is blended with the actual emission of our source. 
When looking at the same galaxies with the deblending function on, in this configuration, the different emissions are well differentiated. 
For this study, we choose to use the deblending function of \photutils.\texttt{SourceFinder}, which will differentiate sources with a 7.5 magnitude difference (contrast=0.001, default value).

\clearpage

\section{\silver\ Category}
\label{ap:B} 

Example of \silver\ category with two detections in the reference segmentation map for each peak.

\begin{figure*}[h!]
\begin{minipage}{0.7\textwidth}
\centering
    \resizebox{\hsize}{!}{\includegraphics{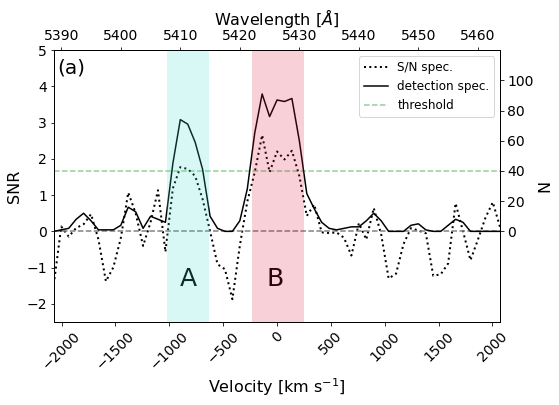}}
\end{minipage}
\begin{minipage}{0.3\textwidth}    
     \centering
         \centering
             \includegraphics[width=\textwidth]{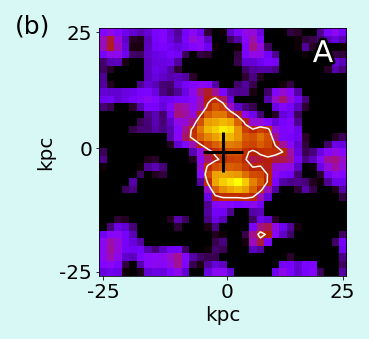}
     \hfill
         \centering
         \includegraphics[width=\textwidth]{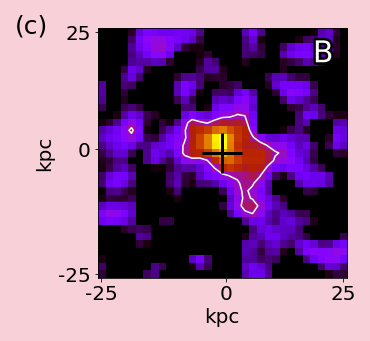}
\end{minipage}
\begin{minipage}{0.5\textwidth}    
     \centering
         \centering
             \includegraphics[width=\textwidth]{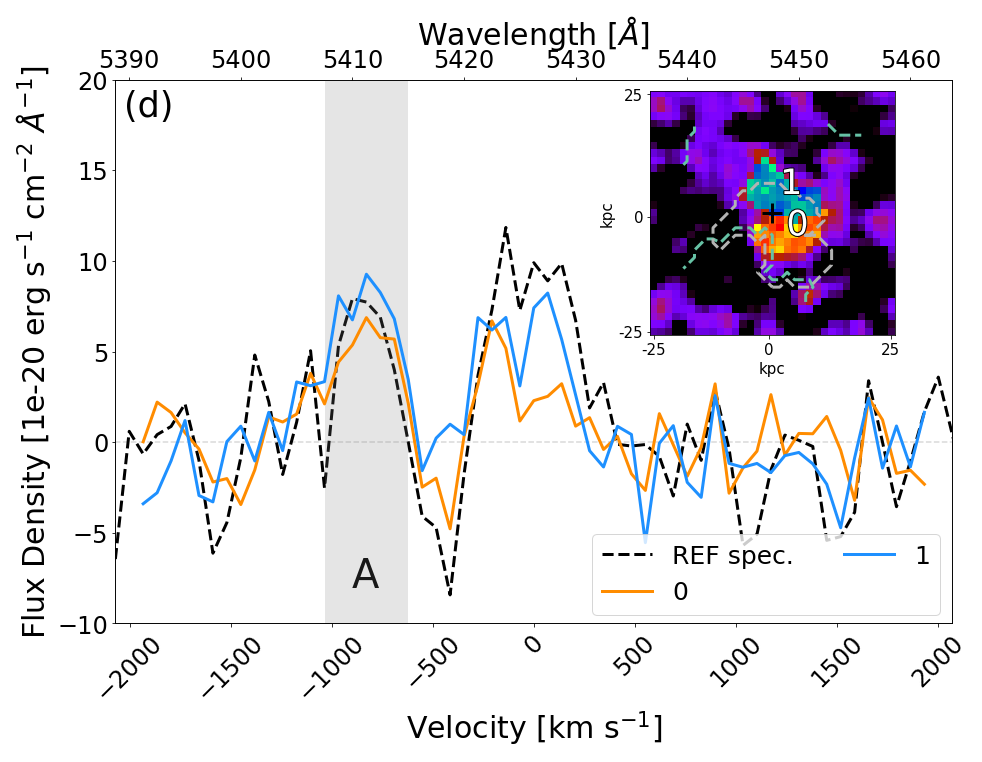}
\end{minipage}
\begin{minipage}{0.5\textwidth}    
     \centering
         \centering
             \includegraphics[width=\textwidth]{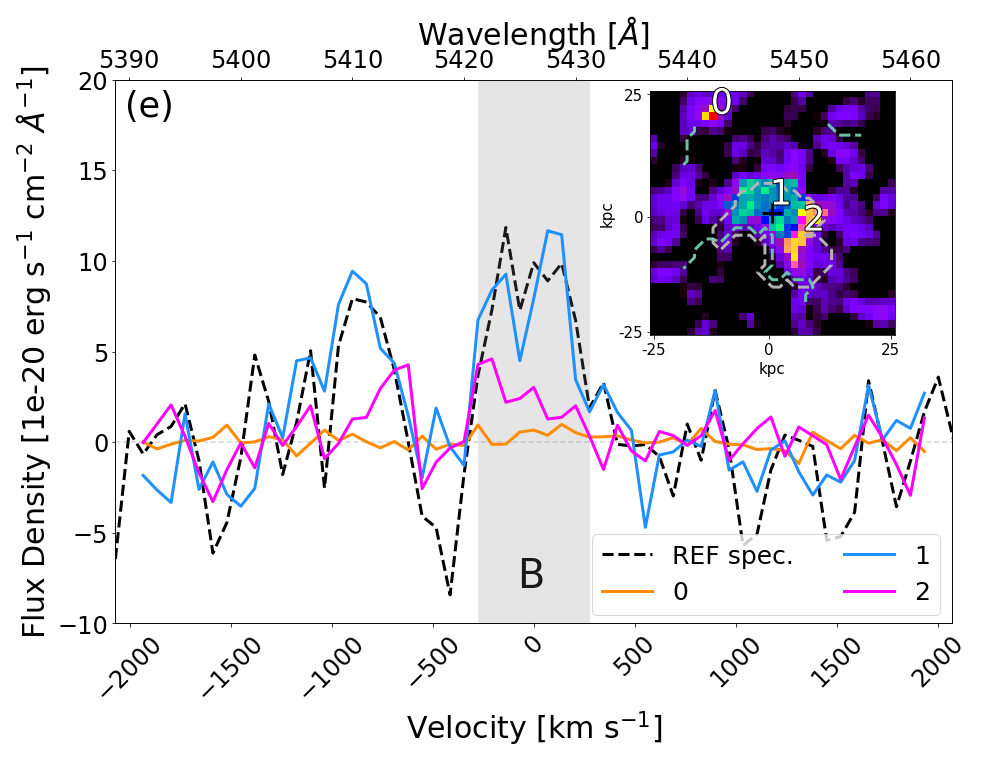}
\end{minipage}
    \caption{ID 163, double-peak, \silver\ category. Same as Fig.~\ref{fig:ID3240_GOLD}. We notice in panel \textit{(b)} two clumps that are detected by \texttt{SourceFinder} in panel \textit{(d)}. The two clumps have the same spectral shape.}
   \label{fig:ID163_SILVER}          
\end{figure*}

\clearpage

\begin{landscape}
\section{Tables}
\label{ap:C}

\begin{table}[h!]
\caption{First ten lines of the single-peaks table. The full table will be available online.}
\label{table:sample_1peak}
\centering
\begin{tabular}{ccccccccccccc}
\hline\hline
$\rm ID$ & z$_{\rm spec}$ &  RA & Dec & $\#$ of peaks & category & pair or triplet & F$_{\rm Ly\alpha}$ & log$_{\rm 10}$(L$_{\rm Ly\alpha}$) & S/N & FWHM \\
& & [deg] & [deg] & (0, 1, 2 or 3) & (G, S or B) & ID & [1e-20 erg s$^{-1}$ cm$^{-2}$ \AA$^{-1}$] & [erg s$^{-1}$] & & [km~s$^{-1}$] \\
\hline
$131$ & $4.04$ & $53.1557$ & $-27.7871$ & $1$ & G & -- & $6.41 \pm 2.23$ & $40.02 \pm 39.57$ & $2.96$ & $306.0 \pm 91.1$ \\
$149$ & $3.72$ & $53.1679$ & $-27.7786$ & $1$ & G & -- & $1553.04 \pm 16.82$ & $42.32 \pm 40.36$ & $91.91$ & $326.4 \pm 5.8$ \\
$279$ & $3.61$ & $53.1592$ & $-27.7841$ & $1$ & G & -- & $196.00 \pm 10.62$ & $41.39 \pm 40.13$ & $18.71$ & $267.7 \pm 40.3$ \\
$280$ & $4.44$ & $53.1577$ & $-27.7822$ & $1$ & G & -- & $60.87 \pm 6.69$ & $41.10 \pm 40.14$ & $9.09$ & $453.0 \pm 88.6$ \\
$306$ & $3.02$ & $53.1713$ & $-27.7807$ & $1$ & G & -- & $682.70 \pm 20.32$ & $41.75 \pm 40.23$ & $32.80$ & $383.0 \pm 37.2$ \\
$313$ & $5.14$ & $53.1709$ & $-27.7824$ & $1$ & G & -- & $418.65 \pm 5.48$ & $42.09 \pm 40.20$ & $74.64$ & $200.9 \pm 5.9$ \\
$338$ & $4.14$ & $53.1569$ & $-27.7862$ & $1$ & G & -- & $47.86 \pm 3.30$ & $40.92 \pm 39.76$ & $14.78$ & $240.0 \pm $ 41.5\\
$391$ & $3.71$ & $53.1618$ & $-27.7745$ & $1$ & G & -- & $98.49 \pm 15.09$ & $41.12 \pm 40.31$ & $6.39$ & $326.9 \pm 86.6$ \\
$430$ & $4.51$ & $53.1521$ & $-27.7813$ & $1$ & G & -- & $137.08 \pm 16.30$ & $41.47 \pm 40.54$ & $8.43$ & $447.3 \pm 73.5$ \\
$518$ & $5.06$ & $53.1685$ & $-27.7828$ & $1$ & G & -- & $178.82 \pm 7.17$ & $41.70 \pm 40.30$ & $26.19$ & $203.5 \pm 25.4$ \\
\hline
\end{tabular}
\tablefoot{
ID: identifier in \citetalias{bacon2023}, 
z$_{\rm spec}$: spectroscopic redshift in \citetalias{bacon2023}, 
RA and Dec: right ascension and declination \citepalias{bacon2023}, 
\# of peaks: 0 for no-peak, 1 for single-peak, 2 for double-peaks or 3 for triple-peaks, 
category: G for \gold, S for \silver\ or B for \bronze,
pair or triplet: ID of the associated galaxy(ies),
F$_{\rm Ly\alpha}$: total flux of the \Lya\ line given in 1e-20 erg s$^{-1}$ cm$^{-2}$ $\AA^{-1}$, 
log$_{\rm 10}$(L$_{\rm Ly\alpha}$): \Lya\ luminosity in erg s$^{-1}$, 
S/N: S/N of the \Lya\ line, 
FWHM: full width at half maximum of each peak of the \Lya\ line in km~s$^{-1}$.
}
\end{table}
\end{landscape}

\begin{landscape}
\begin{table}
\caption{First ten lines of the double-peaks table. The full table will be available online.}
\label{table:sample_2peaks}
\tiny
\centering
\begin{tabular}{ccccccccccccccc}
\hline\hline
$\rm ID$ & z$_{\rm spec}$ &  RA & Dec & $\#$ of peaks & category & pair or triplet & F$_{\rm Ly\alpha}$ & log$_{\rm 10}$(L$_{\rm Ly\alpha}$) & S/N & FWHM 1 & FWHM 2 & B/T & v$_{\rm sep}$\\
& & [deg] & [deg] & (0, 1, 2 or 3) & (G, S or B) & ID & [1e-20 erg s$^{-1}$ cm$^{-2}$ \AA$^{-1}$] & [erg s$^{-1}$] & & [km~s$^{-1}$] & [km~s$^{-1}$] & & [km~s$^{-1}$] \\
\hline
$68$ & $4.94$ & $53.1712$ & $-27.7785$ & $2$ & G & -- & $3877.79 \pm 23.06$ & $43.01 \pm 40.79$ & $166.02$ & $207.6 \pm 32.4$ & $363.3 \pm 15.4$ & $0.058 \pm 0.003$ & $467.1 \pm 21.1$ \\
$106$ & $3.28$ & $53.1637$ & $-27.7791$ & $2$ & G & -- & $1990.39 \pm 23.11$ & $42.30 \pm 40.37$ & $85.72$ & $432.0 \pm 11.7$ & $504.0 \pm 8.5$ & $0.400 \pm 0.006$ & $719.9 \pm 39.8$ \\
$118$ & $3.02$ & $53.1571$ & $-27.7803$ & $2$ & G & 828 & $397.74 \pm 20.10$ & $41.52 \pm 40.22$ & $19.32$ & $230.0 \pm 84.7$ & $460.0 \pm 56.7$ & $0.049 \pm 0.020$ & $536.7 \pm 45.4$ \\
$174$ & $2.99$ & $53.1600$ & $-27.7835$ & $2$ & G & -- & $297.70 \pm 13.94$ & $41.38 \pm 40.05$ & $20.75$ & $231.6 \pm 99.0$ & $463.1 \pm 44.3$ & $0.027 \pm 0.016$ & $540.3 \pm 45.23$ \\
$180$ & $3.46$ & $53.1640$ & $-27.7797$ & $2$ & G & -- & $1000.42 \pm 22.57$ & $42.06 \pm 40.41$ & $43.85$ & $345.6 124.6\pm $ & $483.9 \pm 23.2$ & $0.014 \pm 0.008$ & $622.1 \pm 55.88$ \\
$298$ & $4.21$ & $53.1649$ & $-27.7737$ & $2$ & G & -- & $563.41 \pm 20.71$ & $42.01 \pm 40.58$ & $28.30$ & $177.4 \pm 69.4$ & $236.6 \pm 5.6$ & $0.031 \pm 0.013$ & $354.9 \pm 24.70$ \\
$305$ & $3.04$ & $53.1691$ & $-27.7810$ & $2$ & G & 8364 & $288.19 \pm 16.16$ & $41.38 \pm 40.13$ & $18.20$ & $228.7 \pm 65.7$ & $304.9 \pm 20.9$ & $0.024 \pm 0.017$ & $609.8 \pm 7.97$ \\
$324$ & $3.00$ & $53.1710$ & $-27.7812$ & $2$ & G & -- & $914.45 \pm 23.38$ & $41.87 \pm 40.28$ & $38.00$ & $385.6 \pm 76.9$ & $308.5 \pm 36.8$ & $0.066 \pm 0.013$ & $462.8 \pm 51.96$ \\
$357$ & $3.44$ & $53.1588$ & $-27.7751$ & $2$ & G & -- & $450.04 \pm 22.53$ & $41.70 \pm 40.40$ & $20.25$ & $278.0 \pm 97.1$ & $347.5 \pm 36.1$ & $0.060 \pm 0.023$ & $486.4 \pm 43.35$ \\
$364$ & $3.94$ & $53.1538$ & $-27.7792$ & $2$ & G & -- & $138.16 \pm 15.76$ & $41.33 \pm 40.39$ & $8.85$ & $311.8 \pm 72.7$ & $187.1 \pm 31.2$ & $0.930 \pm 0.047$ & $436.5 \pm 40.15$ \\
\hline
\end{tabular}
\tablefoot{
ID: identifier in \citetalias{bacon2023}, 
z$_{\rm spec}$: spectroscopic redshift in \citetalias{bacon2023}, 
RA and Dec: right ascension and declination \citepalias{bacon2023}, 
\# of peaks: 0 for no-peak, 1 for single-peak, 2 for double-peaks or 3 for triple-peaks, 
category: G for \gold, S for \silver\ or B for \bronze,
pair or triplet: ID of the associated galaxy(ies),
F$_{\rm Ly\alpha}$: total flux of the \Lya\ line given in 1e-20 erg s$^{-1}$ cm$^{-2}$ $\AA^{-1}$, 
log$_{\rm 10}$(L$_{\rm Ly\alpha}$): \Lya\ luminosity in erg s$^{-1}$, 
S/N: S/N of the \Lya\ line, 
FWHM: full width at half maximum of each peak of the \Lya\ line in km~s$^{-1}$,
B/T: blue-to-total flux ratio, 
v$_{\rm sep}$: peak separation in km~s$^{-1}$.
}
\end{table}
\end{landscape}

\begin{landscape}
\begin{table}
\caption{First ten lines of the triple-peaks table. The full table will be available online.}
\label{table:sample_3peaks}
\tiny
\centering
\begin{tabular}{cccccccccccccccccc}
\hline\hline
$\rm ID$ & z$_{\rm spec}$ &  RA & Dec & $\#$ of peaks & category & pair or triplet & F$_{\rm Ly\alpha}$ & log$_{\rm 10}$(L$_{\rm Ly\alpha}$) & S/N & FWHM 1 & FWHM 2 & FWHM 3 \\
& & [deg] & [deg] & (0, 1, 2 or 3) & (G, S or B) & ID & [1e-20 erg s$^{-1}$ cm$^{-2}$ $\AA^{-1}$] & [erg s$^{-1}$] & & [km~s$^{-1}$] & [km~s$^{-1}$] & [km~s$^{-1}$] \\
\hline
$148$ & $3.07$ & $53.1676$ & $-27.7747$ & $3$ & G & -- & $546.29 \pm 31.11$ & $41.67 \pm 40.43$ & $17.24$ & $303.0 \pm 57.3$ & $454.6 \pm 50.1$ & $227.3 \pm 38.8$ \\
$510$ & $3.34$ & $53.1542$ & $-27.7876$ & $3$ & G & -- & $110.32 \pm 21.22$ & $41.06 \pm 40.35$ & $5.36$ & $284.4 \pm 60.1$ & $284.4 \pm 54.6$ & $142.2 \pm 33.2$ \\
$736$ & $4.31$ & $53.1537$ & $-27.7813$ & $3$ & G & -- & $145.91 \pm 18.82$ & $41.45 \pm 40.56$ & $7.86$ & $232.4 \pm 46.5$ & $116.2 \pm 38.8$ & $290.5 \pm 65.2$ \\
$6473$ & $4.55$ & $53.1671$ & $-27.7881$ & $3$ & G & -- & $93.84 \pm 7.24$ & $41.31 \pm 40.20$ & $12.89$ & $222.1 \pm 57.0$ & $222.1 \pm 46.3$ & $111.1 \pm 26.7$ \\
$6696$ & $4.21$ & $53.1659$ & $-27.7847$ & $3$ & G & -- & $385.09 \pm 13.94$ & $41.85 \pm 40.41$ & $26.53$ & $177.4 \pm 29.2$ & $295.6 \pm 25.4$ & $236.5 \pm 47.6$ \\
$7089$ & $4.16$ & $53.1581$ & $-27.7914$ & $3$ & G & -- & $1026.95 \pm 8.84$ & $42.26 \pm 40.19$ & $117.49$ & $239.1 \pm 24.1$ & $358.7 \pm 11.7$ & $119.6 \pm 89.9$ \\
$7167$ & $3.71$ & $53.1742$ & $-27.7900$ & $3$ & G & -- & $990.53 \pm 34.96$ & $42.13 \pm 40.67$ & $27.52$ & $196.4 \pm 70.2$ & $261.8 \pm 44.2$ & $327.3 \pm 31.5$ \\
$7664$ & $3.92$ & $53.1675$ & $-27.7769$ & $3$ & G & -- & $70.69 \pm 11.11$ & $41.04 \pm 40.23$ & $6.36$ & $188.0 \pm 47.2$ & $250.6 \pm 85.1$ & $125.3 \pm 38.2$ \\
$7676$ & $4.77$ & $53.1538$ & $-27.7850$ & $3$ & G & -- & $116.15 \pm 7-09$ & $41.45 \pm 40.24$ & $16.79$ & $160.2 \pm 26.7$ & $480.6 \pm 84.8$ & $106.8 \pm 136.4$ \\
$7847$ & $4.47$ & $53.1733$ & $-27.7843$ & $3$ & G & -- & $71.29 \pm 6.81$ & $41.18 \pm 40.16$ & $10.45$ & $169.0 \pm 39.2$ & $281.7 \pm 52.5$ & $169.0 \pm 35.1$ \\
\hline
\end{tabular}
\tablefoot{
ID: identifier in \citetalias{bacon2023}, 
z$_{\rm spec}$: spectroscopic redshift in \citetalias{bacon2023}, 
RA and Dec: right ascension and declination \citepalias{bacon2023}, 
\# of peaks: 0 for no-peak, 1 for single-peak, 2 for double-peaks or 3 for triple-peaks, 
category: G for \gold, S for \silver\ or B for \bronze,
pair or triplet: ID of the associated galaxy(ies),
F$_{\rm Ly\alpha}$: total flux of the \Lya\ line given in 1e-20 erg s$^{-1}$ cm$^{-2}$ $\AA^{-1}$, 
log$_{\rm 10}$(L$_{\rm Ly\alpha}$): \Lya\ luminosity in erg s$^{-1}$, 
S/N: S/N of the \Lya\ line, 
FWHM: full width at half maximum of each peak of the \Lya\ line in km~s$^{-1}$.
}
\end{table}
\end{landscape}

\clearpage

\section{Spectra}
\label{ap:D}

\begin{figure}[!h]
\centering
   \resizebox{0.9\hsize}{!}{\includegraphics{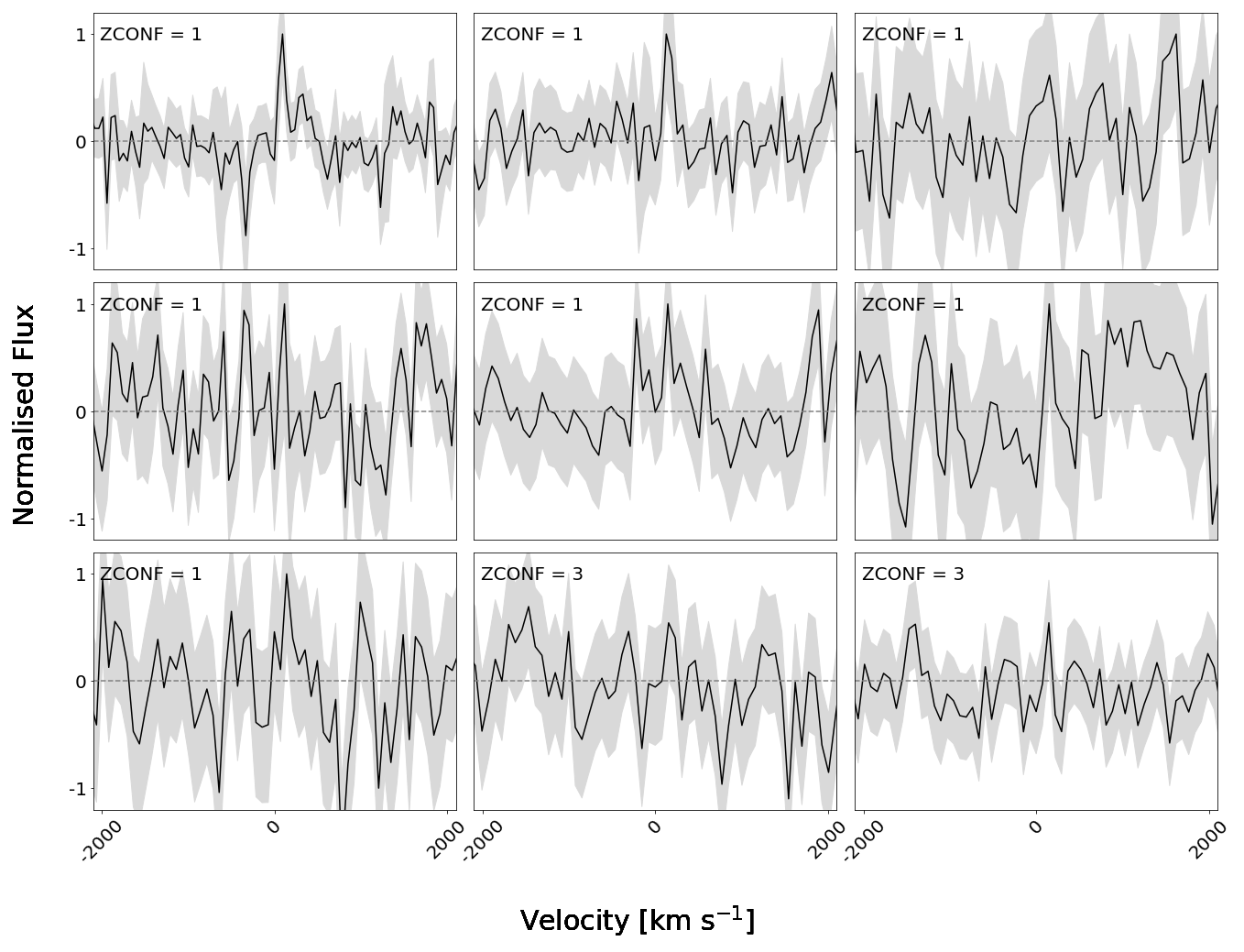}}
    \caption{Spectra of the sources classified as No-peak. The redshift confidence level is mentioned on each spectra. The two sources with a ZCONF of 3 are \Lya\ absorbers (ID 103 and ID 8537). Otherwise ZCONF = 1.}
    \label{fig:parent_sample_0peak}
\end{figure}

The method described in Sect.~\ref{sec:3} has classified a total of 9 objects as No-peak. The spectra of these sources are presented in Fig.~\ref{fig:parent_sample_0peak}. \\

\begin{landscape}
\begin{figure}[!h]
\centering
   \resizebox{0.9\hsize}{!}{\includegraphics{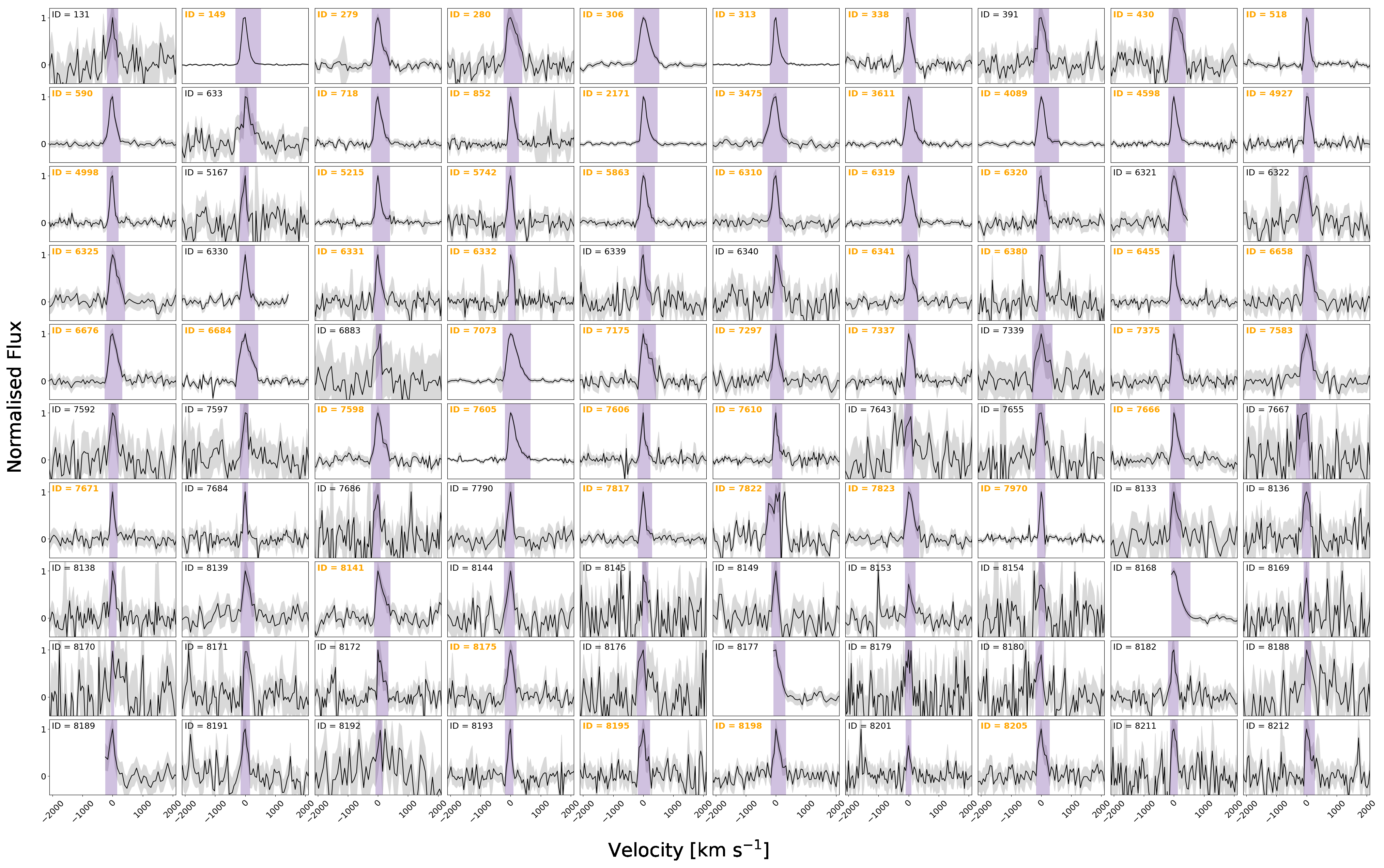}}
    \caption{Spectra of the single-peaked objects display as following the order of Table~\ref{table:sample_1peak}. IDs written in bold orange are part of the unbiased sample.}
    \label{fig:sp_sample_1}
\end{figure}

\begin{figure}[!h]
\centering
   \resizebox{0.9\hsize}{!}{\includegraphics{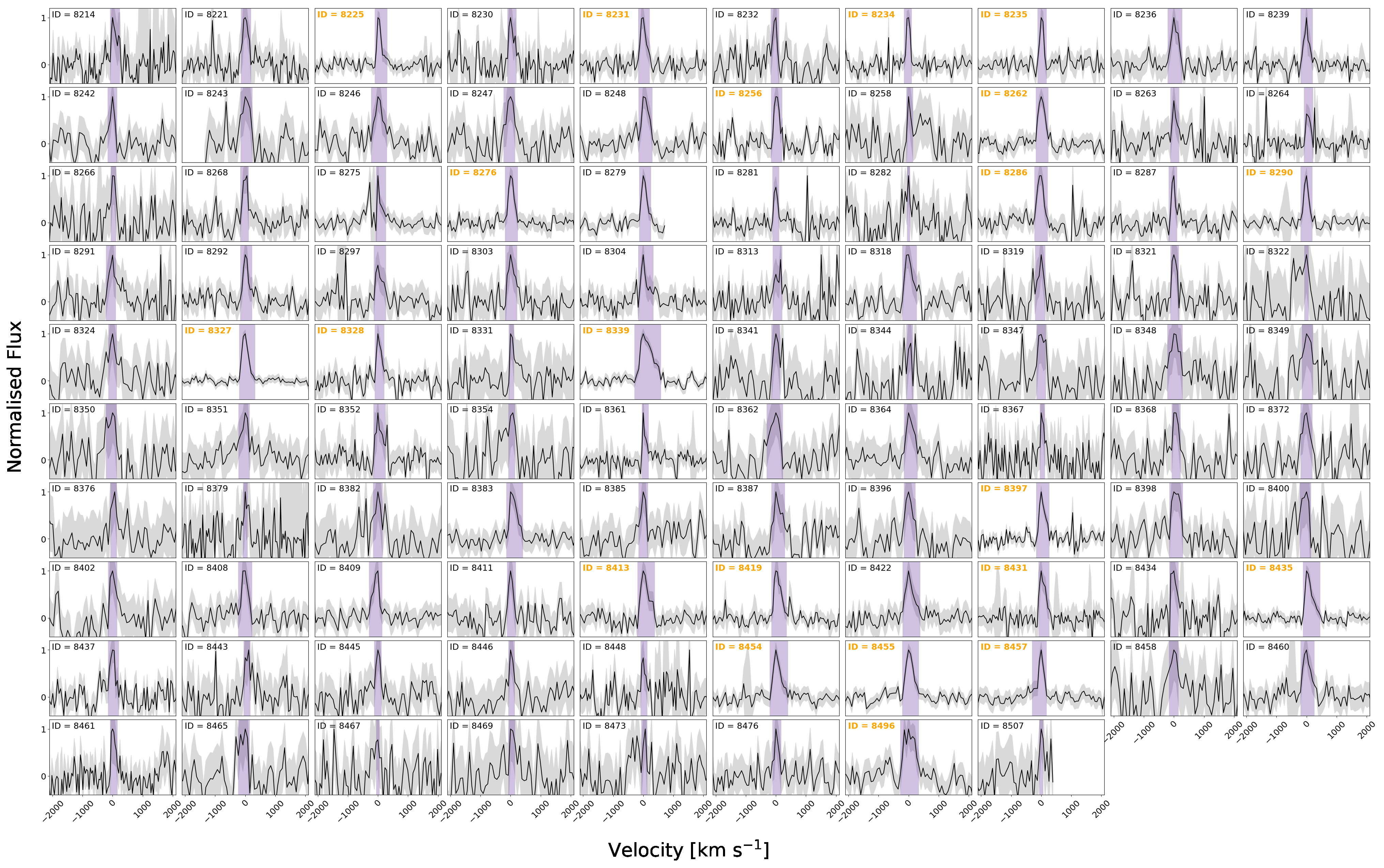}}
    \caption{continued.}
    \label{fig:sp_sample_2}
\end{figure}
\end{landscape}

\begin{landscape}
\begin{figure}[!h]
\centering
   \resizebox{0.9\hsize}{!}{\includegraphics{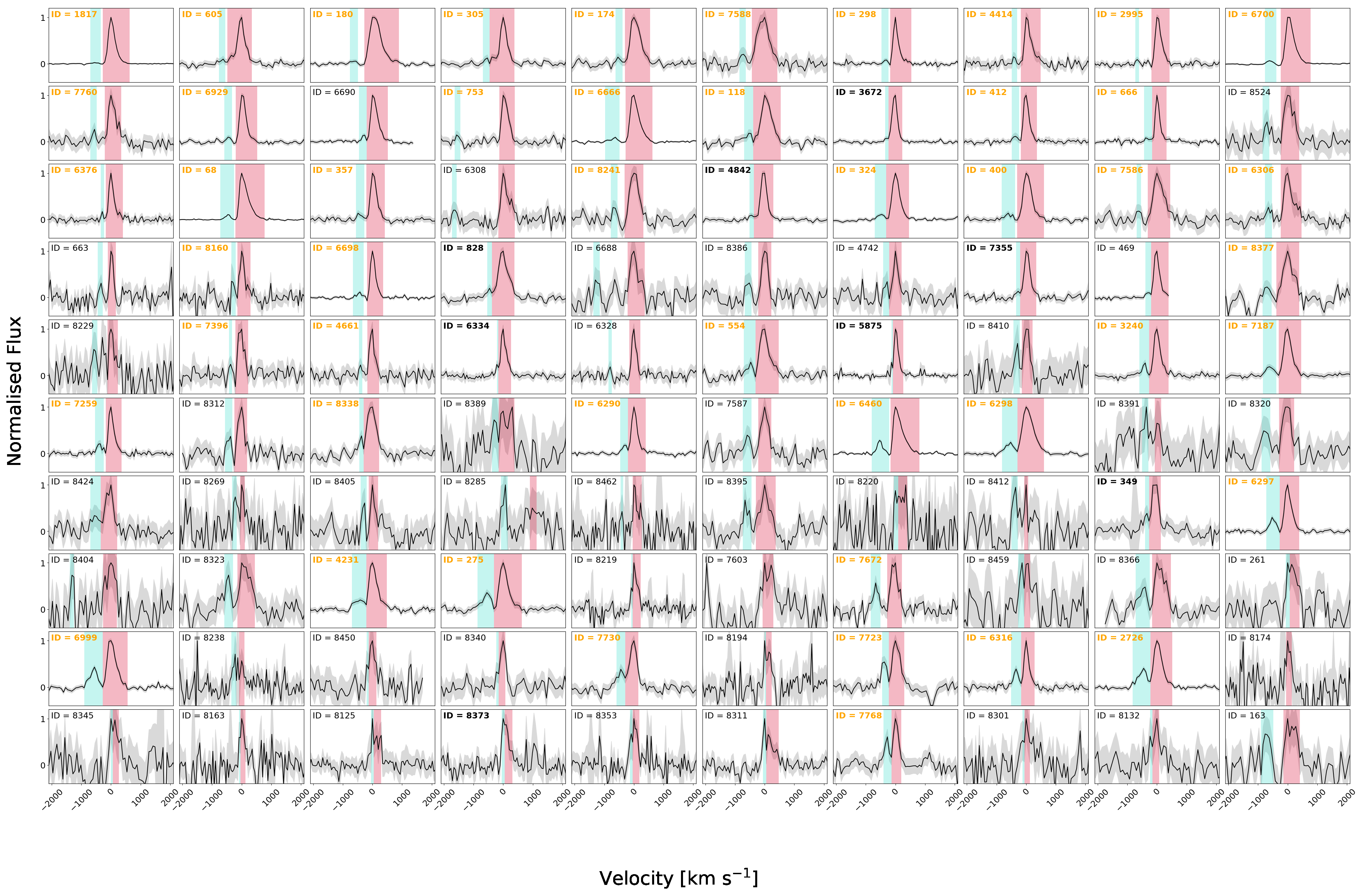}}
    \caption{Spectra of the double-peaked objects ordered by ascending B/T flux ratio. IDs written in bold black are part of the inclusive unbiased double-peak sample and IDs written in bold orange are part of both the inclusive and restrictive unbiased double-peak samples.}
    \label{fig:dp_sample_1}
\end{figure}

\begin{figure}[!h]
\centering
   \resizebox{0.9\hsize}{!}{\includegraphics{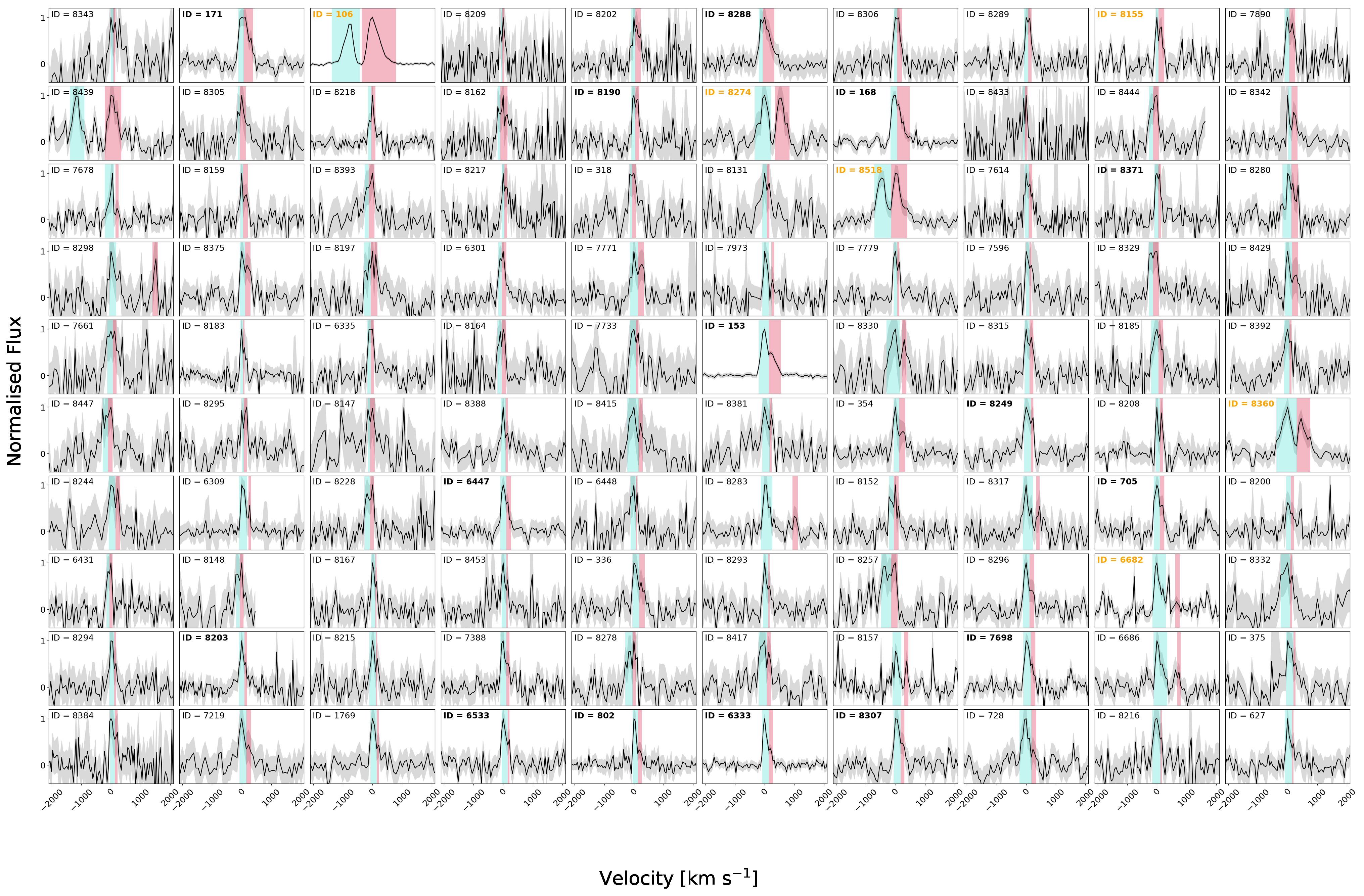}}
    \caption{continued.}
    \label{fig:dp_sample_2}
\end{figure}

\begin{figure}[!h]
\centering
   \resizebox{0.9\hsize}{!}{\includegraphics{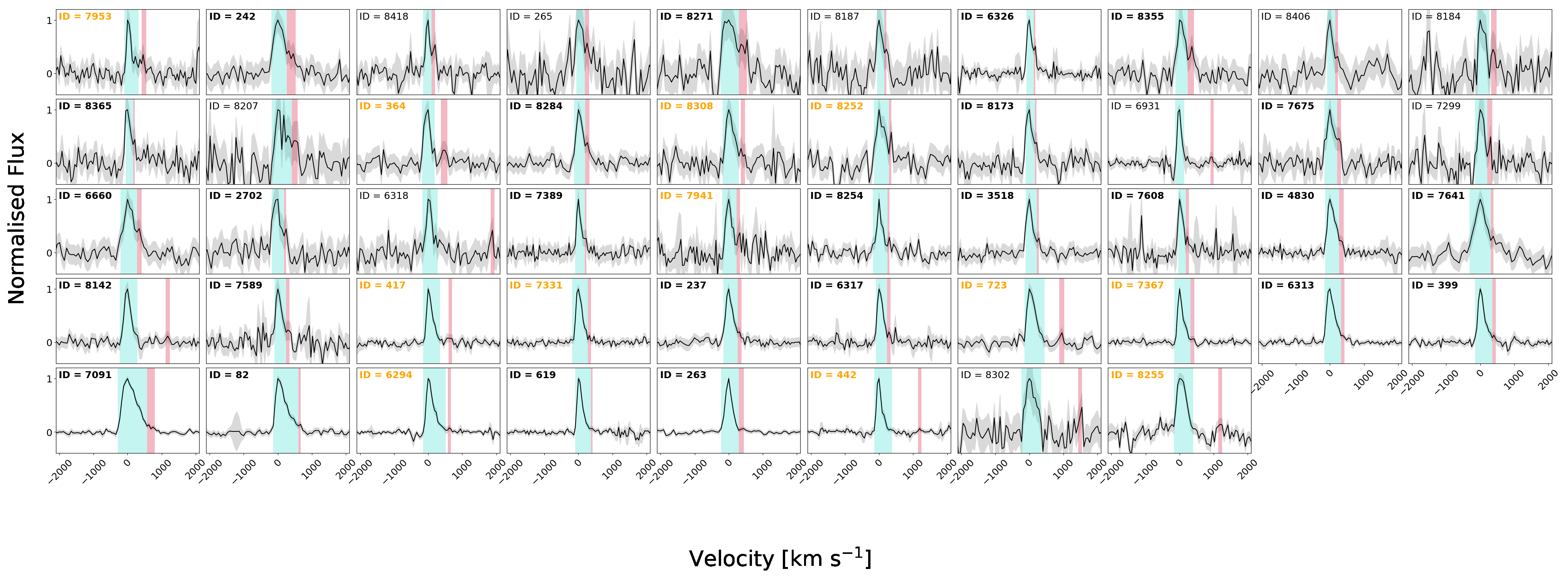}}
    \caption{continued.}
    \label{fig:dp_sample_3}
\end{figure}
\end{landscape}

\begin{landscape}
\begin{figure}[!h]
\centering
   \resizebox{\hsize}{!}{\includegraphics{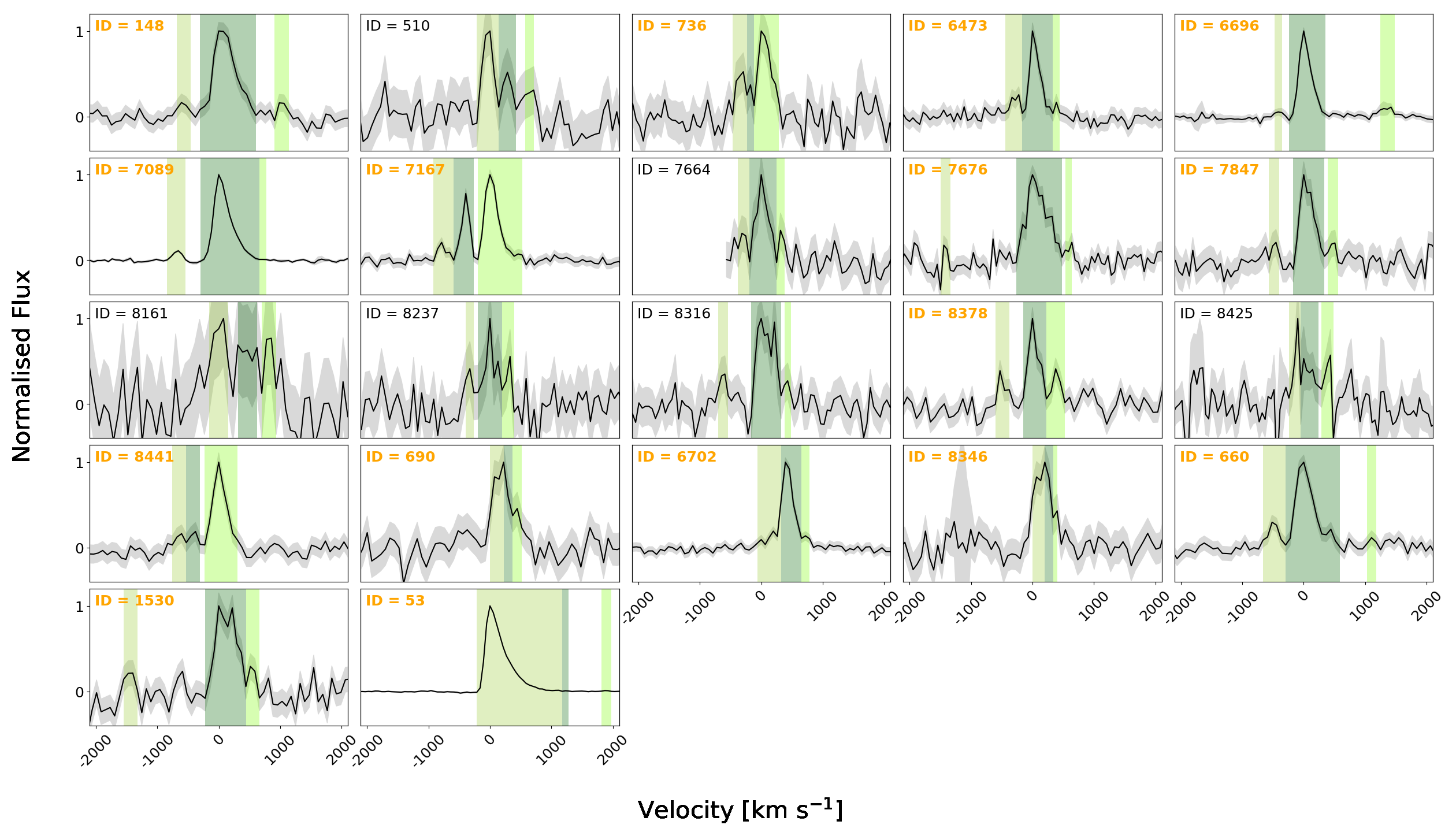}}
    \caption{Spectra of the triple-peaked objects with the respective 3 \Lya\ peaks in different coloured shaded areas. Ordered following Table~\ref{table:sample_3peaks}. IDs written in bold orange are part of the unbiased sample.}
    \label{fig:parent_sample_mp}
\end{figure}
\end{landscape}

\clearpage

\section{Evolution of the fraction of double-peaks for the \gold\ and \silver\ galaxies in the unbiased sample}
\label{ap:E}

As our sample is divided by spatial category (\gold, \silver\ and \bronze) in addition to the spectral ones, it is interesting to see if the evolution of the fraction of double-peaks changes in function of the spatial category of the galaxies. In order to observe this evolution, we use the inclusive unbiased double-peak sample $UDP_{I}$ which contains more double-peaks (i.e. more statistics) than the restrictive unbiased double-peak sample $UDP_{R}$. It is important to note here that the inclusive unbiased double-peak sample $UDP_{I}$ only contains \gold\ and \silver\ galaxies. We refer the reader to Sect.~\ref{sec:413} for more details.

\begin{figure}[!h]
\centering
    \resizebox{0.6\hsize}{!}{\includegraphics{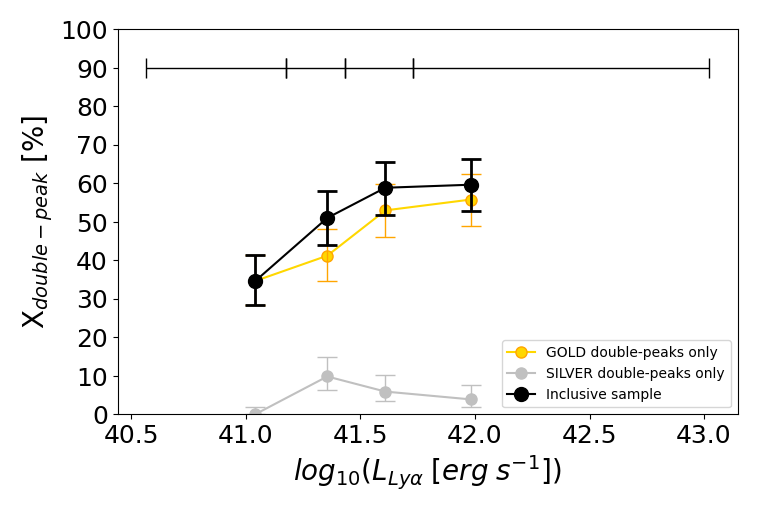}}
    \caption{Fraction of double-peaked LAEs plotted against the logarithmic \Lya\ luminosity. The unbiased sample $U$ has been divided into 4 luminosity bins with the same number of objects (51 or 52). The fraction of double-peaks has been derived in each bin. The results are positioned at the median \Lya\ luminosity of each bin. The black dots represents the total fraction of double-peaks ($N^{UDP_I} / N^{U}$). The $UDP_{I}$ is composed of \gold\ and \silver\ double-peaks. The yellow dots show the fraction of \gold\ double-peaks ($N^{UDP_I \, GOLD} / N^{U}$) among $U$. The fraction of \silver\ double-peaks ($N^{UDP_I \, SILVER} / N^{U}$) is represented by the grey dots. The horizontal black line at the top of the figure shows the size of each \Lya\ luminosity bin.}
    \label{fig:Xdp_per_LumBin}
\end{figure}

To investigate if the fraction of double-peaks of the different categories varies with luminosity, we apply the same procedure as in Sect.~\ref{sec:432}. We divide our unbiased sample into four \Lya\ luminosity bins with the same number of objects. Figure~\ref{fig:Xdp_per_LumBin} shows the fraction of double-peaks for each of the four luminosity bins for $U$ and for the \gold\ and \silver\ categories (in yellow and grey in the figure, respectively). 
The fraction of \gold\ double-peaks evolves from around 34\% for the faintest luminosities to 56\% for the brightest bin (41.7 < log(L$_{Ly\alpha}$ [erg s$^{-1}$]) < 43).
On the other hand, there is no double-peaked \silver\ objects at the faintest luminosities. Their fraction reaches almost 10\% for intermediate \Lya\ luminosities (41.2 < log(L$_{Ly\alpha}$ [erg s$^{-1}$]) < 41.7) before dropping to 4\% the higher values, due to the significant lack of objects in our parent sample at such luminosities. \silver\ double-peaks appear to not be faint objects but rather luminous ones.
The fraction of \gold\ double-peaks shows a similar trend as the fraction of the whole sample (black dots in Fig.~\ref{fig:Xdp_per_LumBin}), mainly explained by the high number of \gold\ galaxies populating the sample compared to the \silver\ ones.

\begin{figure}[!h]
\centering
   \resizebox{0.6\hsize}{!}{\includegraphics{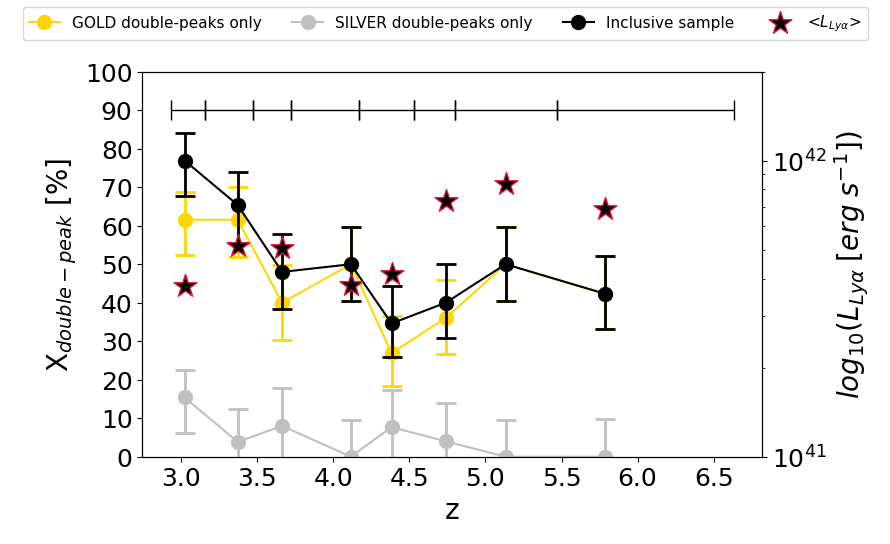}}
    \caption{Fraction of double-peaked LAEs plotted against the redshift. The unbiased sample $U$ has been divided into 8 redshift bins with the same number of objects (25 or 26). The fraction of double-peaks has been derived in each bin. The results are positioned at the median redshift of each bin. The black dots represents the total fraction of double-peaks ($N^{UDP_I} / N^{U}$). The $UDP_{I}$ is composed of \gold\ and \silver\ double-peaks. The yellow dots show the fraction of \gold\ double-peaks ($N^{UDP_I \, GOLD} / N^{U}$) among the unbiased sample $U$. The fraction of \silver\ double-peaks ($N^{UDP_I \, SILVER} / N^{U}$) is represented by the grey dots. The horizontal black line at the top of the figure shows the size of each redshift bin. The black stars surrounded in red represent the mean \Lya\ luminosity of each bin.}
    \label{fig:Xdp_vs_z}
\end{figure}

We also repeat the same procedure to the evolution of the double-peak fraction with redshift. 
Figure~\ref{fig:Xdp_vs_z} shows the evolution of the fraction of double-peaks with redshift for $U$ in black, the \gold\ double-peaks in yellow ($N^{UDP_I \, GOLD} / N^{U}$) and the \silver\ double-peaks in grey ($N^{UDP_I \, SILVER} / N^{U}$).
At low redshift ($z < 3.5$), we see an interesting increase in the double-peak fraction driven by the \silver\ sample. If this trend is confirmed with more data, it could indicate an intrinsic evolution of the LAE population towards cosmic noon, as \silver\ objects are composed of multiple clumps of \Lya\ emission. If such a possible evolution is confirmed, it could make less pertinent the use of the double-peak fraction to probe the opacity of the IGM at this redshift regime.
At redshift above 5, no more \silver\ objects are observed, the double-peak fraction being entirely driven by the \gold\ sample.

\clearpage

\section{Objects with systemic redshift}
\label{ap:F}

\begin{figure}[!h]
\centering
   \resizebox{0.8\hsize}{!}{\includegraphics{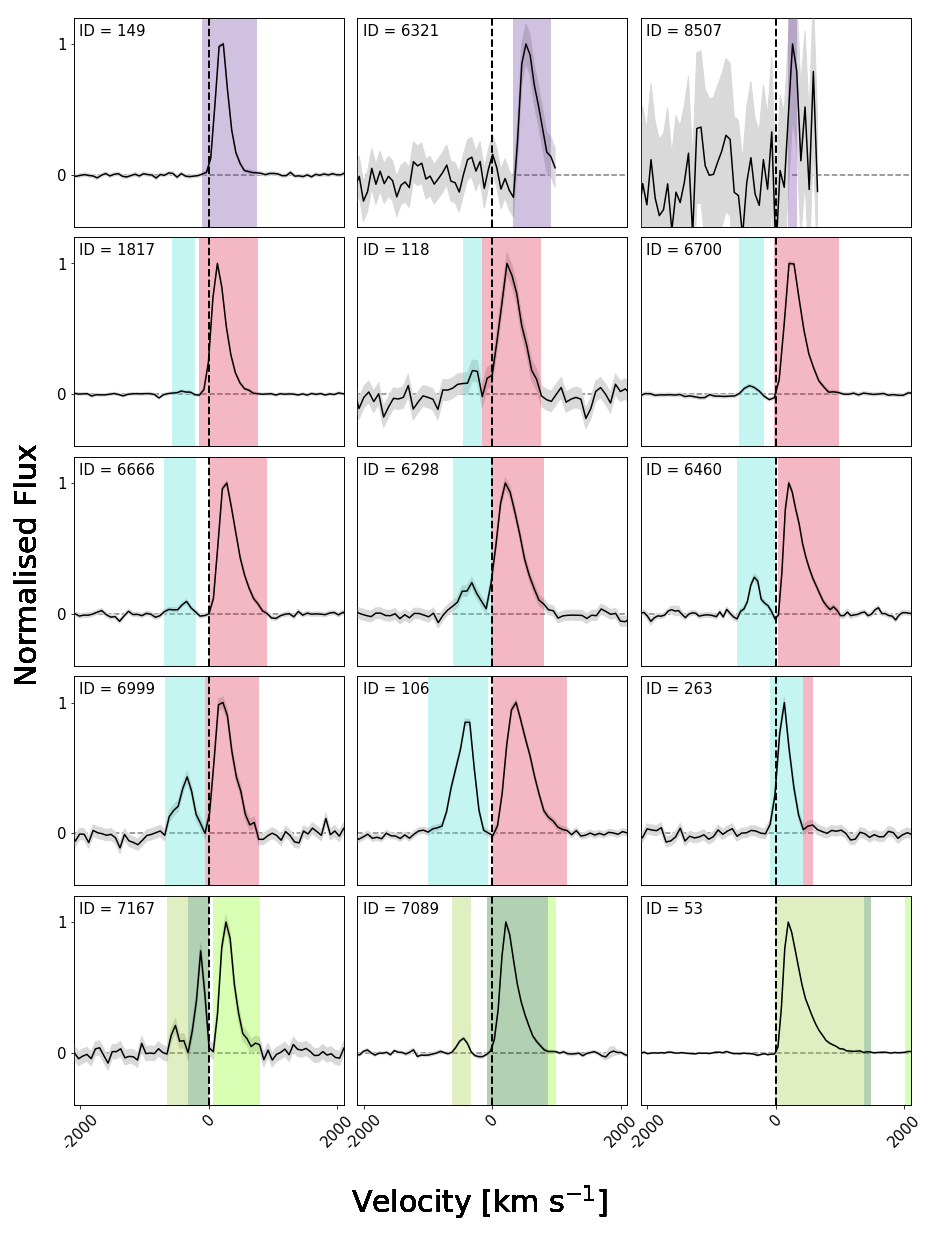}}
    \caption{Spectra of the sources having a systemic redshift. The shaded coloured areas designate the peaks of the \Lya\ line. The black vertical dashed line is the position of the systemic redshift. \textit{First row:} Single-peaks. \textit{Second, third and fourth rows:} Double-peaks. \textit{Last row:} Triple-peaks.}
    \label{fig:zsys_spectra}
\end{figure}

From nebular lines also observed in MUSE spectra, 15 galaxies from the parent sample have a secure systemic redshift. Among them, one galaxy has been studied in detail in \cite{matthee2022} and its systemic redshift has been well constrained (ID 53, last spectrum of Fig.~\ref{fig:zsys_spectra}). We refer the reader to Sect.~\ref{sec:53} for more details. \\

\end{appendix}

\end{document}